\documentclass[prb,aps,twocolumn,notitlepage]{revtex4-1}
\usepackage{amsmath}
\usepackage{amsfonts}
\usepackage{graphicx}
\usepackage{bm}
\usepackage[dvipsnames]{xcolor}

\newcommand{\bb}[1]{\mathbf{#1}}
\newcommand{\m}[1]{\mathcal{#1}}

\begin{document}

\title {Gauge fixing for strongly correlated electrons coupled to quantum light}
\author{Olesia Dmytruk and Marco Schir\'o}
\affiliation{JEIP, USR 3573 CNRS, Coll\`ege de France, PSL Research University, F-75321 Paris, France}
\date{\today}

\begin{abstract}
We discuss the problem of gauge fixing for strongly correlated electrons coupled to quantum light, described by projected low-energy models such as those obtained within tight-binding methods. Drawing from recent results in the field of quantum optics, we present a general approach to write down quantum light-matter Hamiltonian in either dipole or Coulomb gauge which are explicitly connected by a unitary transformation, thus ensuring gauge equivalence even after projection. 
The projected dipole gauge Hamiltonian features a linear light-matter coupling and an instantaneous self-interaction for the electrons, similar to the structure in the full continuum theory. On the other hand, in the Coulomb gauge the photon field enters in a highly non-linear way, through phase factors that dress the electronic degrees of freedom. We show that our approach generalises the well-known Peierls approximation, to which it reduces when only local, on-site orbital contributions to light-matter coupling are taken into account. As an application, we study a two-orbital model of interacting electrons coupled to a uniform cavity mode, recently studied in the context of excitonic superradiance and associated no-go theorems. Using both gauges we recover the absence of superradiant phase in the ground state and show that excitations on top of it, described by polariton modes, contain instead non-trivial light-matter entanglement. Our results highlight the importance of treating the non-linear light-matter interaction of the Coulomb gauge non-perturbatively, to obtain a well-defined ultrastrong coupling limit and to not spoil gauge equivalence.
\end{abstract}

\maketitle

\section{Introduction}

The experimental progress in coupling light and matter at the quantum level achieved with cavity and circuit Quantum Electrodynamics (CQED)~\cite{raimond2001manipulating,wallraff2004strong} has brought forth new platforms for many-body quantum optics where light and matter play equally important roles in collective quantum behavior. Examples include microcavity exciton-polaritons showing non-equilibrium superfluidity,~\cite{carusotto2013quantum} arrays of coupled CQED cavities hosting correlated states of light~\cite{houck2012chip,lehur2016many,fitzpatrick2017observation,ma2019dissipatively} or ultra-cold atoms embedded in high-finesse cavities allowing one to explore the competition between Mott physics and Dicke superradiance.~\cite{klinder2015observation,landig2016quantum,roux2020strongly}

An exciting new frontier is to take advantage of the quantum nature of light in solid state experiments by coupling quantum materials to fluctuating dynamical cavity fields. First experiments have recently appeared, involving two-dimensional electron gases,~\cite{todorov2010ultrastrong,scalari2012ultrastrong,zhang2016collective} van der Waals materials,~\cite{slootsky2014room,liu2015strong,basov2016polaritons} organic semiconductors,~\cite{orgiu2015conductivity} magnetic materials~\cite{mergenthaler2017strong} and, very recently, conventional and High-Temperature superconductors.~\cite{thomas2019exploring} As a result, many theoretical proposals have recently been put forward,  to dress, cool and control selected collective excitations of solids,~\cite{laplace2016proposed,kiffner2019mott} to enhance transport~\cite{schachenmayer2015cavity,hagenmuller2018cavity} or to induce or enhance superconductivity~\cite{schlawin2019cavity,sentef2018cavity,sheikhan2019cavity,curtis2019cavity,allocca2019cavity,li2020manipulating} or ferroelectricity~\cite{ashida2020quantum} by coupling to cavity photons. 
Finally, the phenomenon of Dicke superradiance was predicted in a number of platforms, including spin-Hall insulator coupled to circularly polarized quantized electromagnetic field~\cite{gulacsi2015floquet} and excitonic insulator~\cite{mazza2019superradiant}. Those ground-state realizations of the Dicke superradiance raise a number of conceptual questions, and even in a much simpler context of two-level systems the phenomenon remains elusive and controversial~\cite{nataf2010no,viehmann2011superradiant,andolina2019cavity}. Very recently, evidence for electronic superradiance beyond the no-go theorem has been demonstrated in presence of a spatially-varying electromagnetic field.~\cite{nataf2019rashba,guerci2020superradiant,andolina2020theory}

A fundamental issue for theoretical modeling of those platforms is to write down an Hamiltonian that complies with the guiding principle of gauge invariance, which puts a number of constraints on the form of light-matter interaction and on certain physical properties of the system. The gauge freedom allows one to express light-matter interactions in terms of a scalar and vector potential, as in  the Coulomb gauge often used in the solid-state context, or in terms of displacement and magnetic field through the Power-Zienau-Woolley (PZW) transformation and leading to the dipole gauge relevant in atomic Cavity Quantum ElectroDynamics (CQED) when magnetic interactions are negligible. While the first-principle discussion of gauge invariance in condensed matter system coupled to light is textbook material, several practical and conceptual questions emerge when one tries to write down effective low-energy models describing a subset of degrees of freedom after projecting out irrelevant ones, while preserving gauge invariance.  In this context, the choice of the gauge, so called gauge fixing, becomes crucial. A recent work has addressed this issue in the context of tight-binding models for strongly correlated electrons and demonstrated, for two model systems, that while both gauges converge to the same result when sufficiently many bands are included, at fixed truncation different gauges lead to different results, with the dipole gauge being more accurate.~\cite{li2020electromagnetic} 
 Similar results have been obtained for fundamental models of CQED, such as the Rabi or Dicke models describing one or multiple two-level systems coupled to a single cavity mode, where a breakdown of gauge invariance has been reported~\cite{de2018cavity,de2018breakdown,stokes2019gauge} in the regime of ultrastrong light-matter coupling~\cite{ciuti2005quantum,frisk2019ultrastrong,forndiaz2019ultrastrong,settineri2020gauge}. 
A unitary transformation has been proposed to effectively decouple light and matter degrees of freedom at ultrastrong coupling, thus alleviating the consequences of projecting onto a low-energy manifold~\cite{ashida2020cavity}.
 In the context of cavity-controlled chemistry~\cite{schafer2020relevance} it has been emphasized the importance of ab-initio approaches preserving the gauge invariance of the full microscopic theory. The resolution of these gauge ambiguities has been recently demonstrated for Rabi and Dicke models, leading to a consistent strategy to write down a projected quantum light-matter Hamiltonian which preserves gauge equivalence.~\cite{de2018cavity,de2018breakdown,di2019resolution,garziano2020gauge}

Motivated by these latest developments, in this work, we reconsider the issue of gauge fixing for models describing the coupling between photonic modes and strongly-correlated electronic matter. Following the general idea of Refs.~\onlinecite{di2019resolution,garziano2020gauge}, we present a formalism that allows to write down the Hamiltonian of correlated electrons coupled to photons in the dipole and Coulomb gauges which remain fully equivalent, i.e. related by a unitary transformation, even after projection. We discuss the relation between our approach and the so called Peierls substitution, often used to describe light-matter coupling in tight binding models. We apply our formalism to a two-band model for excitonic insulator coupled to a uniform cavity mode, recently studied in the literature.~\cite{andolina2019cavity}  Using the dipole gauge we confirm the absence of superradiance beyond mean-field theory, in accordance with the recent no-go theorem. We also highlight how to recover such result within the Coulomb gauge where it is crucial to treat light-matter interaction non-perturbatively to all orders. Furthermore, we compute the excitation spectrum of the model, which differently from the ground state contains non-trivial light-matter entanglement in the form of polariton modes. We explicitly show that the polariton spectrum is the same within our projected dipole and Coulomb gauge, a further demonstration of gauge equivalence. 

The paper is organized as follows: In Section~\ref{sec:gaugeinvariant}, we review how to couple electronic many body systems to the electromagnetic field in the continuum field theory second-quantized framework, paying particular attention to the choice of the gauge. 
In Section~\ref{sec:projectedbasis}, we introduce a projected electronic basis in the spirit of tight-binding models for strongly correlated systems, and present a general framework to write down quantum-light matter Hamiltonian which preserves gauge equivalence even after projection. In Section~\ref{sec:twobandmodel}, we provide some examples of our construction in the case of single and two-band models.
In Section~\ref{section:application}, we study in detail the resulting two-band model respectively in the dipole and Coulomb gauge and discuss its polariton spectrum. Section~\ref{sec:conclusions} is devoted to conclusions.

\section{Coupling  Quantum Matter and Light in the Continuum } \label{sec:gaugeinvariant}

We consider a quantum many-body systems of interacting electrons with mass $m$ in presence of a periodic potential $V(\bb{r})$ provided by the ions of the lattice. In the following, we set units such that $\hbar=c=1$.  Within the second quantization, we can write down the Hamiltonian of the system as  $\mathcal{H}_{el} = \m{H}_0 + \m{H}_{ee}$, where the non-interacting part reads
\begin{align}
\m{H}_0 = \int d\mathbf{r}\, \psi^\dag(\bb{r})h_0(\bb{r})\psi(\bb{r}),
\label{eq:H0second0}
\end{align}
with
\begin{align}\label{eq:h0}
h_0(\bb{r}) = -\dfrac{\nabla^2}{2m} + V(\bb{r})
\end{align}
while the electron-electron interactions Hamiltonian in general form is given by
\begin{align}
\m{H}_{ee}= \int d\bb{r}\ d\bb{r'}  \psi^\dag(\bb{r})\psi^\dag(\bb{r'})U(\bb{r} - \bb{r'})\psi(\bb{r}')\psi(\bb{r}).
\label{eq:Helel0}
\end{align}
The electronic problem is invariant under a local phase transformation $\psi(\bb{r})\rightarrow e^{i\chi}\psi(\bb{r})$ and the associated (Noether) current reads
\begin{equation}
\bb{J}(\bb{r})=\frac{1}{m}\psi^{\dagger}(\bb{r})\left(-i\nabla\right)\psi(\bb{r})+\mbox{h.c.}
\end{equation}

Next, we derive the Hamiltonian that describes the quantum matter coupled to the electromagnetic field. We start by deriving the continuum light-matter Hamiltonian in the Coulomb gauge, i.e. we consider a purely transverse vector potential $\bb{A}(\bb{r})$ such that $\nabla\cdot \bb{A}=0$. The Hamiltonian of the electromagnetic field reads
\begin{equation}
\m{H}_{ph}=\int d\bb{r}\left[\bb{\Pi}^2+
\left(\nabla\times\bb{A}\right)^2\right],
\end{equation}
where we have introduced the conjugate field $\bb{\Pi}(\bb{r})$ associated to the transverse component of the electric displacement. In this work, we consider a single mode decomposition of the fields 
\begin{align}
\bb{A}(\bb{r}) = \bb{A}_0(\bb{r})(a+a^\dag),\\
\bb{\Pi}(\bb{r}) = i\bb{\Pi}_0(\bb{r})(a-a^\dag),
\end{align}
where $a^\dag$ ($a$) are photon creation (annihilation) operators satisfying $[a,a^{\dag}]=1$ while $\bb{A}_0(\bb{r}),\bb{\Pi}_0(\bb{r})$ are the mode functions. In terms of this single mode decomposition, the photon Hamiltonian reads
$$
H_{ph} = \omega_c a^\dag a,
$$
where $\omega_c$ is the mode frequency.

The light-matter interaction can be introduced via the minimal coupling scheme by replacing the momentum operator as 
\begin{equation}\label{eqn:minimalcoupling}
-i\nabla \rightarrow -i\nabla +e\bb{A}(\bb{r}),
\end{equation}
where $e>0$ is the elementary charge and $\bb{A}(\bb{r})$ is the vector potential of the electromagnetic field. Employing a minimal coupling scheme, the total light-matter Hamiltonian in the Coulomb gauge reads
\begin{align}
\m{H}_C= \int d\mathbf{r}\, \psi^\dag(\bb{r})h_c(\bb{r})\psi(\bb{r}) + \m{H}_{ee} + \m{H}_{ph},
\label{eq:H0p}
\end{align}
where
\begin{align}\label{eq:hc}
h_c(\bb{r}) = \dfrac{\left(-i\nabla +e\bb{A}\right)^2}{2m} + V(\bb{r}),
\end{align}
which by construction satisfies gauge invariance. In fact, we can perform a transformation on the electronic and electromagnetic fields
\begin{align}
&\psi(\bb{r})\rightarrow e^{i\Lambda(\bb{r})}\psi(\bb{r}),\\
&\bb{A}(\bb{r})\rightarrow \bb{A}(\bb{r})-\frac{1}{e}\nabla\Lambda(\bb{r}),
\end{align}
which leaves invariant Eq.~(\ref{eq:H0p}).

 We note that the minimal coupling replacement Eq.~\eqref{eqn:minimalcoupling} could be also implemented by performing a transformation on the matter degrees of freedom \emph{only}~\cite{di2019resolution}. This is not surprising: the standard way to convert a field theory, which has a certain global symmetry (in our case $U(1)$ due to charge conservation), into a gauge theory is to promote the symmetry to a local one. This naturally leads to fluctuating gauge fields minimally coupled to the matter. In the present case, we can therefore define the unitary operator
\begin{align}
\m{U}\left(\chi\right) = \exp\left[i\left(a+a^{\dagger}\right)\int d\bb{r} \ \psi^\dag(\bb{r}) \chi(\bb{r}) \psi(\bb{r})\right],
\label{eq:Ucontinuum}
\end{align}
which transforms the electronic field as $\m{U}^{\dag}\psi(\bb{r})\m{U}=e^{i(a+a^\dag)\chi(\bb{r})}\psi(\bb{r})$. Applying this transformation to the electronic Hamiltonian and choosing 
\begin{equation}\label{eq:chi_def}
\nabla\chi(\mathbf{r})=e\bb{A}_0(\bb{r})\,,
\end{equation}
we obtain the minimal coupling Hamiltonian, Eq.~(\ref{eq:H0p}), i.e. we have
\begin{align}
\m{H}_C &= \m{H}_{ph} + \m{U}^{\dag}\left(\m{H}_{0} + \m{H}_{ee} \right) \m{U}\notag\\
&=\m{H}_{ph} + \m{H}_{ee} +\m{U}^{\dag}\m{H}_{0}\m{U}.
\label{eq:HCunitary}
\end{align}
We note that, in the last step, we have used the fact that the electron-electron interactions Hamiltonian remains invariant under a global or local phase rotation, $\m{H}_{ee} = \m{U}^{\dagger}\m{H}_{ee} \m{U}$ (see Appendix~\ref{transformeeinteractions} for the details of the derivation).

In the Coulomb gauge, the continuum Hamiltonian of the coupled electron-photon system, Eq.~(\ref{eq:H0p}) has a linear term in the vector potential and a quadratic one, called the diamagnetic term, obtained by expanding $h_c$ in Eq.~(\ref{eq:hc}). The physical current operator that corresponds to the Hamiltonian $\mathcal{H}_C$, Eq.~\eqref{eq:H0p}, can be defined as
\begin{align}
J_A(\bb{r}) = -e\dfrac{\delta \m{H}_C}{\delta \bb{A}(\bb{r})}=-e\bb{J}(\bb{r})-\frac{e^2}{m}\psi^\dag(\bb{r})\psi(\bb{r})\bb{A}(\bb{r}),
\end{align}
 and has also two contributions: the usual paramagnetic and the diamagnetic one. Conservation of the electron charge imposes a  constraint on the paramagnetic and diamagnetic coefficients, such that the physical current-current correlation function vanishes in the static limit ($\omega=0$, $\textbf{q}\rightarrow0$). This also implies the electronic Thomas-Reiche-Kuhn (TRK) sum rule~\cite{reiche1925uber,kuhn1925unter}, recently extended to strongly coupled light-matter quantum optical systems~\cite{savasta2020trk}.

A different choice of gauge can be performed which explicitly eliminates the quadratic term in the vector potential. This is implemented through a unitary transformation on the entire system, as we are going to discuss next.


As mentioned in the introduction, it is possible to write down an equivalent formulation of electrodynamics and light-matter interaction which does not rely on the vector potential $\bb{A}(\bb{r})\sim (a+a^\dag)$ but uses its conjugate moment, the displacement field $\bb{\Pi}(\bb{r})\sim i(a-a^\dag)$, as fundamental degree of freedom. This so called dipole gauge Hamiltonian can be obtained by performing a unitary transformation of the PZW type on the entire system Hamiltonian Eq.(\ref{eq:HCunitary}), i.e.
\begin{equation}
\m{H}_D = \m{T}^\dag \m{H}_C \m{T},
\end{equation} 
where $\m{T}$ is defined as
\begin{equation}
\m{T} =\exp\left[-i\left(a+a^{\dagger}\right)\int d\bb{r} \ \psi^\dag(\bb{r}) V_{\perp}(\bb{r}) \psi(\bb{r})\right].
\end{equation}
Following Ref.~\onlinecite{cottet2015electron}, we have introduced a photonic pseudopotential 
\begin{equation}\label{eq:Vperp}
V_{\perp}(\bb{r})=e\int_{\gamma} \bb{A}_0(\bb{r'})\cdot d\bb{r'},
\end{equation}
where $\gamma$ is a path ending in $\bb{r}$. Within the electric dipole approximation, we can write $\nabla V_{\perp}\simeq e\bb{A}_0(\bb{r})$, i.e. disregard the magnetic contribution coming from the flux of $\nabla\times \bb{A}_0$ such that we can pose $V_{\perp}(\bb{r})=\chi(\bb{r})$
and therefore identify~\cite{di2019resolution}
$$
\mathcal{T}=\mathcal{U}^{\dagger},\;\;\qquad
\mathcal{T}^{\dagger}=\mathcal{U}.
$$
Therefore the dipole gauge Hamiltonian can be equivalently obtained by applying the inverse unitary transformation of Eq.(\ref{eq:Ucontinuum}) to the photon system only,  i.e,
\begin{equation}\label{eq:dipolegaugecontinuum1}
\m{H}_D = \m{T}^\dag \m{H}_C \m{T}  =\m{U}\m{H}_{ph}\m{U}^{\dag} +  \m{H}_{0} + \m{H}_{ee},
\end{equation} 
where in the second equation we have used Eq.~\eqref{eq:HCunitary}. The result for the Hamiltonian in the Dipole Gauge reads
\begin{align}
\m{H}_D &=  \m{H}_{ph} + i  \omega_c (a- a^\dag )\int d\bb{r} \psi^\dag(\bb{r}) \chi(\bb{r}) \psi(\bb{r})\notag\\
&+\omega_c\left(\int d\bb{r} \psi^\dag(\bb{r}) \chi(\bb{r}) \psi(\bb{r})\right)^2 + \m{H}_{0} + \m{H}_{ee},
\label{eq:dipolegaugecontinuum}
\end{align}
where we have used the fact that under the action of $\mathcal{T}$ the photon field transforms as
\begin{align}\label{eq:Tphoton}
\m{T}^{\dag}\,a\,\m{T}=\m{U}\,a\,\m{U}^{\dag}=a-i\int d\bb{r}\psi^\dag(\bb{r}) \chi(\bb{r}) \psi(\bb{r}).
\end{align}
We see that in the dipole gauge the photon field couples to the matter only linearly, through the other quadrature of the field corresponding to the displacement, but the price to pay is the presence of a self-interaction term for the matter fields which is also due to the photon. We note Eq.~(\ref{eq:dipolegaugecontinuum}) does not contain \emph{magnetic} couplings between electrons and photons, as a result of the electric dipole approximation done below Eq.~(\ref{eq:Vperp}). While in principle it is possible to add higher order corrections, namely magnetic dipole interactions, in the dipole gauge Hamiltonian Eq.~(\ref{eq:dipolegaugecontinuum}) we leave this for future work. The general form of the light-matter coupling Hamiltonian in the PZW (or multipolar) gauge that includes coupling between magnetic field $\mathbf{B}$ and magnetization $\mathbf{M}$ can be found in the literature, for example~\cite{babiker1983derivation}.

In the next section, we discuss how the structure of the Coulomb and dipole gauge Hamiltonian change when the electronic degrees of freedom are projected onto a restricted set of modes and how to enforce gauge equivalence between them.

\section{Gauge Invariant Light-Matter Coupling  in a Projected Electronic Basis}\label{sec:projectedbasis}

In the theoretical discussion of strongly correlated electron systems, one usually cannot deal with the full complexity of the solid but rather focuses on an effective model which deals with a restricted (typically low-energy) subset of degrees of freedom. 
For example, in many transition metal oxides, the electronic states of interest lie in relatively narrow bands which are to a good extent separated from the rest of the spectrum. The low-energy Hamiltonian can be obtained, at least formally, by integrating out the degrees of freedom corresponding to higher energy bands, or more formally by perfoming a unitary transformation which (perturbatively) decouples the low and high energy sectors, followed by a projection operator. The resulting projected models have the advantage of being more accessible to many-body approaches than the full continuum theory.  On the other hand, a highly non-trivial question is how to properly couple electromagnetic fields to these projected models in order to preserve gauge invariance. 

In fact, as it has been long known, projection to a restricted set of bands violates the fundamental commutation relation between position and momentum operator in the first quantization, $[\bb{r}_a,\bb{p}_b]=i\delta_{ab}$ and transforms a local potential depending only on position, such as $V(\bb{r})$ in Eq.(\ref{eq:h0}), into a non-local one depending on both position and momentum~\cite{starace1971length,bassani1977choice,girlanda1981twophoton,ismail2001coupling,boykin2001electromagnetic}. As emphasized recently~\cite{di2019resolution}, a straightforward projection of the Coulomb gauge Hamiltonian~(\ref{eq:H0p}) obtained through minimal coupling misses the contribution to light-matter interaction coming from this non-local potential. To overcome this problem it has been recently suggested to proceed differently~\cite{di2019resolution,garziano2020gauge}, namely first project the matter Hamiltonian and then perform the minimal coupling substitution through the action of the unitary transformation~(\ref{eq:Ucontinuum}), which is itself consistently projected onto the selected manifold of degrees of freedom.

In this section, we present this approach in detail for models of strongly correlated electrons coupled to quantum light. First, in Section~\ref{sec:project_el}, we write down the electronic Hamiltonian $\mathcal{H}_{el}$, introduced in Section~\ref{sec:gaugeinvariant}, in terms of a restricted subset of Wannier orbital. This takes the form of a tight-binding model plus local interactions, relevant for many strongly correlated electron systems. In Section~\ref{sec:project_unitary}, we write down the projected unitary transformation Eq.~(\ref{eq:Ucontinuum}) and discuss its action on the electronic and photonic degrees of freedom. Using these results we write down the quantum light-matter Hamiltonian in the Coulomb gauge (Section~\ref{sec:coulombgauge}) and discuss its relation with the so called Peierls substitution, often employed in the solid-state context to discuss the coupling of classical and quantum light to electrons within tight-binding models. In Section~\ref{sec:dipolegauge} we obtain the projected dipole gauge Hamiltonian and finally,  in Section~\ref{sec:gauge_equiv}, we prove explicitly the gauge equivalence between the projected dipole and Coulomb gauge Hamiltonian.

\subsection{Projected Electronic Hamiltonian}\label{sec:project_el}

We start by considering the electronic sector and project over a set of low energy states
\begin{align}
\Psi(\bb{r})=P\psi(\bb{r})P = \sum_{\bb{R}\mu}\phi_{\bb{R}\mu} (\bb{r})c_{\bb{R}\mu},
\label{eq:fermifield}
\end{align}
where $c_{\bb{R}\mu}$ ($c_{\bb{R}\mu}^\dag$) are the fermionic annihilation (creation) operators that satisfy canonical anticommutation relations. Here, as a basis set of single-particle wavefunctions we choose the Wannier functions $\phi_{\bb{R}\mu}(\bb{r})$ that are localized 
around a lattice site $\bb{R}$, and $\mu$ labels the orbital. In terms of these modes the projected electronic Hamiltonian reads
\begin{align}\label{eq:H0second}
H_{el} \equiv P \mathcal{H}_{el}P=\sum_{\bb{R},\bb{R'}} \sum_{\mu,\mu'}t^{\mu\mu'}_{\bb{R},\bb{R'}}
c^{\dagger}_{\bb{R}\mu}c_{\bb{R'}\mu'}+\notag\\
+\sum_{\bb{R}}\sum_{\mu_1\ldots\mu_4}
U^{\mu_1\mu_2\mu_3\mu_4}
c^{\dagger}_{\bb{R}\mu_1}
c^{\dagger}_{\bb{R}\mu_2}c_{\bb{R}\mu_3}c_{\bb{R}\mu_4}.
\end{align}
The parameters entering this Hamiltonian are defined in terms of expectation values over Wannier functions, respectively as
\begin{align}
t^{\mu\mu'}_{\bb{R},\bb{R'}} = \int d\mathbf{r}\, \phi^*(\bb{r})_{\bb{R}\mu}h_0(\bb{r})\phi(\bb{r})_{\bb{R'}\mu'},
\label{eq:tnmR1R2}
\end{align}
including both hopping (typically next-neighbors) and on-site energies, while for the interaction we consider only local (same site) terms so we obtain
\begin{align}
U^{\mu_1\mu_2\mu_3\mu_4}&= \int d\bb{r}\ d\bb{r'} \phi^*_{\bb{R}\mu_1}(\bb{r})\phi^*_{\bb{R}\mu_2}(\bb{r'})\times\notag
\\ &\times U(\bb{r} - \bb{r'})\phi_{\bb{R}\mu_3}(\bb{r})
\phi_{\bb{R}\mu_4}(\bb{r'}).
\label{eq:UniRi}
\end{align}
We note that in general there is a certain freedom in choosing the Wannier basis, which can be exploited for example to minimize the real-space extension of the functions $\phi_{\bb{R}\mu}(\bb{r})$ leading to the so called Maximally Localized Wannier functions~\cite{marzari2012maximally}, or to define orbitals with well-defined angular momentum character which usually leads to simplification in the evaluation of interaction matrix elements~\cite{lecherman2006dynamical,amadon2008plane}.
For the current discussion, we can omit these details and limit ourselves to the expansion in Eq.~(\ref{eq:fermifield}), leaving specific examples to Section~\ref{sec:twobandmodel}.

\subsection{Projected Unitary Transformation}\label{sec:project_unitary}

We now consider the unitary operators $\m{U}(\chi)$ and $\m{T}(\chi)$, introduced in Section~\ref{sec:gaugeinvariant} respectively to generate the Coulomb and dipole gauge Hamiltonian, and write them down in the projected subspace, in terms of projected degrees of freedom only. This quite generically reads
\begin{equation}
U\left(\chi\right)\equiv P \m{U} P=\exp\left(i\left(a+a^{\dagger}\right)\sum_{\bb{R}\bb{R'}}\sum_{\mu\mu'}c^{\dagger}_{\bb{R}\mu}\chi^{\mu\mu'}_{\bb{R}\bb{R'}}
c_{\bb{R'}\mu'}\right),
\label{eq:Omega1}
\end{equation}
where
\begin{equation}
\chi^{\mu\mu'}_{\bb{R}\bb{R'}}= \int d\mathbf{r}\phi^*_{\bb{R}\mu}(\mathbf{r})\chi(\mathbf{r}) \phi_{\bb{R'}\mu'}(\mathbf{r}).
\label{eq:chimumu'}
\end{equation}
is the matrix element of the local phase $\chi(\bb{r})$, directly related to the vector potential through Eq.~(\ref{eq:chi_def}), between Wannier states and satisfies $\left(\chi^{\mu\mu'}_{\bb{R}\bb{R'}}\right)^*=\chi^{\mu'\mu}_{\bb{R'}\bb{R}}$. It is useful to discuss the transformation of electronic operators under the action of $U\left(\chi\right)$. This reads
\begin{equation}\label{eqn:transform_field}
U^{\dagger}\left(\chi\right)c_{\bb{R}\mu}U\left(\chi\right)=\sum_{\bb{R}\mu'}\left(e^{i\left(a+a^{\dag}\right)\chi}\right)^{\mu\mu'}_{\bb{R}\bb{R'}}c_{\bb{R'}\mu'}.
\end{equation}
We see therefore that the unitary transformation entangles the electronic degrees of freedom with the photonic ones through generalised phase factors that have a non-trivial structure in real and orbital space. As we are going to discuss, these factors will appear in the Coulomb gauge Hamiltonian through Eq.~(\ref{eq:HCunitary}). 
Similarly, we obtain for the projected PZW transformation $T(\chi)\equiv P \m{T} P$
\begin{equation}
T\left(\chi\right)=\exp\left(-i\left(a+a^{\dagger}\right)\sum_{\bb{R}\bb{R'}}\sum_{\mu\mu'}c^{\dagger}_{\bb{R}\mu}\chi^{\mu\mu'}_{\bb{R}\bb{R'}}
c_{\bb{R'}\mu'}\right),
\end{equation}
which satisfies $T^{\dagger}(\chi)=U(\chi)$. The action of the unitary transformation on the photonic degree of freedom, needed to evaluate the Hamiltonian in the dipole gauge through Eq.~(\ref{eq:dipolegaugecontinuum}), reads therefore also in the projected case as a simple shift, see Eq.~(\ref{eq:Tphoton})
\begin{equation}\label{eqn:transform_field_ph}
U\left(\chi\right)a\,U^{\dagger}\left(\chi\right)=
a-i\sum_{\bb{R}\bb{R'}}\sum_{\mu\mu'}c^{\dagger}_{\bb{R}\mu}\chi^{\mu\mu'}_{\bb{R}\bb{R'}}
c_{\bb{R'}\mu'}.
\end{equation}
As we are going to discuss next, the different way in which photonic and electronic degrees of freedom are dressed by the projected unitary transform is at the origin of the radically different structure of light-matter interaction in the projected dipole and Coulomb gauge.

\subsection{Projected Hamiltonian in the Coulomb Gauge}\label{sec:coulombgauge}

We start discussing the construction of the projected Coulomb gauge Hamiltonian. As discussed before (see also Refs.~\onlinecite{di2019resolution,garziano2020gauge,savasta2020gauge}), this is obtained by applying the projected unitary operator, Eq.~\eqref{eq:Omega1}, to the projected electronic Hamiltonian $H_{el}$, Eq.~\eqref{eq:H0second}, i.e.
\begin{align}\label{eq:Hc_proj}
H_C = H_{ph} + U^\dag\left(\chi\right) H_{el} U\left(\chi\right).
\end{align}
Using the action of the unitary transformation on the fermionic operators, Eq.~\eqref{eqn:transform_field}, we can write
\begin{align}\label{eq:Coulombprojgeneral}
&H_C = H_{ph} + \sum_{\bb{R},\bb{R'}} \sum_{\mu,\mu'}\tilde{t}^{\mu\mu'}_{\bb{R},\bb{R'}}
c^{\dagger}_{\bb{R}\mu}c_{\bb{R'}\mu'}\notag\\
&+\sum_{\bb{R}_1\ldots \bb{R}_4}\sum_{\mu_1\ldots\mu_4}
\tilde{U}^{\mu_1\mu_2\mu_3\mu_4}_{\bb{R}_1\bb{R}_2\bb{R}_3\bb{R}_4}
c^{\dagger}_{\bb{R}\mu_1}
c^{\dagger}_{\bb{R}\mu_2}c_{\bb{R}\mu_3}c_{\bb{R}\mu_4}.
\end{align}
where the hopping and interaction parameters have been dressed as result of the unitary transformation and they now read respectively as
\begin{equation}
\tilde{t}^{\mu\mu'}_{\bb{R},\bb{R'}}=
\sum_{\bb{R}_1\bb{R}_2
\alpha_1,\alpha2}
\left(e^{-i\left(a+a^{\dag}\right)\chi}\right)^{\mu\alpha_1}_{\bb{R}\bb{R}_1}t^{\alpha_1\alpha_2}_{\bb{R}_1\bb{R}_2} \left(e^{i\left(a+a^{\dag}\right)\chi}\right)^{\alpha_2\mu'}
_{\bb{R}_2\bb{R'}}
\end{equation}
and a similar, yet more involved, expression for the interaction that we give in Appendix~\ref{transformeeinteractions} for completeness.

An important point is worth to be stressed concerning the final result of the projected Coulomb gauge Hamiltonian. In the continuum, the Coulomb gauge Hamiltonian contains the vector potential at most to quadratic order, see Eq.~(\ref{eq:H0p}). On the other hand, in Eq~(\ref{eq:Coulombprojgeneral}), the photon field enters in a highly non-linear way, through the phase factors that arise from the projected unitary transform $U(\chi)$. While it would be tempting to expand the Hamiltonian~(\ref{eq:Coulombprojgeneral}) to lowest orders and recover the conventional paramagnetic and diamagnetic contributions to light-matter interaction, as it is sometimes done in the literature in the context of the Peierls substitution, we will explicitly show later in this paper that this can lead to inconsistencies in the regime of ultrastrong light-matter coupling. A natural question at this point is how to connect our result for the projected Coulomb gauge Hamiltonian, Eq.~(\ref{eq:Coulombprojgeneral}), with what is usually obtained within the Peierls substitution, often used in the literature in the context of tight-binding models coupled to the electromagnetic field. We discuss this important issue in the next section.

\subsection{Comparison with Peierls Substitution}\label{sec:Peierls}

For tight-binding models the Peierls substitution is a standard approach to couple electronic degrees of freedom to light. This amounts to dress the hopping terms entering the electronic Hamiltonian $H_{el}$ in Eq.~(\ref{eq:H0second}) as
\begin{align}
t^{\mu\mu'}_{\bb{R},\bb{R'}} \rightarrow t^{\mu\mu'}_{\bb{R},\bb{R'}}e^{ie\int_{\bb{R}}^{\bb{R'}}d\bb{r} \cdot \bb{A}(\bb{r})}.  
\end{align}
We already see from the above expression that within this approach the vector potential only couples non-local hopping elements, i.e. intra-atomic orbital transitions are absent.

In order to see how the Peierls substitution emerges within our approach it is useful to go back to the projected unitary transformation in Eq.~(\ref{eq:Omega1}) and  expand $\chi(\mathbf{r})$ around a lattice site $\mathbf{R}$, assuming the electromagnetic field varies slowly on the scale of the lattice spacing (electric dipole approximation) to obtain
\begin{equation}
\chi^{\mu\mu'}_{\bb{R}\bb{R'}} = \chi(\mathbf{R})\delta_{\bb{R}\bb{R'}}\delta_{\mu\mu'}+\partial_{\mathbf{r}}
\chi \vert_{\bb{R}}\mathbf{L}_{\bb{R}\bb{R'}}^{\mu\mu'},
\end{equation}
where the connection coefficients are defined as
\begin{equation}
\mathbf{L}_{\bb{R}\bb{R'}}^{\mu\mu'}=\int d\mathbf{r}\phi^*_{\bb{R}\mu}(\mathbf{r})\left(\bb{r}-\bb{R}\right) \phi_{\bb{R'}\mu'}.
\end{equation}
One can readily see that the Peierls substitution is equivalent to setting the connection coefficients to zero~\cite{paul2003thermal}. Indeed we have in this case
\begin{equation}
U^{\dagger}c_{\bb{R}\mu}U=e^{i(a+a^\dag)\chi(\bb{R})} \,c_{\bb{R}\mu},
\label{eq:peierlsgauge}
\end{equation}
which gives rise to the well-known Peierls dressing of the hopping terms. In other words, the Peierls substitution is invariant under a \emph{restricted} gauge transformation, Eq.~(\ref{eq:peierlsgauge}), that ignores the connections~\cite{paul2003thermal}. 

In this respect, as we are going to discuss further in the next sections, our projected Coulomb gauge Hamiltonian does not assume any specific structure in orbital space for $\chi^{\mu\mu'}_{\bb{R}\bb{R'}}$ and it is able to account for non-trivial connection coefficients. Furthermore, since by construction the same function $\chi^{\mu\mu'}_{\bb{R}\bb{R'}}$ enters in the projected Coulomb and dipole gauge Hamiltonians in Eqs.~(\ref{eq:Coulombprojgeneral}) and (\ref{eq:HDprojected}), this guarantees gauge equivalence: a given choice on the structure of $\chi$ will immediately translate into a dipole and Coulomb Hamiltonian related by a unitary transformation.

We note that related issues with the Peierls substitution (or Peierls approximation) emerge in other contexts and are not specific to the quantum light-matter case. In fact, similar problems already emerge when trying to derive the appropriate second quantized current operator for a projected tight-binding model. Setting to zero the connection coefficients amounts to approximating the matrix elements of the momentum operator between Wannier states, which results in an expression for the Peierls current depending in general on the choice of Wannier basis (and in general on the local interaction for multi-orbital problems)~\cite{paul2003thermal}.
A related discussion appears in the context of calculations of optical conductivity, which depends on the momentum operator matrix element. In that context it is indeed well known that the Peierls substitution disregards local intra/inter-band processes, exactly those encoded by the connection coefficients, and that this can have effects on calculations of transport properties~\cite{millis2001optical,tomczak2009optical,wissgott2012dipole,schuler2021gauge}.  We notice that another issue with Peierls substitution and gauge invariance has been recently reported~\cite{misuse2020skolimowski}.

\subsection{Projected Hamiltonian in the Dipole Gauge}\label{sec:dipolegauge}

We now discuss the form of the projected Hamiltonian in the dipole gauge. To proceed we apply the projected unitary transformation $U\left(\chi\right)$ to the photonic Hamiltonian only, according to Eqs.(\ref{eq:dipolegaugecontinuum1}-\ref{eq:dipolegaugecontinuum}), i.e. 
\begin{align}\label{eq:Hd_proj}
H_D=U\left(\chi\right)H_{ph}U^{\dagger}\left(\chi\right)+H_{el}.
\end{align}
Using Eq.~(\ref{eqn:transform_field_ph}) we obtain
\begin{eqnarray}
H_D &=&H_{el} + i\omega_c \left(a-a^{\dagger}\right)
\sum_{\bb{R}\bb{R'}}\sum_{\mu\mu'}
c^{\dagger}_{\bb{R}\mu}\chi^{\mu\mu'}_{\bb{R}\bb{R'}}c_{\bb{R'}\mu'}\nonumber\\
&&+\omega_c\left(\sum_{\bb{R}\bb{R'}}\sum_{\mu\mu'}
c^{\dagger}_{\bb{R}\mu}\chi^{\mu\mu'}_{\bb{R}\bb{R'}}c_{\bb{R'}\mu'}\right)^2 +\omega_c a^{\dagger}a.
\label{eq:HDprojected}
\end{eqnarray}
As in the continuum formulation, we see that within the dipole gauge the light field couples linearly to the matter through the displacement, $\left(a-a^{\dagger}\right)$, rather than through the vector potential. Depending on the spatial dependence of $\chi(\bb{r})$ and the resulting structure in real and orbital space of $\chi^{\mu\mu'}_{\bb{R}\bb{R'}}$, the cavity photon can mediate shifts in the orbital energies, corresponding in the second term of Eq.~(\ref{eq:HDprojected}) to terms where $\bb{R}=\bb{R'}$ and $\mu=\mu'$, or dipole-like couplings between different orbitals (when $\bb{R}=\bb{R'}$ and $\mu\neq \mu'$) as well as photon-mediated hopping terms.
In addition, the cavity also gives rise to an instantaneous self-interaction term for the electronic sector.
As we are going to discuss in Section~\ref{section:application} in the context of a concrete model example, this term plays an important role in renormalizing the bare electronic interaction, an effect which is often called depolarisation shift~\cite{todorov2012intersubband}. As such this term cannot be dropped, especially in the ultrastrong light-matter coupling regime~\cite{schafer2020relevance}. It is worth stressing the difference between projecting directly Eq.~(\ref{eq:dipolegaugecontinuum}) in the continuum, which would have lead to a self-interaction term written as 
$$
P\left(\int d\bb{r} \psi^\dag(\bb{r}) \chi(\bb{r}) \psi(\bb{r})\right)^2P,
$$
and applying the projected unitary transformation, which leads to the square of the polarisation operator. Finally, we notice that the construction of a projected dipole gauge Hamiltonian has been discussed before, in the context of Mesoscopic Cavity QED~\cite{cottet2015electron} and multimode Cavity QED coupled to Quantum Materials ~\cite{li2020electromagnetic}, and that our results coincide with those presented in those works when a single mode of the cavity is retained.

\subsection{Gauge Equivalence of Projected Hamiltonians}\label{sec:gauge_equiv}

We conclude this section by discussing explicitly the gauge equivalence of the projected Coulomb and dipole gauge Hamiltonian that we have derived above. It is worth emphasizing that, as compared to the full continuum theory discussed in section~\ref{sec:gaugeinvariant}, such equivalence is not obvious a priori given the structure of the two projected Hamiltonians, Eq.~(\ref{eq:Coulombprojgeneral}) and Eq.~(\ref{eq:HDprojected}). Indeed while the projected dipole gauge retain a similar structure of light-matter coupling with respect to the continuum theory (namely a linear term and a self-interaction), the projected Coulomb gauge acquires a highly non-linear form, with the photon field entering to all orders. Despite this difference the two gauge formulations are  fully equivalent, i.e. they are related by a unitary transformation.  In fact, if we apply the projected unitary transformation $T^{\dagger}(\chi)$ to the Coulomb gauge Hamiltonian, Eq.~(\ref{eq:Hc_proj}-\ref{eq:Coulombprojgeneral}), and use the fact that $T^{\dagger}(\chi)=U(\chi)$ we recover the Hamiltonian in the dipole gauge, given by Eq.~\eqref{eq:HDprojected}, i.e.
\begin{equation}\label{eqn:gaugeequiv}
T^{\dag}\left(\chi\right) H_C T\left(\chi\right)=U\left(\chi\right) H_{ph} U^\dag\left(\chi\right) +H_{el}  \equiv H_D.
\end{equation}
As a result, the gauge equivalence is fully preserved in our formulation and calculations performed on the two models will yield the same answers for physical, gauge invariant, quantities, such as for example the energy spectrum. In addition, one can use the above strategy to compare predictions for gauge dependent operators, by applying the same unitary transformation also to the observable of interest.

We emphasize that in order for gauge equivalence to hold one needs massive cancellations on the left-hand side of Eq.~(\ref{eqn:gaugeequiv}), order by order in the light-matter coupling, since the right-hand side has only linear and quadratic (self-interaction) contributions. This suggests that the truncation of the projected Coulomb gauge Hamiltonian to lowest orders in the light-matter coupling has to be performed with care if one wants to preserve gauge equivalence. We will come back to this issue in Section~\ref{section:application} in the context of a specific two-orbital model.

\section{Examples} \label{sec:twobandmodel}

We will now provide two concrete examples to further clarify the general results obtained in the previous section. First we consider a single band Hubbard model, for which we demonstrate that our projected Coulomb Hamiltonian recovers the one obtained through the Peierls substitution. Then we move to a two-orbital problem, recently studied in the literature~\cite{mazza2019superradiant,andolina2019cavity,lenk2020collective}, where the non-trivial orbital structure of the unitary transform makes clear the importance of properly treating the connection coefficients in order to obtain a Coulomb gauge which is equivalent to the dipole one. We will discuss the physics of this model in detail in Section~\ref{section:application}.

\subsection{Single Band Hubbard Model}

For a single orbital Hubbard model the Hamiltonian reads
\begin{align}
H=-\sum_{\langle\bb{R}\bb{R'}\rangle}\sum_{\sigma}t_{\bb{R}\bb{R'}}\left(c^{\dagger}_{\bb{R}\sigma}c_{\bb{R'}\sigma}+hc\right)+U\sum_{\bb{R}}
n_{\bb{R}\uparrow}
n_{\bb{R}\downarrow}
\end{align}

In this case, the connections coefficients are identically zero and the Peierls substitution is correct. In fact we can write

\begin{equation}
\chi^{\sigma\sigma'}_{\bb{R}\bb{R'}}=\delta_{\sigma\sigma'}\int d\mathbf{r}\phi^*_{\bb{R}}(\mathbf{r})\chi(\mathbf{r}) \phi_{\bb{R'}}(\mathbf{r})\simeq \delta_{\sigma\sigma'}\delta_{\bb{R}\bb{R'}}\chi(\bb{R}).
\end{equation}
As a result projected Coulomb gauge Hamiltonian obtained through our approach reads
\begin{align}
H_C &= \omega_c a^\dag a +U\sum_{\bb{R}}
n_{\bb{R}\uparrow}
n_{\bb{R}\downarrow}\notag\\
&-\sum_{\langle\bb{R}\bb{R'}\rangle}\sum_{\sigma}
t_{\bb{R}\bb{R'}}\left(e^{ig(a+a^\dag)}
c^{\dagger}_{\bb{R}\sigma}c_{\bb{R'}\sigma}+\text{h. c.}\right),
\end{align}
and coincides with the one obtained within the Peierls substitution~\cite{sentef2020quantum}.

It is useful to write the Hamiltonian in the Dipole Gauge, which reads
\begin{eqnarray}
H_D &=&H_{el} +\omega_c a^{\dagger}a + i\omega_c \left(a-a^{\dagger}\right)
\sum_{\bb{R}\sigma}\chi_{\bb{R}}
c^{\dagger}_{\bb{R}\sigma}c_{\bb{R}\sigma}+\nonumber\\
&&+\omega_c\left(\sum_{\bb{R}\sigma}\chi_{\bb{R}}
c^{\dagger}_{\bb{R}\sigma}c_{\bb{R}\sigma}\right)^2.
\end{eqnarray}
We emphasize again that in order to keep the gauge equivalence intact all the way into the strong light-matter coupling regime it is crucial to keep all the terms in the Peierls phase, as recently done in Ref.~\onlinecite{li2020manipulating,sentef2020quantum}.

\subsection{Two-orbital model}

We now consider a  model of spinless electrons hopping on an inversion-symmetric crystal with two atomic orbitals with opposite parity
(such as $s$ and $p_z$, denoted as $\alpha=1,2$ in the following) and interacting with local Coulomb repulsion. The model has been introduced before in the literature in the context of electronic superradiance~\cite{andolina2019cavity,lenk2020collective}.

We consider a one dimensional chain with lattice sites $\bb{R}=j\bb{x}$, where $j$ is an integer (we set the lattice constant $a_l = 1$), and periodic boundary conditions. The electronic Hamiltonian reads
\begin{align}
H_{el}=\left(E_g/2\right)\sum_{j}\Psi^{\dagger}_{j}\sigma^z\Psi_{j}+U\sum_j n_{j1}n_{j2}\nonumber\\
-\sum_{j} \Psi^{\dagger}_{j}\left(t_s\sigma^z -i\tilde{t}\sigma^y\right)\Psi_{j+1}+\mbox{h.c.},
\label{eq:H0second2orbital}
\end{align}
where we have defined electronic spinor operators 
\begin{equation}\label{eqn:spinor}
\Psi^{\dagger}_j=\left(
\begin{array}{l}
c^{\dagger}_{j1}\;\;c^{\dagger}_{j2}
\end{array}
\right),
\;\qquad
\Psi_j=\left(
\begin{array}{l}
c_{j1}\\
c_{j2}
\end{array}\right)
\end{equation}
satisfying standard anticommutation rules $\left\{c_{i\alpha},c^{\dagger}_{j\beta}\right\}=\delta_{ij}\delta_{\alpha\beta}$ and introduced the Pauli matrices $\sigma^{a}$. Here, $E_g$ is the local atomic energy, $t_s$ ($\tilde{t}$) describes interband (intraband) next neighbor hopping, and $U$ is the local density-density repulsion among orbitals, with $n_{j\alpha}=c^{\dagger}_{j\alpha}c_{j\alpha}$.
For what concerns the electromagnetic field we consider a single cavity mode with a uniform vector potential polarized along the chain, i.e. $\bb{A}=\bb{u}_xA_0 (a+a^{\dagger})$, which gives rise to a photonic pseudo-potential $\chi(x)=eA_0 x$. The photon Hamiltonian reads $H_{ph}=\omega_c a^{\dagger}a$.

We now write down the projected unitary operator, Eq.~(\ref{eq:Omega1}), for our two-orbital case. We assume the matrix element $\chi^{\alpha\alpha'}_{jj'}$ to be local in space and completely off-diagonal in orbital space, i.e. we consider only the leading local dipole interband matrix element,
\begin{align}
\chi^{\alpha\alpha'}_{jj'}=\gamma\delta_{jj'}\sigma^x_{\alpha\alpha'},
\end{align}
where we have introduced the light-matter coupling 
\begin{align}
\gamma= eA_0\,x_{12},
\end{align}
with $x_{12}=\int dx\  \phi_1^*(x)x\phi_2(x)$ the dipole matrix element between Wannier orbitals. Then we get for the projected unitary transformation, Eq.~(\ref{eq:Omega1}), the form
\begin{align}\label{eq:Uchi2band}
U=e^{i\gamma\left(a + a^\dag\right)\sum_j\sigma^x_{j}},
\end{align}
where we have introduced the pseudo-spin operators
\begin{equation}\label{eqn:pseudospin}
\sigma^a_j = \Psi^{\dagger}_j\sigma^a \Psi_j
\end{equation}
satisfying the algebra  $\left[ \sigma^a_j, \sigma^b_{j'}\right]=2i\epsilon^{abc}\sigma^c_j\,\delta_{jj'}$. Similarly we can define the projected PZW transformation as
\begin{align}
T=e^{-i\gamma\left(a + a^\dag\right)\sum_j\sigma^x_{j}}.
\end{align}
Using the projected unitary transformation, Eq.~(\ref{eq:Uchi2band}), and Eq.~(\ref{eq:Hd_proj}) we obtain the dipole Gauge Hamiltonian in the form
\begin{align}
H_D &=  H_{el}+ \omega_c a^\dag a + i  \omega_c \gamma (a - a^\dag)\sum_j\sigma^x_j
\notag\\
&+\omega_c \gamma^2\left(\sum_j\sigma^x_j\right)^2.
\label{eq:Hdipoletwoband}
\end{align}
We note that this result coincides with the dipole Hamiltonian discussed in Ref.~\onlinecite{lenk2020collective} for a related model for excitonic insulator coupled to a single mode cavity.

To evaluate the Coulomb gauge Hamiltonian we follow the recipe discussed in Section~\ref{sec:projectedbasis}. First, we evaluate the action of the projected unitary transform on the electronic operators, Eq.~(\ref{eqn:transform_field}), which reads
\begin{align}\label{eq:Ufermions_2band}
&U^{\dagger}c_{j\alpha}U=\sum_{\beta}\left(e^{i\gamma\left(a + a^\dag\right)\sigma^x}\right)_{\alpha\beta}c_{j\beta},\\
&U^{\dagger} c^{\dagger}_{j\alpha}U=\sum_{\beta}c^{\dagger}_{j\beta}\left(e^{-i\gamma\left(a + a^\dag\right)\sigma^x}\right)_{\beta\alpha}.
\label{eq:Ufermions_2band_dag}
\end{align}
Plugging these results into Eq.~(\ref{eq:Coulombprojgeneral}), and using the transformation for the pseudo-spin components
\begin{align}
&U^{\dag} \sigma^x_j  U= \sigma^x_j,\nonumber\\
&U^{\dag} \sigma^y_j U= \cos\left[2 \gamma \left(a+a^\dag\right)\right]\sigma^y_j+\sin\left[2 \gamma \left(a+a^\dag\right)\right]
\sigma^z_j,\nonumber\\
&U^{\dag} \sigma^z_j U =  \cos\left[2 \gamma \left(a+a^\dag\right)\right]\sigma^z_j - \sin\left[2 \gamma \left(a+a^\dag\right)\right]
\sigma^y_j,\nonumber
\end{align}
we obtain the Coulomb gauge Hamiltonian for a two-orbital system
\begin{align}
&H_C = \omega_c a^\dag a +U\sum_j n_{j1}n_{j2}+\nonumber\\
&+\sum_{j}\Psi^{\dagger}_{j}\left(E_g/2\right)\left(\cos(2\gamma\left(a + a^\dag\right))\sigma^z-\sin(2\gamma\left(a + a^\dag\right))\sigma^y\right)\Psi_{j}\nonumber\\
&- \sum_{j} \Psi^{\dagger}_{j}
\cos(2\gamma\left(a + a^\dag\right))\left(t_s\sigma^z -i\tilde{t}\sigma^y\right)\Psi_{j+1}\nonumber\\
&+\sum_{j} \Psi^{\dagger}_{j}\sin(2\gamma\left(a + a^\dag\right))\left(t_s\sigma^y +i\tilde{t}\sigma^z\right)
\Psi_{j+1} +\text{h.c.}
\label{eq:Hcoulombtwoband}
\end{align}
We emphasize that, as in the continuum case, the local  Hubbard interaction $H_{ee}=U \sum_{j}n_{j1}n_{j2}$ is not affected by the electromagnetic field, i.e.
\begin{equation}
U^{\dagger} H_{ee}U=H_{ee},
\end{equation}
a result that we explicitly prove in Appendix~\ref{transformeeinteractions}.

Finally, as a consistency check we can explicitly verify that the derived Coulomb gauge and dipole gauge Hamiltonian, even for the truncated model, are related by a unitary transformation. Indeed we have, using the fact that $T^{\dag}=U$
\begin{align}
T^{\dag} H_C T=
UH_{ph}U^{\dagger}+H_{el}=H_D.
\end{align}

We note that the obtained Hamiltonian in the Coulomb gauge significantly differs from the one typically used when describing a material coupled to quantum light as it contains the photonic operators $a$ and $a^\dag$  up to all orders. However, having this complicated structure is important to have a well-defined ultrastrong coupling limit in the tight-binding model. 

\subsection{Comparison with Peierls Substitution}

Before concluding this section it is instructive to compare, for the specific model under consideration, our projected Coulomb gauge Hamiltonian in Eq.~(\ref{eq:Hcoulombtwoband}) with the one obtained through the Peierls substitution, which has been studied for example in Ref.~\onlinecite{andolina2019cavity}. The resulting Hamiltonian reads in real space
\begin{align}
&H_{P}= \omega_c a^\dag a +\sum_{j}\Psi^{\dagger}_{j}\left(E_g/2\right)\sigma^z\Psi_{j}+U\sum_j n_{j1}n_{j2}+\nonumber\\
&-\sum_{j} \Psi^{\dagger}_{j}e^{ig\left(a+a^\dagger\right)}\left(t_s\sigma^z -i\tilde{t}\sigma^y\right)\Psi_{j+1} +\mbox{h.c.},
\label{eq:HtwobandPeierls}
\end{align}
where $g=e A_0$ is the light-matter coupling. We can immediately see that this Peierls Hamiltonian differs from the projected Coulomb gauge Hamiltonian we have obtained in Eq.~(\ref{eq:Hcoulombtwoband}). We can trace back this difference to the fact that within the Peierls approximation each hopping term in Eq.~(\ref{eq:H0second2orbital}) is dressed by the same phase factor, which therefore does not account for local orbital transitions mediated by the photon, as we discussed in Section~\ref{sec:Peierls}. While this can describe a different physical situation, depending on the structure of local orbitals chosen for the projection, it is important to stress that in order to preserve gauge equivalence all the way into the ultrastrong coupling regime it is crucial to treat the Peierls phase to all orders, as we are going to discuss more in detail in the next section. Another important difference among our Coulomb gauge and Eq.~(\ref{eq:HtwobandPeierls}) is that within Peierls the light matter coupling $g$ is completely fixed by the strength of the field and does not really depend on any material property. This is not surprising after all since, as we discussed, the Peierls substitution can be equivalently seen as an approximation to the momentum operator matrix element which is completely determined by tight-binding parameters. As we are going to discuss in the next section this will have physical consequences for example on the polariton spectrum of the system.

\section{Application: Two-Orbital Model Coupled to Cavity} 
\label{section:application}

In this section, we study in more detail the two-orbital model introduced in the previous section. First, using the dipole gauge Hamiltonian we derive an electron-only effective action after integrating out exactly the cavity photon and show that, even beyond mean-field theory, the light-matter coupling goes to zero at low frequency, i.e. the ground state is factorized and no superradiance is possible~\cite{lenk2020collective}. Then we re-derive this result within the Coulomb gauge, solving for the ground state within mean-field theory. We emphasize the crucial role played by photon non-linearities and the danger associated with expanding the Coulomb gauge Hamiltonian in light-matter coupling. Finally, we compute the polariton spectrum of the problem and show that, although the ground state of the problem is factorized in the thermodynamic limit, excitations on top of it are actually entangled. We show explicitly how polariton frequencies are the same within our projected Coulomb gauge and dipole gauge, as expected from the gauge equivalence.

\subsection{Dipole Gauge Hamiltonian: Effective Action for Electrons and Asymptotic Decoupling} \label{sec:absencesuperdipole}

The dipole gauge Hamiltonian, Eq.~(\ref{eq:Hdipoletwoband}), has the nice feature that the photon mode only enters linearly. Therefore we can integrate it out  \emph{exactly} within a path integral formulation and obtain an effective action for the electronic sector only. We start from the partition function associated to the dipole gauge Hamiltonian, which reads
\begin{align}
Z =  \int \prod_j\mathcal{D}\left[p,\Psi_j,\Psi_j^*\right]e^{-S_{ph}  - S_{el}  - S_{el-ph}},
\label{eq:partitionfunct}
\end{align}
where we separated the different contributions to the total action $S$: $S_{ph}$ describes to the photonic fields (see Appendix~\ref{photonicaction} for the details of the derivation), $S_{el}$ corresponds to the electronic system, and $S_{el-ph}$ describes to the electron-photon interaction,
\begin{align}
&S_{ph} = \int_{0}^{\beta}d\tau \ d\tau '\  p(\tau)\mathcal{D}^{-1}(\tau - \tau ')p(\tau '),\\
&S_{el} = \int_{0}^{\beta}d\tau  \ d\tau '\  \sum_{j,j'} \Psi_j^*(\tau)\mathcal{G}_{jj'}^{-1}(\tau - \tau ') \Psi_{j'}(\tau ')\notag\\
 &+ \int_{0}^{\beta}d\tau \  \omega_c \gamma^2\left(\sum_j\sigma^x_j(\tau)\right)^2 + \int_{0}^{\beta} \mathcal{H}_{ee},\\
&S_{el-ph} = -\int_{0}^{\beta}d\tau  \ \sqrt{2\omega_c}\gamma p(\tau)\sum_j\sigma^x_j(\tau).
\end{align}
Here, 
\begin{align}
\mathcal{D}^{-1}(\tau - \tau ') =  \dfrac{1}{2\omega_c^2}\delta(\tau - \tau ')\left(\omega_c^2 - \partial_\tau^2\right)
\end{align}
is the photonic Green's functions and $\mathcal{G}^{-1}_{jj'}(\tau - \tau ')$ is the non-interacting electronic Green's functions. After performing the Gaussian integration over $p(\tau)$, the partition function given by Eq.~\eqref{eq:partitionfunct} becomes

\begin{align}
Z[p] =  \int\prod_j \mathcal{D}\left[\Psi_j,\Psi_j^*\right]e^{-S_{eff}[\Psi_j,\Psi_j^*]},
\label{eq:partitionfunct2}
\end{align}
with the effective action given by

\begin{align}
&S_{eff} = \int_{0}^{\beta}d\tau \ d\tau ' \sum_{j,j'}\Psi_j^*(\tau)\mathcal{G}^{-1}_{jj'}(\tau - \tau ') \Psi_{j'}(\tau ') 
\notag\\
& + \omega_c \gamma^2  \int_{0}^{\beta}d\tau \ \left(\sum_j\sigma^x_j(\tau)\right)^2 + \int_{0}^{\beta} \mathcal{H}_{ee}\notag\\
&- \frac{\omega_c}{2}\gamma^2 \int_{0}^{\beta}d\tau  \ d\tau ' \ \sum_{jj'}\sigma^x_j(\tau)\mathcal{D}(\tau - \tau ')\sigma^x_{j'}(\tau ').
\end{align}
We see that in the effective electronic action there is now an additional term proportional to $\gamma^2$, a retarded electron-electron interaction arising from the exact integration out of the photonic mode
Defining the Fourier transform as $\psi(\tau) = \sum_{\omega_n}e^{-i\omega_n\tau}\psi(i\omega_n)/\sqrt{\beta}$ and calculating the photonic Green's function 
\begin{align}
\mathcal{D}(i\omega_n) =  \dfrac{2\omega_c^2}{\omega_c^2 -\left(i\omega_n\right)^2},
\end{align}
the effective action becomes
\begin{align}
&S_{eff} =\sum_{\omega_n}\sum_{j,j'}\Psi_j^*(i\omega_n)\mathcal{G}^{-1}_{jj'}(i\omega_n) \Psi_{j'}(i\omega_n)
+ \notag\\
& +\mathcal{H}_{ee} +
\sum_{jj'}\sum_{\omega_n}J_{\rm eff}(i\omega_n)
\sigma^x_j(i\omega_n)\sigma^x_{j'}(i\omega_n).
\label{eq:Seffmatsubara}
\end{align}
where we have introduced the overall effective electron-electron 
$$
J_{\rm eff}(i\omega_n)=\omega_c \gamma^2\left(1-\dfrac{\omega_c^2}{\omega_c^2 -(i\omega_n)^2}\right)
$$
After analytic continuation, $i\omega_n\rightarrow \omega +i\eta$ we see that this effective interaction vanishes in the static limit $\omega\rightarrow0$
 and the effective action is given only by the matter Hamiltonian and independent of the light-matter coupling strength,
\begin{align}
S_{eff}\left(\omega \rightarrow 0 \right) = \sum_{j,j'}\Psi_j^*(\omega)\mathcal{G}^{-1}_{jj'}(\omega) \Psi_{j'}(\omega) +  \mathcal{H}_{ee}.
\label{eq:Seffmatsubara2}
\end{align}
This result shows that at low frequency electrons and photons are fully decoupled. Since a putative equilibrium superradiant phase transition would emerge as zero frequency criticality of the coupled electron-photon system the above result shows that the system remains always in the normal symmetric phase, at least for what concerns the photon. The electronic sector can in principle break a symmetry due to the local Hubbard-like electron-electron interaction in Eq.~(\ref{eq:Seffmatsubara}) but this does not lead to any photonic order parameter. We have further checked this result by solving the problem within mean-field theory (see Appendix~\ref{dipolegauge}).
Finally, we note that while at zero frequency the two sectors are decoupled, excitations at finite frequency can carry non-trivial light-matter entanglement. We show this explicitly in Section~\ref{sec:polaritons}, where we discuss the polariton spectrum.

\subsection{Coulomb gauge Hamiltonian: Mean-field solution} \label{sec:absencesupercoulomb}
In this section, we show how the result of the previous section, the absence of superrandiant phase, can be obtained in the Coulomb gauge, i.e. from the Hamiltonian in Eq.~(\ref{eq:Hcoulombtwoband}). In order for this to work it is crucial to keep the structure of cosine and sine intact. In fact, as we are going to show explicitly below, expanding the Coulomb gauge Hamiltonian and keeping only linear and quadratic couplings leads to a breakdown of the model in the ultrastrong coupling limit, both within our Coulomb gauge and within the Peierls substitution.

Before proceeding it is convenient to introduce Fourier modes
\begin{align}
c_{n,j} = \dfrac{1}{\sqrt{N}}\sum_k e^{i k j}c_{n,k},
\label{eq:reciprocal}
\end{align}
where $N$ is the number of lattice sites and $k$ belongs to reciprocal lattice, and to rewrite the pseudo-spin operators~Eq.~(\ref{eqn:pseudospin}) in momentum space
\begin{equation}\label{eqn:pseudospin_k}
\sigma^{a}_k =\Psi^{\dagger}_k \sigma^{\alpha} \Psi_k,
\end{equation}
where $\sigma^{a}$, with $a = x,y,z$, are Pauli matrices and $\Psi_k$ is the Fourier transform of the spinor defined in Eq.~(\ref{eqn:spinor}). Thus,
the Coulomb gauge Hamiltonian~Eq.~(\ref{eq:Hcoulombtwoband}) reads in a more compact form
\begin{align}\label{eq:Hcoulombtwoband_kspace}
&H_C=  \omega_c a^\dag a + U\sum_j n_{j1}n_{j2} +\notag\\
&+\sum_k \Big[\{\varepsilon_k\cos[2\gamma(a+a^\dag)] - 2\tilde{t}\sin(k)\sin[2\gamma(a+a^\dag)]\}\sigma^z_k\notag\\
&-\{2\tilde{t}\sin(k)\cos[2\gamma(a+a^\dag)] +\varepsilon_k\sin[2\gamma(a+a^\dag)]\}\sigma^y_k
\Big],
\end{align}
where $\varepsilon_k=E_g/2 - 2 t_s\cos(k)$.
Next, we study the Coulomb gauge Hamiltonian $H_C$ in mean field that corresponds to neglecting correlations between the cavity modes and electrons, 
$$
|\Psi\rangle = |\psi\rangle |\phi\rangle.
$$
Here, $|\phi\rangle$ is a coherent state, $ a |\phi\rangle = \alpha \sqrt{N} |\phi\rangle$, with $\alpha$ being the photonic order parameter that could have both real and imaginary parts, $\alpha = \alpha' + i\alpha''$. Finite value of $\alpha$ corresponds to the superradiant phase, while $\alpha$ is always zero in the normal phase. As a result of the mean-field decoupling, we have to solve a photonic problem with Hamiltonian
\begin{align}
&H_{ph}^{mf} =  \omega_c a^\dag a + A\cos[2\gamma(a+a^\dag)] +
 B\sin[2\gamma(a+a^\dag)],
 \label{eq:meanfieldphoton}
\end{align}
where
\begin{align}
&A=\sum_k \Big[ \varepsilon_k\langle \psi| \sigma^z_k |\psi\rangle - 2\tilde{t}\sin(k)\langle \psi|\sigma^y_k|\psi\rangle \Big],\\
&B=\sum_k \Big[ -2\tilde{t}\sin(k) \langle \psi| \sigma^z_k |\psi\rangle -\varepsilon_k\langle \psi|\sigma^y_k|\psi\rangle \Big],
\end{align}
and the electronic mean-field Hamiltonian
\begin{align}
&H_{el}^{mf} = U\sum_j n_{j1}n_{j2} +
\sum_k \left(\epsilon_k A_1- 2\tilde{t}\sin(k)A_2\right)\sigma^z_k+\nonumber\\
&-\sum_k \left( 2\tilde{t}\sin(k)A_1 + \varepsilon_k A_2 \right)\sigma^y_k,
\label{eq:Helel3}
\end{align}
where we introduced the expectation values of the photonic operators over the coherent state $|\phi\rangle$
\begin{align}
&A_1 = \langle \phi |\cos[2\gamma(a+a^\dag)] |\phi\rangle,\label{eq:A1}\\
&A_2 = \langle \phi |\sin[2\gamma(a+a^\dag)] |\phi\rangle.
\label{eq:A2}
\end{align}

Making the Hartree-Fock approximation, the electron-electron interactions Hamiltonian $H_{ee}$, Eq.~\eqref{eq:Helel3}, becomes~\cite{andolina2019cavity,mazza2019superradiant}
\begin{align}
H_{ee} &= -U\sum_k\left(\dfrac{m}{2}\sigma^z_k + \mathcal{I}'\sigma^x_k  - \mathcal{I}'' \sigma^y_k\right)\notag\\
&+ U N \left(\dfrac{m^2}{4} + |\mathcal{I}|^2\right),
\label{eq:hartreefockmeanf}
\end{align}
where $m = \left(1/N\right)\sum_q\langle\sigma^z_q\rangle$ and $\mathcal{I} = \left(1/N\right)\sum_q \langle c^\dag_{q,2}c_{q,1}\rangle \equiv \mathcal{I}' +i\mathcal{I}''$, and the electronic mean-field Hamiltonian can be written as
\begin{align}
H_{el}^{mf}= \sum_{a=x,y,z} h^a_k\sigma^a_k.
\label{eq:coulombelectmeanf}
\end{align}
Here, the coefficients $h^a_k$, with $a = x,y,z$, are given by
\begin{align}
&h^x_k = - U\mathcal{I}',\label{eq:hxcoulombelel}\\
&h^y_k =  -2A_1 \tilde{t}\sin(k) -A_2 \epsilon_k+ U\mathcal{I}'',\label{eq:hycoulombelel}\\
&h^z_k =  A_1 \epsilon_k - 2 A_2 \tilde{t}\sin(k)- U\dfrac{m}{2}\label{eq:hzcoulombelel}.
\end{align}
The resulting Hamiltonian can be easily diagonalized by a Bogoliubov transformation. At zero temperature we find that $\langle \psi |\sigma^a_k|\psi \rangle$ $=$ $-h^a_k/E_k$ and $\mathcal{I}$ $=$  $-\left(1/2\right)\sum_k\left(h^x_k - ih^y_k\right)/E_k$, where $E_k$ $=$ $\sqrt{\sum_{a}\left(h^a_k\right)^2}$.

\begin{figure}[t!] 
\includegraphics[width=0.95\linewidth]{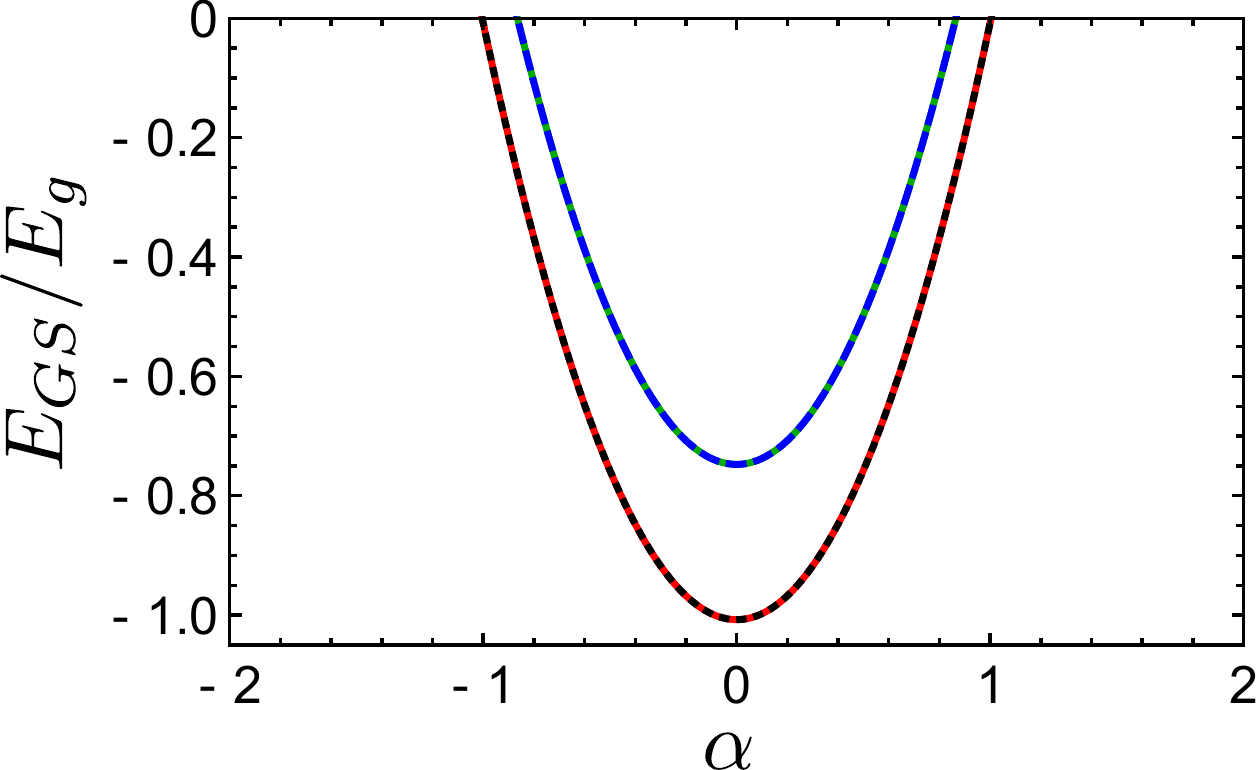}
\caption{Ground-state energy $E_{GS}/E_g$ of the Coulomb gauge Hamiltonian, given by Eq.~\eqref{eq:groundenergycoulombinter}, as a function of the photonic order parameter $\alpha$. In the absence of the electron-electron interactions, $U = 0$, $E_{GS}/E_g$ has a parabolic shape, with a single minimum at $\alpha = 0$ [green dotdashed line corresponds to $\gamma_0 = 0.1$ and blue dashed line corresponds to $\gamma_0 = 0.95$]. Note that $E_{GS}$ is independent of the light-matter coupling. For finite electron-electron interactions, $U/E_g = 2$, $E_{GS}/E_g$ has a shape of a parabola and is independent of the values of $\gamma_0$ [black dotted line corresponds to $\gamma_0 = 0.1$ and red solid line corresponds to $\gamma_0 = 0.95$].
Other parameters are chosen as $\omega_c/E_g = 1$, $t_s/E_g= 0.5$, $\tilde{t}/E_g = 0.1$. }
\label{fig:groundstateenergy}
\end{figure}

Next, we find that $A_1 = e^{-2\gamma_0^2/N}\cos\left(4\gamma_0\alpha\right)$ and $A_2 = e^{-2\gamma_0^2/N}\sin\left(4\gamma_0\alpha\right)$, where $\gamma_0 = \gamma\sqrt{N}$. In the limit $N\rightarrow \infty$, we find that $A_1 = \cos\left(4\gamma_0\alpha\right)$ and $A_2 = \sin\left(4\gamma_0\alpha\right)$. The ground-state energy is given by the expectation value of $\mathcal{H}_C$ over $|\Psi\rangle$. In presence of interactions, the ground-state energy reads
\begin{align}
\dfrac{E_{GS}}{N} = \omega_c\left|\alpha\right|^2 - \dfrac{1}{N}\sum_k \sqrt{\sum_{a}\left(h^a_k\right)^2} + U\left(\dfrac{m^2}{4} + |\mathcal{I}|^2\right),
\label{eq:groundenergycoulombinter}
\end{align}
which reduces for $U=0$ to the result
\begin{align}
\dfrac{E_{GS}}{N} = \omega_c\left|\alpha\right|^2 - \dfrac{1}{N}\sum_k E_k,
\label{eq:groundenergycoulomb}
\end{align}
where $E_k  = \sqrt{\epsilon_k^2 + 4\tilde{t}^2\sin^2(k)}$. We plot in Fig.~\ref{fig:groundstateenergy} the behavior of $E_{GS}$ as a function of the photonic order parameter $\alpha$ for the non-interacting case as well as for $U\neq0$. We see that in both cases the ground-state energy has a well-defined minimum at $\alpha=0$, which is perfectly consistent with the saddle point equation
\begin{align}
&\omega_c \alpha  = \dfrac{2\gamma_0}{N}\sum_k \Big[\langle \psi |\sigma_z^k|\psi \rangle \left(2 A_1\tilde{t}\sin(k) +A_2 \epsilon_k \right)\notag\\
&+ \langle \psi |\sigma_y^k|\psi \rangle\left(A_1 \epsilon_k -2 A_2 \tilde{t}\sin(k)\right)\Big].
\label{eq:saddlepointcoulomb}
\end{align}
Introducing the expectation values of the electronic operators into Eq.~\eqref{eq:saddlepointcoulomb}, we find that the right-hand side of the saddle point equation is zero. Thus, $\alpha = 0$ is the only solution, which corresponds to the absence of superradiance in the system. We also notice that the electronic contribution to the ground-state energy does not depend on $\alpha$, i.e.
$$
\frac{\partial E^{el}_{GS}}{\partial \alpha}=0,
$$
from which we conclude, in analogy with similar arguments for the Peierls substitution~\cite{guerci2020superradiant}, that the TRK sum-rule is satisfied for our projected Coulomb Hamiltonian. In fact we can show this in quite some generality using gauge equivalence. First, we rewrite the Coulomb Hamiltonian in Eq.~(\ref{eq:Hcoulombtwoband_kspace}) as
\begin{equation}
H_C=H_{ph}+\tilde{H}_{el},
\end{equation}
where the dressed electronic Hamiltonian reads by construction
$\tilde{H}_{el}=U^{\dagger}H_{el}U$. The physical current operator in our theory is given by the derivative of this dressed electronic Hamiltonian with respect to the field, i.e.
\begin{equation}
J=\frac{\partial \tilde{H}_{el}}{\partial A},
\end{equation}
where $A=A_0(a+a^{\dagger})$ is now treated as a classical field. The average value of the current on the Coulomb gauge ground state can be written using Hellmann-Feynman theorem as
\begin{equation}
\langle J\rangle_C=\langle \frac{\partial \tilde{H}_{el}}{\partial A}\rangle_C=
\frac{\partial \langle\tilde{H}_{el}\rangle_C}{\partial A},
\end{equation}
where $\langle \rangle_C$ indicates average over the ground state of Coulomb gauge Hamiltonian. We can evaluate the expectation value of the dressed electronic Hamiltonian using gauge equivalence. Indeed, we have
\begin{equation}
\langle \tilde{H}_{el}\rangle_C= \langle U^{\dagger}H_{el}U\rangle_C=
\langle H_{el}\rangle_D,
\end{equation}
where $\langle \rangle_D$ indicates average over the ground state of dipole gauge Hamiltonian. Here we used the fact that the ground state of Coulomb and dipole gauges are related by a unitary transformation. We therefore conclude that the average current in the Coulomb gauge is given by the derivative with respect to the field of the (undressed) electronic ground-state energy in the dipole gauge. Since however, as we have shown in section~\ref{sec:absencesuperdipole}, in the dipole gauge electrons and photons decouple at low energy we conclude that $\langle H_{el}\rangle_D$ does not in fact depend on the field and therefore
\begin{equation}
\langle J\rangle_C=
\frac{\partial \langle\tilde{H}_{el}\rangle_D}{\partial A}=0\,,
\end{equation}
i.e. a static uniform vector potential does not produce a finite current in the system. We notice that within linear response theory a static uniform current is related to the static limit of the current-current correlation function $Q(\omega,q)$, i.e.
\begin{equation}
\langle J\rangle_C=Q(0,q\rightarrow0)A_0=0,
\end{equation}
which indeed is a manifestation of the TRK sum rule.

\subsection{Discussion: Expanding the Coulomb Gauge Hamiltonian}

The results of previous two sections, i.e. the fact that the photon field always remains incoherent in the ground state for any value of the light-matter coupling, both in the dipole and in the Coulomb gauge, does not come as a surprise at first. Indeed recent works have proven\cite{andolina2019cavity}, under very general hypotheses, a no-go theorem for superradiance in presence of static uniform vector potential in the Coulomb gauge. Crucially, this result has been obtained within the continuum model, where vector potential enters through paramagnetic and diamagnetic contribution, and relies on the TRK sum rule and gauge invariance.  

We now show that in order to correctly reproduce this result within a projected tight-binding model it is crucial to treat the non-linear light-matter coupling of the Coulomb gauge Hamiltonian non-perturbatively. On the other hand expanding the light-matter interaction to the second order, as done recently in the literature in the context of Peierls approximation, would lead to a breakdown of the model at ultrastrong coupling. As we are going to discuss, this is true both for the Peierls approximation as well as for our Coulomb gauge.

To see this we consider the Hamiltonian discussed in Ref.~\onlinecite{andolina2019cavity}, which corresponds to the Peierls approximation Eq.~(\ref{eq:HtwobandPeierls}) expanded to second order and that we rewrite here for completeness.
\begin{align}
H_{P} = H_{el} + \omega_c a^\dag a +\dfrac{g_0}{\sqrt{N}}j_p \left(a + a^\dag\right) - \dfrac{g_0^2}{2N}\mathcal{T} \left(a+a^\dag\right)^2,
\label{eq:polinimodel}
\end{align}
where
\begin{align}
j_p = 2t_s\sin(k)\sigma^z_k-2\tilde{t}\cos(k)\sigma^y_k
\end{align}
and
\begin{align}
\mathcal{T} = -2t_s\cos(k)\sigma^z_k-2\tilde{t}\sin(k)\sigma^y_k.
\end{align}
are the paramagnetic and diamagnetic terms and $g_0 = g\sqrt{N}$ is the light-matter coupling.
\begin{figure}[t!] 
\includegraphics[width=0.95\linewidth]{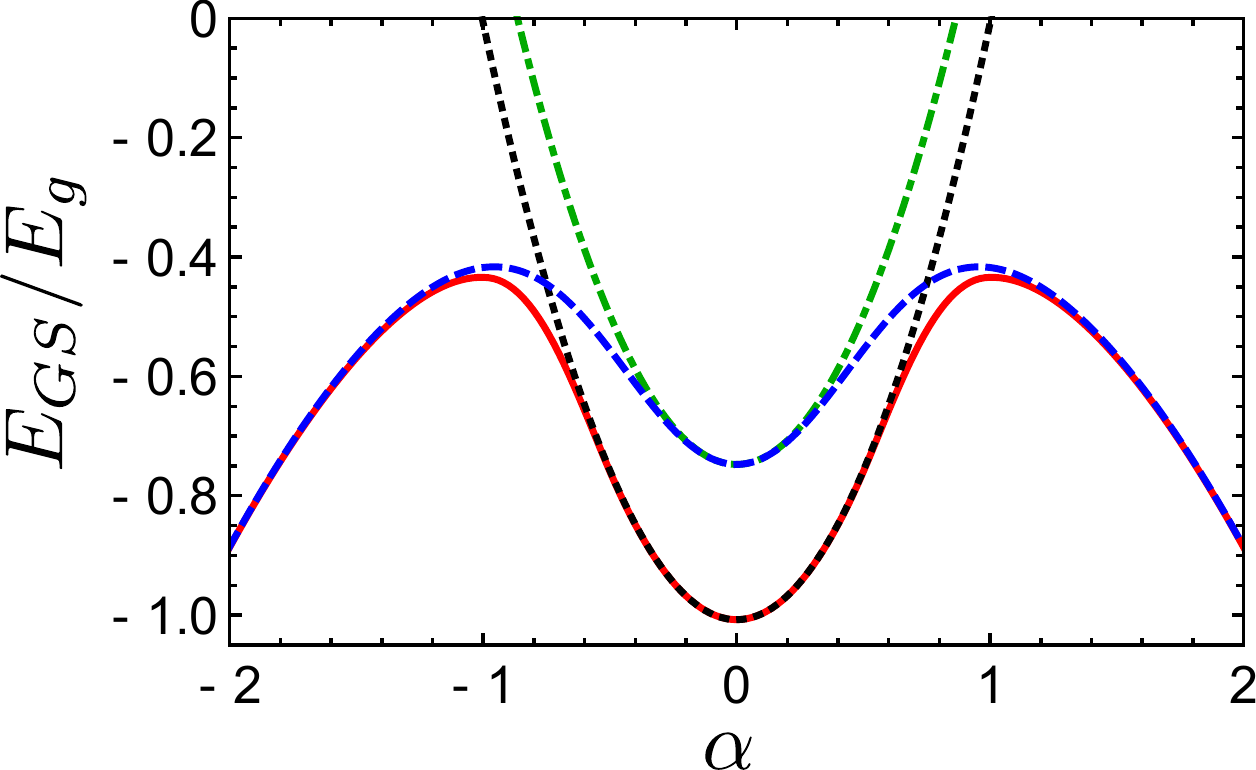}
\caption{Ground-state energy $E_{GS}/E_g$ of the expanded Peierls Hamiltonian as a function of the photonic order parameter $\alpha$ for different values of $g_0$ and $U$.
In the absence of the electron-electron interactions, $U = 0$, for  small value of the light-matter coupling, $g_0 = 0.1$ (green dotdashed line), $E_{GS}/E_g$ has a shape of a parabola with a single minimum at $\alpha = 0$. For large values of the light-matter coupling, $g_0 = 0.95$ (blue dashed line), the ground-state energy, in addition to minimum at $\alpha = 0$, develops two maxima at finite values of $\alpha$. In the presence of the electron-electron interactions, $U/E_g = 2$, the ground-state energy has a single minimum for $g_0 = 0.1$ (black dotted line corresponds) and two additional maxima for $g_0 = 0.95$ (red solid line).
Other parameters are the same as in Fig.~\ref{fig:groundstateenergy}.}
\label{fig:groundstateenergy2}
\end{figure}

Solving the problem within mean field, through a similar calculation as the one sketched before (see also Ref.~\onlinecite{andolina2019cavity} for details), gives a ground-state energy as a function of the photonic order parameter $\alpha$, which we plot for different values of light matter coupling in Fig.~\ref{fig:groundstateenergy2}. We see that for small light-matter coupling $g_0$ the energy has the expected parabolic behavior with a well-defined minimum at $\alpha=0$. However, upon increasing $g_0$ the shape of the ground-state energy changes qualitatively. In particular, while the $\alpha=0$ solution remains a local minimum, the system develops two additional maxima at finite $\alpha$ and, more importantly, a negative curvature for finite $\alpha$, which implies the $\alpha=0$ solution is not the global minimum anymore. We emphasize that while solving for the small $\alpha$ behavior does indeed allow one to predict the absence of superradiance, as reported in Ref.~\onlinecite{andolina2019cavity}, the behavior of the ground-state energy plotted in Fig.~\ref{fig:groundstateenergy2} suggests that the Hamiltonian Eq.~(\ref{eq:polinimodel}) is not well-defined at ultrastrong coupling. This problem is readily solved by treating exactly the Peierls phase. In fact, for $U = 0$ we obtain
\begin{align}
\dfrac{E_{GS}}{N} = \hbar \omega_c \alpha^2 - \dfrac{1}{2\pi}\int_{-\pi}^{\pi}dk E_{2\alpha  g_0+k}.
\label{eq:groundstatePeierls}
\end{align}
The integral $\int_{-\pi}^{\pi}dk E_k$ does not depend on $\alpha$ or $g_0$, as we obtained for our Coulomb gauge Hamiltonian. It is therefore important to stress that the problem here is not the Peierls substitution \emph{per se}. In fact, performing the same expansion a priori in our Coulomb gauge Hamiltonian would have led to the same issue. This clarifies that expanding a projected Coulomb gauge Hamiltonian into linear (paramagnetic) and quadratic (diamagnetic) terms, akin to the structure in the continuum field theory, is a particularly dangerous operation at ultrastrong coupling. It could lead to inconsistencies which could be particularly relevant in models which admit a good superradiant phase. The importance of taking into account all terms in the Peierls substitution was pointed out recently~\cite{li2020manipulating,sentef2020quantum}.

\subsection{Polariton Spectrum} \label{sec:polaritons}

In the previous section, we found that the photonic order parameter $\alpha$ is zero, and our system is always in the normal phase. However, even in the normal phase there are polaritons in the system, that give rise to non-zero optical response. Below we present two different approaches to obtain the polariton excitation. First, we develop an effective spin wave theory which allows one to introduce quantum fluctuations on top of the mean field giving rise to a simple bosonic Hamiltonian describing polariton formation. Then, we compute the photon propagator of the full model including Gaussian $1/N$ fluctuations on top of mean field. In the following section, we put $U=0$ for simplicity. However, the effect of interactions on the polariton spectrum is an interesting question that we leave for future work.

\subsubsection{Effective Spin Wave Theory}

We start by considering the dipole gauge Hamiltonian, Eq.~(\ref{eq:Hdipoletwoband}), that we rewrite in momentum space through the pseudo-spin operators Eq.~(\ref{eqn:pseudospin_k}). Due to the uniform nature of the vector potential we notice that the photon field only couples to the $k=0$ (global) electronic polarization, also entering the self-interaction term,  and therefore we can write the dipole gauge Hamiltonian as
\begin{align}
H_D&= \omega_c a^\dag a -\dfrac{\omega_x}{2}\sigma^z_{k=0} + i\gamma\omega_c \left( a - a^\dag\right)\sigma^x_{k=0} \notag\\
&+ \gamma^2\omega_c\left(\sigma^x_{k=0}\right)^2 -\sum_{k\neq 0} \left(\epsilon_k\sigma^z_k-2\tilde{t}\sin(k)\sigma^y_k\right),
\label{eq:Hdipolek=0}
\end{align}
where we introduced $\omega_x = 2\left(2 t_s - E_g/2\right)$. This writing suggests, as first approximation, to disregard the finite momentum electronic modes and focus on the $k=0$ sector, which in the thermodynamic limit can be treated semi classically with quantum fluctuations of the order $1/N$ described by harmonic bosons leading to polariton modes.

Introducing a classical spin vector and a classical coherent field for the photon
$\vec{\sigma}$ $=$ $\left(\sigma^x_{k=0}, \sigma^y_{k=0}, \sigma^z_{k=0}\right)$ $=$ $\left(\rho \sin{\theta}\cos{\phi}, \rho \sin{\theta}\sin{\phi},\rho\cos{\theta}\right)$, $a = \alpha' + i\alpha''$, $a^\dag = \alpha' - i\alpha''$ into Eq.~\eqref{eq:Hdipolek=0}, we find for the classical energy
\begin{align}
&E\left(\rho,\theta,\phi,\alpha',\alpha''\right) = \left(E_s-2 t_s\right)\rho\cos{\theta} + \omega_c\left(\alpha'^2+\alpha''^2\right)\notag\\
&- 2\omega_c\gamma\alpha''\rho\sin{\theta}\cos{\phi} + \omega_c\gamma^2\rho^2\sin^2{\theta}\cos^2{\phi}.
\end{align}
From $\partial E/ \partial\alpha' = 0$ we find that $\alpha'=0$. From $\partial E/\partial\alpha'' = 0$ we find that $\alpha''=  \gamma\rho\sin{\theta}\cos{\phi}$. Using the previous expression for $\alpha''$ we find from $\partial E/\partial\theta = 0$ that $\sin{\theta} = 0$. Thus, for the classical spin we obtain that $\sigma^x_{k = 0} = 0$, $\sigma^y_{k=0} = 0$, $\sigma^z_{k = 0} = \rho$, $\alpha' = 0$, $\alpha''=0$. We find that the ground-state energy is given by $E_{GS} = \left(E_g/2-2 t_s\right) \rho$.

Next, we calculate the spectrum of the lowest excitations above the ground states $E_{GS}$ using  Holstein-Primakoff transformation, which can be written as~\cite{lerose2019impact}
\begin{align}
&\sigma^x_{k = 0} = \sqrt{N}\left(b + b^\dag\right),\notag\\
&\sigma^y_{k = 0} = -i\sqrt{N}\left(b - b^\dag\right)\notag,\\
&\sigma^z_{k = 0} = N-2 b^\dag b,
\label{eq:sigmak=0}
\end{align}
where 
$\left[\sigma^x_{k=0},\sigma^y_{k=0}\right] = 2 i N$, $\left[\sigma^x_{k=0},\sigma^z_{k=0}\right] = -2 i \sigma^y_{k=0}$.

\begin{figure}[t!] 
\includegraphics[width=0.95\linewidth]{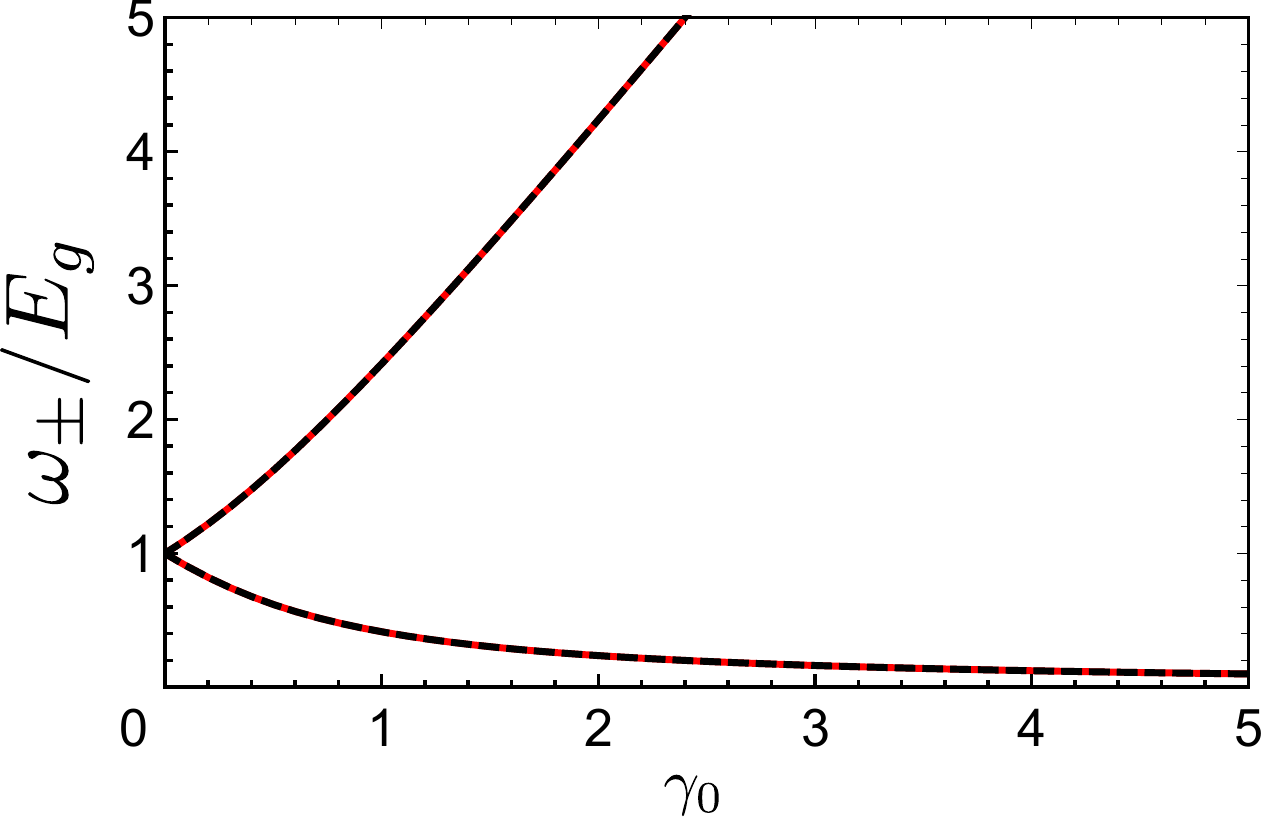}
\caption{
The frequencies $\omega_{\pm}/E_g$ of the polariton modes as a function of the coupling $\gamma_0$ for two different light-matter Hamiltonians: Red solid lines correspond to $\omega_{d,\pm}$, black dashed lines correspond to $\omega_{C,\pm}$. 
The parameters are fixed as $\omega_c/E_g = 1$,  and $t_s/E_g = 0.5$. The polariton frequencies are independent of $\tilde{t}$.}
\label{fig:polariton01}
\end{figure}

Introducing Eqs.~\eqref{eq:sigmak=0} into Eq.~\eqref{eq:Hdipolek=0} and taking the limit $N \rightarrow \infty$, we obtain in the dipole gauge
\begin{align}
\tilde{H}_D &= \omega_c a^\dag a + \omega_x b^\dag b+ i \gamma_0\omega_c\left(a - a^\dag\right)\left(b+b^\dag\right)\notag\\
& + \gamma_0^2 \omega_c \left(b+b^\dag\right)^2 -\dfrac{\omega_x}{2}N.
\end{align}
Here, the last term corresponds to the classical energy.
We note that in the thermodynamic limit $N\rightarrow \infty$, the dipole gauge Hamiltonian contains only the terms proportional to $\gamma_0^2$ and is described by the Hamiltonian of two coupled harmonic oscillators. Performing the Bogoliubov-Hopfield transformation~\cite{deliberato2014light}, we find two modes
\begin{align}
\omega^2_{d,\pm} &= \dfrac{1}{2}\Big(\omega_c^2+\omega_x^2 +4 \gamma_0^2 \omega_c \omega_x\notag\\
&\pm \sqrt{\left(\omega_c^2+\omega_x^2 + 4 \gamma_0^2 \omega_c \omega_x\right)^2-4 \omega_c^2 \omega_x^2}\Big),
\end{align}
which we plot as function of light-matter coupling in Fig.~\ref{fig:polariton01}. We find a lower polariton branch that is strongly suppressed by light-matter coupling while the upper one increases.

It is instructive to repeat the same analysis for the Coulomb gauge Hamiltonian, Eq.~(\ref{eq:Hcoulombtwoband}). Specifically, splitting the $k=0$ sector from the finite momentum modes and disregarding the latter, we obtain
\begin{align}
\tilde{\mathcal{H}}_C &= \omega_c a^\dag a -\dfrac{\omega_x}{2}\cos\left(2\gamma(a+a^\dag)\right)\sigma^z_{k=0}\notag\\
&+\dfrac{\omega_x}{2}\sin\left(2\gamma(a+a^\dag)\right)\sigma^y_{k=0}.
\label{eq:HCoulombk=0}
\end{align}
Using the expressions for the classical spin, we find for the energy 
\begin{align}
&E = \omega_c\left(\alpha'^2 + \alpha''^2\right) - \dfrac{\omega_x}{2}\rho\cos{\theta}\cos(4\gamma\alpha') \notag\\
&+ \dfrac{\omega_x}{2}\rho\sin{\theta}\sin{\phi}\sin(4\gamma\alpha').
\end{align}
As for the dipole gauge Hamiltonian, we find that $\alpha' = \alpha'' = 0$ and  $\sin{\theta} = 0$. After performing the Holstein-Primakoff transformation, we obtain 
\begin{align}
&\tilde{H}_C = \omega_c a^\dag a + \omega_x b^\dag b- i \gamma_0\omega_x \left(a+a^\dag\right)\left(b-b^\dag\right) \notag\\
&+ \gamma_0^2 \omega_x \left(a+a^\dag\right)^2 -\dfrac{\omega_x}{2}N,
\end{align}
where we neglected the terms of the order $1/N$.  Diagonalizing $\tilde{H}_C$, we find that there are two polariton branches with frequencies
\begin{align}
\omega^2_{C,\pm} &= \dfrac{1}{2} \Big(\omega_c^2+\omega_x^2 + 4 \gamma_0^2 \omega_c \omega_x\notag\\
&\pm \sqrt{\left(\omega_c^2+\omega_x^2 + 4 \gamma_0^2 \omega_c \omega_x\right)^2-4 \omega_c^2 \omega_x^2}\Big).
\label{eq:omegaCoulomb}
\end{align}
As expected we find that 
 $\omega^2_{d,\pm} = \omega^2_{C,\pm}$, which immediately follows from the fact that the dipole and Coulomb gauge Hamiltonians are related by a unitary transformation. 

At this point a natural question is to compare the polariton modes we have obtained so far with those that can be obtained from the Peierls Hamiltonian, Eq.~(\ref{eq:HtwobandPeierls}), through the very same calculation. Using Eq.~\eqref{eq:sigmak=0}, we obtain 
\begin{align}
\tilde{H}_P &= \omega_c a^\dag a + \omega_x b^\dag b+ 2 i g_0\tilde{t}\left(a+a^\dag\right)\left(b-b^\dag\right)\notag\\
 &+ g_0^2 t_s \left(a+a^\dag\right)^2 -\dfrac{\omega_x}{2}N.
\end{align}
We note that $\tilde{H}_C$ and $\tilde{H}_P$ both describe the system of two coupled harmonic oscillators, but with different coupling strength.

For the Peierls Hamiltonian $\tilde{H}_P$ the polariton frequencies read
\begin{align}
&\omega^2_{P,\pm} = \dfrac{1}{2} \Big(\omega_x^2 + \tilde{\omega}_c^2\notag\\
&\pm \sqrt{\left(\tilde{\omega}_c^2 - \omega_x^2\right)^2+64 g_0^2 \tilde{t}^2 \omega_c \omega_x},
\Big),
\end{align}
where $\tilde{\omega}_c = \sqrt{\omega_c\left(\omega_c + 4 g_0^2 t_s\right)}$.
Quite interestingly, we see that the light-matter coupling $g_0$, that within the Peierls substitution only amounts to the vector potential amplitude $A_0$, enters always in front of a hopping term. We can understand this result by recalling that within the Peierls substitution the effective momentum matrix element is given by the hopping operator itself. This has some interesting consequence. In contrast to the dipole (or Coulomb) gauge Hamiltonian, $\omega_{P,\pm}$ depends to the hopping amplitude $\tilde{t}$. Moreover, we emphasize that the light-matter coupling  $\gamma_0 = e A_0 x_{12}\sqrt{N}$ in the polariton energy of the dipole gauge Hamiltonian and $g_0 = e A_0 \sqrt{N}$ in the Peierls Hamiltonian are different. We note that $\gamma_0$ depends on the dipole matrix element between Wannier orbitals, being dependent on the material properties, while $g_0$ is completely independent of the material.
Thus, we note that by fine-tuning $g_0$ and $\tilde{t}$ we can match the polariton frequencies obtained from the dipole gauge and Peierls Hamiltonians. Moreover, we find that $\omega_{P,-}$ goes to zero at $g_0^* = \sqrt{\omega_c \omega_x}/\left(2 \sqrt{4 \tilde{t}^2-t_s \omega_x}\right)$,
provided that $\tilde{t}>\sqrt{t_s\omega_x}/2$. A mode softening within the normal phase is usually associated with a superradiance transition. However, in our case, the mode softening comes from making the approximation of taking into account only $ k = 0$ mode. We checked that the saddle point $\alpha = 0$ of the ground-state energy of the expanded Peierls Hamiltonian  calculated at $k=0$ changes from minimum to maximum for $g_0>g_0^*$.

\subsubsection{Fluctuations Corrections to Photon Spectral Function}

A different approach to obtain polariton modes is to compute the photon Green's function and look at its poles. As we are going to see, the advantage of this method is that we also get information about polariton life-time, which was missed in the simple spin-wave theory of the $k=0$ sector. Since at the leading order in $N\rightarrow\infty$ photons and electrons decouple, we have to include Gaussian fluctuations at $1/N$ order. To this extent we expand the action up to second order in photonic fields to include the Gaussian fluctuations in the normal phase~\cite{mazza2019superradiant}. Introducing the Nambu representation of the photon fields as $\Phi^\dag (\tau)= \left(\phi^*(\tau),\phi(\tau)\right)$, the expanded action becomes
\begin{align}
\tilde{S}_{eff} = \dfrac{1}{2}\int d\tau d\tau' \Phi^\dag(\tau)\left[\mathcal{D}_0^{-1}(\tau - \tau ') - \Pi(\tau-\tau')\right]\Phi(\tau'),
\end{align}

where $\mathcal{D}_0^{-1}(\tau - \tau ')$ is the bare photon Green's function given by

\begin{align}
\mathcal{D}_0^{-1}(\tau - \tau ') = 
\begin{pmatrix}
\delta(\tau - \tau ')\left(\partial_\tau + \omega_c\right)& 0\\
0 & -\delta(\tau - \tau ')\left(\partial_\tau - \omega_c\right) 
\end{pmatrix}
\end{align}

and $\Pi(\tau-\tau')$ is the polarization,

\begin{align}
\Pi(\tau - \tau ') = 
\begin{pmatrix}
\dfrac{\delta^2 \log Z_0[\Phi,\Phi^*]}{\delta\phi^*(\tau)\delta \phi(\tau')} & \dfrac{\delta^2 \log Z_0[\Phi,\Phi^*]}{\delta\phi^*(\tau)\delta \phi^*(\tau')}\\
\dfrac{\delta^2 \log Z_0[\Phi,\Phi^*]}{\delta\phi(\tau)\delta \phi(\tau')} & \dfrac{\delta^2 \log Z_0[\Phi,\Phi^*]}{\delta\phi(\tau)\delta \phi^*(\tau')}
\end{pmatrix}\Bigg|_{\Phi(\tau) = \alpha = 0}.
\label{eq:polarization}
\end{align}

\begin{figure}[t!] 
\includegraphics[width=0.95\linewidth]{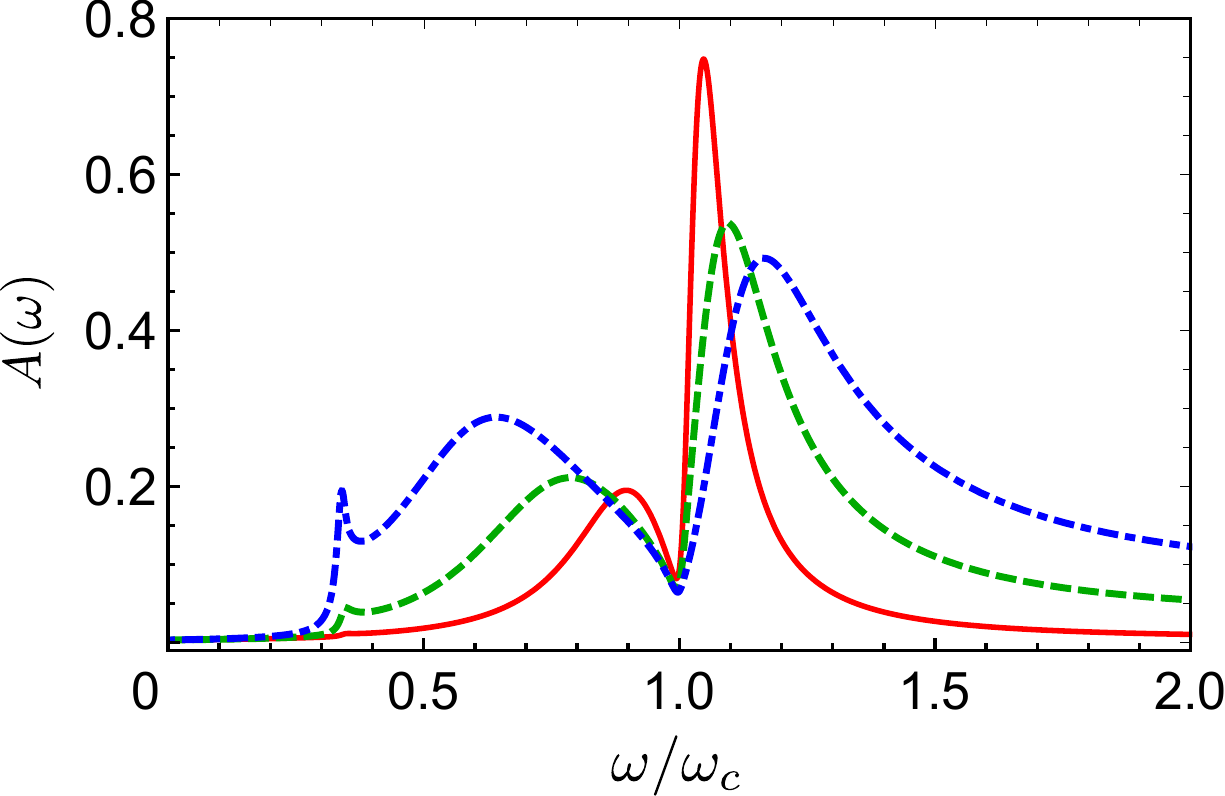}
\caption{Spectral function $A(\omega)$ as a function of frequency $\omega/\omega_c$ for $\omega_c/E_g = 1$, $t_s/E_g = 0.5$, $\tilde{t}/E_g = 0.1$, and $\eta/E_g = 0.01$. Red solid line corresponds to $\gamma_0 = 0.1 $, green dashed line corresponds to $\gamma_0= 0.3$, and blue dotdashed line corresponds to $\gamma_0/E_g = 0.6$. There are three peaks in the spectral function, where the peak around $\omega/\omega_c\approx 0.34 (=2E_k^{min})$ comes the energy gap.}
\label{fig:Aomega}
\end{figure}

Here,
\begin{align}
&Z_0 =  \int\prod_k\mathcal{D}\left[c_{k,s},c_{k,p},c^*_{k,s},c^*_{k,p}\right]e^{- S_{el-ph}},
\\
&S_{el-ph} =  \int_{0}^{\beta}d\tau \Big[\cos[2\gamma(\phi(\tau)+\phi^*(\tau))] \notag\\
&\sum_k \left(\epsilon_k\sigma^z_k
-2\tilde{t}\sin(k) \sigma^y_k\right)-\sin[2\gamma(\phi(\tau)+\phi^*(\tau))] \notag\\
&\sum_k \left(2\tilde{t}\sin(k)\sigma^z_k+\epsilon_k \sigma^y_k\right)\Big].
\end{align}

From Eq.~\eqref{eq:polarization} we find that the polarization reads
\begin{align}
&\Pi(\omega) = 
\begin{pmatrix}
1 & 1\\
1 & 1
\end{pmatrix}\chi(\omega),
\label{eq:polarization3}
\end{align}
where $\chi(\omega) \equiv K(\omega) + \langle J_d \rangle$ is the current-current correlator that has  paramagnetic and diamagnetic contributions, 
\begin{align}
&K(\tau - \tau') = \langle T_c J_p(\tau) J_p(\tau') \rangle,\\
&J_p =2\gamma \sum_k\left(\epsilon_k\sigma^y_k + 2\tilde{t}\sin(k)\sigma^z_k\right),\\
&J_d = \left(2\gamma\right)^2\sum_k\left(\epsilon_k\sigma^z_k- 2\tilde{t}\sin(k)\sigma^y_k\right).
\end{align}

Next, we find that
\begin{align}
&\chi'(\omega) = -\dfrac{2\gamma_0^2}{\pi}\Big[\int_{-\pi}^{\pi}dk \ E_k \notag\\
&+\mathcal{P} \int_{-\pi}^{\pi}dk \ E_k^2\Big(\dfrac{1}{\omega - 2 E_k} - \dfrac{1}{\omega + 2 E_k}\Big)\Big],\\
&\chi''(\omega) = \gamma_0^2\int_{-\pi}^{\pi}dk \ E_k^2\left[\delta\left(E_k - \dfrac{\omega}{2}\right) - \delta\left(E_k + \dfrac{\omega}{2}\right)\right].
\end{align}
Moreover, we note that the current-current response functions vanishes at zero frequency, $\chi(\omega = 0) = 0$.

From the dressed photon Green's function 
\begin{align}
\mathcal{D}^{-1}(\omega) = \mathcal{D}_0^{-1}(\omega) - \Pi(\omega),
\end{align}

we find that the polariton spectral function reads
\begin{align}
A(\omega) = \dfrac{1}{\pi}\text{Im}\left[\mathcal{D}_{11}(\omega)\right].
\label{eq:Aomega}
\end{align}

 In the limit $\eta\rightarrow 0$, we arrive at
\begin{align}
A(\omega)= \dfrac{1}{\pi}\dfrac{\chi''(\omega) (\omega +\omega_c)^2}{(\omega^2-\omega_c^2 +2 \omega_c\chi'(\omega))^2 + (2\omega_c \chi''(\omega) )^2}.
\label{eq:Aomegaapp}
\end{align}

We plot the resulting spectral function in Fig.~\ref{fig:Aomega}  for different values of light matter interaction $\gamma_0$. We see two peaks which move far apart as $\gamma_0$ increases and further broadens. Moreover, there are now three peaks due to the different shape of $\chi(\omega)$.

\begin{figure}[t!] 
\includegraphics[width=0.95\linewidth]{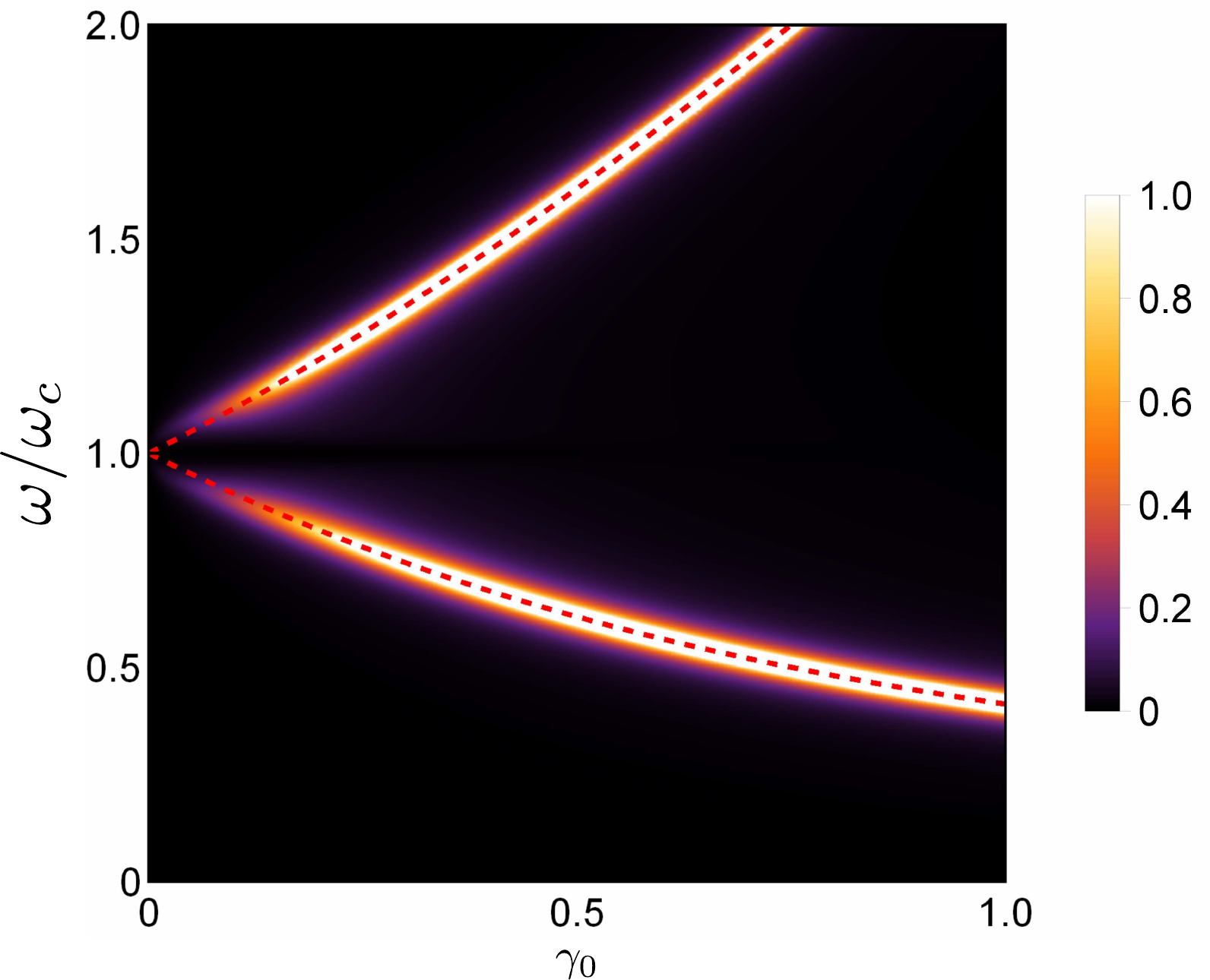}
\caption{Spectral function $A^{k=0}(\omega)$ at $k = 0$ as a function of $\gamma_0$ and $\omega/\omega_c$. The parameters are fixed as $\omega_c/E_g = 1$, $t_s/E_g = 0.5$, $\tilde{t}/E_g = 0.1$, and $\eta/E_g = 0.01$. Red dashed lines correspond to the frequencies of the polariton modes given by Eq.~\eqref{eq:omegaCoulomb}. There is an excellent agreement between the two.
}
\label{fig:A3Dk0}
\end{figure}

To make the connection with the previous section, we consider only the $k = 0$ contribution to the spectral function  Eq.~\eqref{eq:Aomegaapp}. We note that in this case both the real and imaginary parts of the polarization $\chi^{k=0}(\omega)$ have a single peak at $\omega/E_g = 1$, and, as a result, $A^{k=0}(\omega)$ has two branches as a function of the light-matter coupling (see Fig.~\ref{fig:A3Dk0}). In Fig.~\ref{fig:A3Dk0} we compare those branches with the analytical result and find perfect agreement.

It is instructive to compare the analytical estimate with the calculation here. Given the photon Green's function Eq.~\eqref{eq:Aomegaapp} the polariton frequencies are approximately given by the equation
\begin{align}
\omega^2 \approx \omega_c\left(\omega_c -2 \chi'(\omega)\right).
\end{align}

\begin{figure}[t!] 
\includegraphics[width=0.95\linewidth]{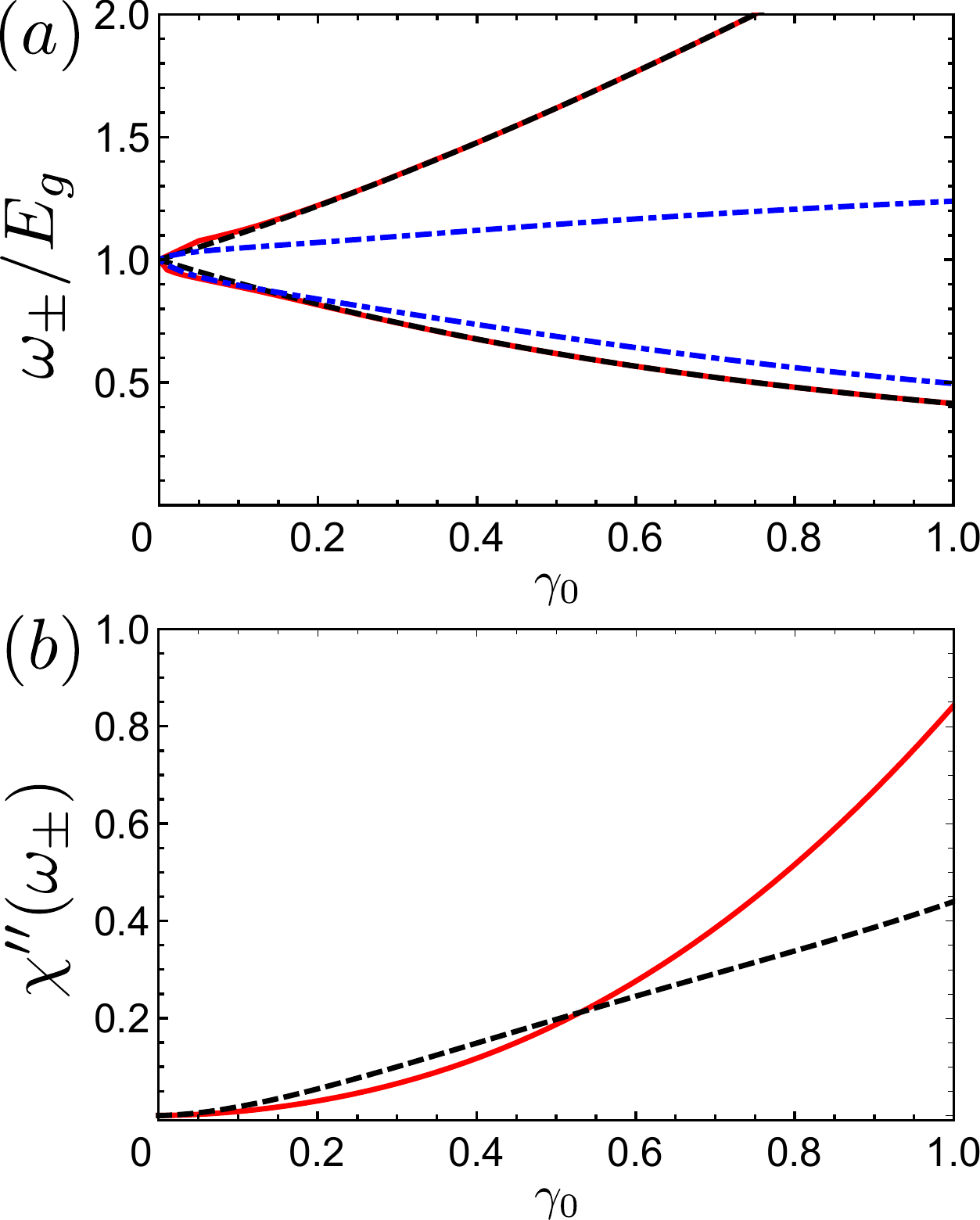}
\caption{(a) Polariton branches for our Coulomb gauge Hamiltonian: Red solid lines correspond to the maximum in $A^{k=0}(\omega)$, black dashed lines correspond to the analytical solution given by Eq.~\eqref{eq:omegaCoulomb}, and blue dotdashed solid lines correspond to the maximum of the spectral function $A(\omega)$.
(b) Polariton lifetime $\chi''(\omega_{\pm})$ as a function of the light-matter coupling $\gamma_0$. Red solid line corresponds to the lifetime of the upper polariton branch $\omega_{+}$, while black dashed line corresponds to the lifetime of the lower polariton branch $\omega_{-}$. The parameters are fixed as $\omega_c/E_g = 1$,  $t_s/E_g = 0.5$, $\tilde{t}/E_g = 0.1$, and $\eta/E_g = 0.01$.}
\label{fig:polaritoncomp}
\end{figure}

Finally, we compare the maximum of the full spectral function $A(\omega)$, reduced to $k = 0$ contribution spectral function $A^{k=0}(\omega)$ and the polariton frequencies obtained analytically $\omega_{C,\pm}$ in Fig.~\ref{fig:polaritoncomp}~(a). As already noted, there is an excellent agreement between  and the maximum of $A^{k=0}(\omega)$, while the maximum of $A(\omega)$ is quite shifted. This shift comes from the finite width of the peaks in the full spectral function as it contains contributions from all modes, and not only $k = 0$ mode. The width of the polariton branches is plotted in Fig.~\ref{fig:polaritoncomp}~(b). We note that for small values of the light-matter coupling, the width of the lower polariton branch, $\chi''(\omega_-)$, is larger than for the upper polariton branch, $\chi''(\omega_+)$, while for large values of $\gamma_0$, $\chi''(\omega_+)$ is much larger than $\chi''(\omega_-)$.

\section{Conclusions} \label{sec:conclusions}

In this work, we have discussed the issue of gauge fixing and gauge equivalence in models of strongly correlated electrons coupled to quantum light. In particular, we have presented a general formalism to write down quantum light-matter Hamiltonian for projected degrees of freedom, either in the Coulomb or dipole gauge, which remain fully equivalent under a change of gauge, i.e. related by a unitary transformation. While this is naturally implemented in a full microscopic description of light-matter interactions its extension to projected models introduce a number of conceptual and practical subtleties and have recently spurred significant interest, both in the solid-state and cavity QED communities.

The central idea of our approach, which generalises to the case of strongly correlated electrons the recent developments obtained for well-known quantum optics models such as Rabi or Dicke models~\cite{di2019resolution,garziano2020gauge}, is that projection onto a subset of degrees of freedom should be done before coupling matter and light and that appropriate electron-photon coupling should be generated by applying a unitary transformation to the matter-only or to the photon-only degrees of freedom, depending on the chosen gauge, which as a result become entangled.

Our result for the dipole gauge Hamiltonian, containing a linear coupling of the photon field to the electrons and an instantaneous self-interaction term for the latter similar to the continuum case, matches recent results obtained in the literature~\cite{cottet2015electron,li2020electromagnetic}. The projected Hamiltonian in the Coulomb gauge instead comes with new features, in particular a highly non-linear photon-electron coupling which is a genuine feature of working with a projected model. The non-linear structure of the light-matter coupling emerges through phase factors dressing the electronic degrees of freedom, which generalise the well-known Peierls phases often used in solid-state context. We show that our projected Coulomb gauge Hamiltonian reduces to the one obtained through the Peierls substitution when disregarding the contribution of local on-site orbital degrees of freedom to the light-matter coupling.  Despite the radically different structure of the projected Hamiltonian in the Coulomb and dipole gauge we explicitly show their gauge equivalence, i.e. how one could move from one to the other by a unitary transformation.
These has two important consequences. First, it implies that physical, gauge invariant, quantities are enforced to be the same when computed using different Hamiltonian. Furthermore, it highlights the importance of treating the non-linear light-matter coupling of the Coulomb gauge Hamiltonian non-perturbatively and that uncontrolled weak coupling expansions can lead to problems with gauge invariance in the ultrastrong coupling regime.
As first application of our formalism, we study an interacting two-orbital model coupled to a single mode cavity with uniform vector potential, recently introduced in the context of excitonic superradiance and related no-go theorems. Working in the dipole gauge, in which photons only enter linearly, we derive an effective action for the electronic degrees of freedom and show that light and matter become fully decoupled in the limit $\omega \rightarrow 0$, thus preventing ground-state superradiance in accordance with a general no-go theorem. We recover the same result within our Coulomb gauge Hamiltonian, that we solve by decoupling electrons and photons in mean-field theory. Interestingly, we show that, within a Coulomb gauge formulation, in order to obtain well-defined results all the way into the ultrastrong coupling regime it is crucial to treat the light-matter coupling non-perturbatively. In fact we explicitly show that expanding the Coulomb gauge Hamiltonian to lowest order, as often done in the context of the Peierls substitution, leads to an unbounded ground-state energy for sufficiently strong light-matter coupling. Finally, we compute the polariton spectrum of the model and show that while the ground state of the system factorizes and lacks any entanglement between light and matter, finite frequency excitations (polaritons) depend on light-matter coupling, as expected from the results obtained within the dipole gauge. We show explicitly that polariton excitations obtained within our projected dipole and Coulomb gauge Hamiltonian are identical for any value of light-matter coupling, a further demonstration of gauge equivalence. 
This work suggests several possible extensions. From one side it would be interesting to broaden our model to consider a spatially-varying vector potential, following the recent prediction of superradiance in such a setup~\cite{guerci2020superradiant,andolina2020theory}. Another promising direction would be to explore the residual light-matter coupling of finite frequency excitations and the possibility of turning them superradiant using a combination of drive and dissipation, as done in other non-equilibrium contexts.

\section*{Acknowledgments} 
We thank F. Nori and S. Savasta for enlightening discussions and the warm hospitality of RIKEN Theoretical Quantum Physics Laboratory.
We acknowledge support from Jeunes Equipes de l'Institut de Physique du Coll\'ege de France and from the ANR grant ”NonEQuMat”  (ANR-19-CE47-0001). This project has received funding from the
 European Union’s Horizon 2020 research and innovation programme under the Marie Skłodowska-Curie Grant Agreement No. 892800.
\appendix

\section{Electron-electron interactions in the dipole gauge} \label{transformeeinteractions}

The dipole gauge Hamiltonian is obtained  by performing the unitary transformation, $\mathcal{H}_D = \mathcal{U} \mathcal{H}_C \mathcal{U}^\dag$. Here, we present details on the derivation for the electron-electron interactions term $\mathcal{H}_{ee}$. To simplify the calculation, we rewrite the unitary operator $\mathcal{U}$ in the form $\mathcal{U} = e^{is}$, where
\begin{align}
s =  \left(a+a^{\dagger}\right)\int d\bb{r} \ \psi^\dag(\bb{r}) \chi(\bb{r}) \psi(\bb{r}).
\end{align}

By using the Campbell-Hausdorff formula, we obtain 
\begin{align}
&\mathcal{U} \mathcal{H}_{ee} \mathcal{U}^\dag = e^{is} \mathcal{H}_{ee} e^{-is} = \mathcal{H}_{ee} + \left[is, \mathcal{H}_{ee}\right]\notag\\
&+\dfrac{1}{2!}\left[is, \left[is, \mathcal{H}_{ee}\right]\right] + \dots
\end{align}

Since $\left[s, \mathcal{H}_{ee}\right] = 0$, we note that the electron-electron interactions term remains the same after performing the gauge transformation
\begin{align}
\mathcal{U} \mathcal{H}_{ee} \mathcal{U}^\dag = \mathcal{H}_{ee}.
\end{align}

We now show that the same is true in the projected two-band model discussed in the main text. Specifically we show that, given $H_{ee}=U \sum_{j}n_{j1}n_{j2}$  and the projected unitary  $U =e^{i\gamma\left(a + a^\dag\right)\sum_j\sigma_x^{j}}$
we have
\begin{equation}
U^{\dagger} H_{ee}U=H_{ee}
\end{equation}
To show this we use the transformation rules of the fermionic and pseudospin operators, given in the main text in Eq.(\ref{eq:Ufermions_2band}-\ref{eq:Ufermions_2band_dag}). We first rewrite the density electrons at site $j$ and orbital $\alpha=1,2$ as
\begin{align}
n_{j\alpha}=
\frac{1-(-1)^{\alpha}\sigma^z_j}{2}
\end{align}
which transforms under the action of $U^{\dagger}$ as
\begin{align}
U^{\dagger}n_{j\alpha}U=
\left(\frac{1-(-1)^{\alpha}\left(\cos(2A)\sigma^z_j-\sin(2A)\sigma^y_j\right)}{2}\right)
\end{align}
with $A=\gamma(a+a^{\dagger})$. The transformed Hubbard interaction therefore reads, in terms of pseudo spin operators
\begin{align}
& U^{\dagger}n_{j1}UU^{\dagger}n_{j2}U=\nonumber\\
&=\frac{1}{4}\left(n^2_j-\cos^2(2A)\left(\sigma^z_j\right)^2-\sin^2(2A)\left(\sigma^y_j\right)^2\right)=\nonumber\\
&=n_{j1}n_{j2}
\end{align}
where in the last step we we have used the fact that $[n_j,\sigma^\alpha_j]=0$ and $\left\{\sigma^\alpha_j,\sigma^{\beta}_{j'}\right\}=2\delta_{\alpha\beta}\delta_{jj'}$ as well as that we can rewite the square of the pseudospin operators only in terms of the density, i.e. $\left(\sigma^z_j\right)^2=\left(\sigma^y_j\right)^2=n_j-2n_{j1}n_{j2}$.

\section{Photonic action}\label{photonicaction}

The photonic Hamiltonian $H_{ph} = \omega_c a^\dag a$ is equivalent to a one-dimensional harmonic oscillator. To make the connection explicit, we rewrite the photonic operators $a$ ($a^\dag$) as

\begin{align}
&a = \sqrt{\dfrac{\omega_c}{2}} x + i\sqrt{\dfrac{1}{2\omega_c}}p,\label{eq:aoperator}\\
&a^\dag = \sqrt{\dfrac{\omega_c}{2}} x - i\sqrt{\dfrac{1}{2\omega_c}}p\label{eq:adagoperator},
\end{align}
where $x$ and $p$ are the position and momentum operators, respectively, and we find that $H_{ph} = \left(1/2\right)\left(\omega_c^2 x^2 + p^2\right)$.

The photonic action reads
\begin{align}
S_{ph} = \frac{1}{2}\int_{0}^{\beta}d\tau\left[\omega_c^2x(\tau)^2 + p(\tau)^2 + 2i x(\tau)\dot{p}(\tau)\right],
\end{align}
where the full derivatives $x^2$, $p^2$ and $x p$ were omitted. Performing the Gaussian integration over $x(\tau)$, we obtain for the photonic action
\begin{align}
S_{ph}[p] = \frac{1}{2}\int_{0}^{\beta}d\tau\left[ p_x(\tau)^2  +  \dfrac{\dot{p}(\tau)^2}{\omega_c^2}\right].
\end{align}

\section{Mean-field solution: Dipole gauge Hamiltonian}\label{dipolegauge}
Here, we present details of the mean-field solution of the light-matter Hamiltonian in the dipole gauge, Eq.~\eqref{eq:Hdipoletwoband}. We assume that there are no correlations between the electronic and photonic systems. This allows us to do the factorization of the wavefunction as
 $|\Psi\rangle = |\psi\rangle |\phi\rangle$, where $|\psi\rangle$ ($|\phi\rangle$) corresponds to the electronic (photonic) system. Moreover, we assume that $|\phi\rangle$ is a coherent state, such that the photonic order parameter $\alpha$ could be introduced as $\langle \alpha | a |\phi\rangle = \alpha \sqrt{N}$. Also, we note that in general $\alpha$ has both real and imaginary parts, thus it could be written as $\alpha = \alpha' + i\alpha''$. To treat $\left(\sum_k\sigma_x^k\right)^2$ term in the dipole gauge Hamiltonian given by Eq.~\eqref{eq:Hdipoletwoband}, we employ a mean-field approximation as

\begin{align}
&\left(\sum_k\sigma_x^k\right)^2 = \sum_k\left(\langle \sigma_x^k\rangle + [\sigma_x^k - \langle \sigma_x^k\rangle]\right)\notag\\
&\times \sum_q\left(\langle \sigma_x^q\rangle + [\sigma_x^q - \langle \sigma_x^q\rangle]\right) = 2 M \sum_k\sigma_x^k - M^2,
\label{eq:sigmakq}
\end{align}
where $M = \sum_q\langle\sigma_x^q\rangle$. And the electron-electron interactions Hamiltonian $H_{ee}$ could be approximated by using Eq.~\eqref{eq:hartreefockmeanf}.

We start by solving the photonic mean-field Hamiltonian that reads
\begin{align}
H_{D,ph}^{mf} =
\omega_c a^\dag a + i\omega_c \gamma M (a - a^\dag).
\label{eq:dipolephotonmeanf}
\end{align}
Rewriting $a$ and $a^\dag$ in terms of the position $x$ and momentum $p$ operators as in Appendix~\ref{photonicaction}, we arrive at
\begin{align}
H_{D,ph}^{mf} = \dfrac{1}{2}\left[\omega_c^2 x^2  + \left(p - \sqrt{2\omega_c}\gamma M\right)^2\right],
\end{align}
which is the Hamiltonian for a one-dimensional harmonic oscillator. Using that $\langle \phi| x |\phi\rangle  = 0$ and $\langle \phi| p - \sqrt{2\omega_c}\gamma M|\phi\rangle  = 0$, we obtain that $\alpha' = 0$ and $\alpha''$ is given by
\begin{align}
\alpha'' = \gamma_0\dfrac{M}{N}.
\label{eq:eqforalpha}
\end{align}

However, in the Coulomb gauge we obtained that the photonic order parameter is zero for any two-orbital model. To make the connection between the expectation value of the photonic operators in the Coulomb and dipole gauge,  we should apply the unitary transformation $U\left(\chi\right)$ to the photonic annihilation (creation) operators. This brings us to
\begin{align}
&U\left(\chi\right)^\dag a U\left(\chi\right) = a + i\gamma\sum_k\sigma_x^k.
\end{align}

We write the electronic mean-field Hamiltonian in the form $H_{D,el}^{mf}$ $=$ $\sum_{a} h^a_k\sigma^a_k$,  with $a = x,y,z$, 
and the coefficients $h^a_k$ are given by
\begin{align}
&h^x_k = -2 \omega_c \gamma \sqrt{N}\left( \alpha'' - \dfrac{\gamma M}{\sqrt{N}}\right) - U\mathcal{I}',
\label{eq:hxdipoleelel}\\
&h^y_k = 2\tilde{t}\sin(k) + U\mathcal{I}'',\label{eq:hydipoleelel}\\
&h^z_k = \epsilon_k - U\dfrac{m}{2}.\label{eq:hzdipoleelel}
\end{align}

Next, we find the ground-state energy of the dipole gauge Hamiltonian. To simplify the calculations we put $U = 0$ and we obtain
\begin{align}
&\dfrac{E_{GS}}{N}= \omega_c\left[\left(\alpha'\right)^2 + \left(\alpha'' - \dfrac{\gamma M}{\sqrt{N}}\right)^2\right] \notag\\
&- \dfrac{1}{N}\sum_k  \sqrt{\epsilon_k^2 + 4\tilde{t}\sin^2(k)}.
\label{eq:groundenergydipole}
\end{align}
We note that, as in the case of the ground-state energy calculated in the Coulomb gauge, Eq.~\eqref{eq:groundenergycoulomb}, $E_{GS}$ is separated into a sum of the energy of the photonic system and electronic system, respectively.


\begin{thebibliography}{99}%
	\makeatletter
	\providecommand \@ifxundefined [1]{%
		\@ifx{#1\undefined}
	}%
	\providecommand \@ifnum [1]{%
		\ifnum #1\expandafter \@firstoftwo
		\else \expandafter \@secondoftwo
		\fi
	}%
	\providecommand \@ifx [1]{%
		\ifx #1\expandafter \@firstoftwo
		\else \expandafter \@secondoftwo
		\fi
	}%
	\providecommand \natexlab [1]{#1}%
	\providecommand \enquote  [1]{``#1''}%
	\providecommand \bibnamefont  [1]{#1}%
	\providecommand \bibfnamefont [1]{#1}%
	\providecommand \citenamefont [1]{#1}%
	\providecommand \href@noop [0]{\@secondoftwo}%
	\providecommand \href [0]{\begingroup \@sanitize@url \@href}%
	\providecommand \@href[1]{\@@startlink{#1}\@@href}%
	\providecommand \@@href[1]{\endgroup#1\@@endlink}%
	\providecommand \@sanitize@url [0]{\catcode `\\12\catcode `\$12\catcode
		`\&12\catcode `\#12\catcode `\^12\catcode `\_12\catcode `\%12\relax}%
	\providecommand \@@startlink[1]{}%
	\providecommand \@@endlink[0]{}%
	\providecommand \url  [0]{\begingroup\@sanitize@url \@url }%
	\providecommand \@url [1]{\endgroup\@href {#1}{\urlprefix }}%
	\providecommand \urlprefix  [0]{URL }%
	\providecommand \Eprint [0]{\href }%
	\providecommand \doibase [0]{http://dx.doi.org/}%
	\providecommand \selectlanguage [0]{\@gobble}%
	\providecommand \bibinfo  [0]{\@secondoftwo}%
	\providecommand \bibfield  [0]{\@secondoftwo}%
	\providecommand \translation [1]{[#1]}%
	\providecommand \BibitemOpen [0]{}%
	\providecommand \bibitemStop [0]{}%
	\providecommand \bibitemNoStop [0]{.\EOS\space}%
	\providecommand \EOS [0]{\spacefactor3000\relax}%
	\providecommand \BibitemShut  [1]{\csname bibitem#1\endcsname}%
	\let\auto@bib@innerbib\@empty
	\bibitem [{\citenamefont {Raimond}\ \emph {et~al.}(2001)\citenamefont
		{Raimond}, \citenamefont {Brune},\ and\ \citenamefont
		{Haroche}}]{raimond2001manipulating}%
	\BibitemOpen
	\bibfield  {author} {\bibinfo {author} {\bibfnamefont {J.-M.}\ \bibnamefont
			{Raimond}}, \bibinfo {author} {\bibfnamefont {M.}~\bibnamefont {Brune}}, \
		and\ \bibinfo {author} {\bibfnamefont {S.}~\bibnamefont {Haroche}},\
	}\href@noop {} {\bibfield  {journal} {\bibinfo  {journal} {Rev. Mod. Phys.}\
		}\textbf {\bibinfo {volume} {73}},\ \bibinfo {pages} {565} (\bibinfo {year}
		{2001})}\BibitemShut {NoStop}%
	\bibitem [{\citenamefont {Wallraff}\ \emph {et~al.}(2004)\citenamefont
		{Wallraff}, \citenamefont {Schuster}, \citenamefont {Blais}, \citenamefont
		{Frunzio}, \citenamefont {Huang}, \citenamefont {Majer}, \citenamefont
		{Kumar}, \citenamefont {Girvin},\ and\ \citenamefont
		{Schoelkopf}}]{wallraff2004strong}%
	\BibitemOpen
	\bibfield  {author} {\bibinfo {author} {\bibfnamefont {A.}~\bibnamefont
			{Wallraff}}, \bibinfo {author} {\bibfnamefont {D.~I.}\ \bibnamefont
			{Schuster}}, \bibinfo {author} {\bibfnamefont {A.}~\bibnamefont {Blais}},
		\bibinfo {author} {\bibfnamefont {L.}~\bibnamefont {Frunzio}}, \bibinfo
		{author} {\bibfnamefont {R.-S.}\ \bibnamefont {Huang}}, \bibinfo {author}
		{\bibfnamefont {J.}~\bibnamefont {Majer}}, \bibinfo {author} {\bibfnamefont
			{S.}~\bibnamefont {Kumar}}, \bibinfo {author} {\bibfnamefont {S.~M.}\
			\bibnamefont {Girvin}}, \ and\ \bibinfo {author} {\bibfnamefont {R.~J.}\
			\bibnamefont {Schoelkopf}},\ }\href@noop {} {\bibfield  {journal} {\bibinfo
			{journal} {Nature (London)}\ }\textbf {\bibinfo {volume} {431}},\ \bibinfo
		{pages} {162} (\bibinfo {year} {2004})}\BibitemShut {NoStop}%
	\bibitem [{\citenamefont {Carusotto}\ and\ \citenamefont
		{Ciuti}(2013)}]{carusotto2013quantum}%
	\BibitemOpen
	\bibfield  {author} {\bibinfo {author} {\bibfnamefont {I.}~\bibnamefont
			{Carusotto}}\ and\ \bibinfo {author} {\bibfnamefont {C.}~\bibnamefont
			{Ciuti}},\ }\href@noop {} {\bibfield  {journal} {\bibinfo  {journal} {Rev.
				Mod. Phys.}\ }\textbf {\bibinfo {volume} {85}},\ \bibinfo {pages} {299}
		(\bibinfo {year} {2013})}\BibitemShut {NoStop}%
	\bibitem [{\citenamefont {Houck}\ \emph {et~al.}(2012)\citenamefont {Houck},
		\citenamefont {T{\"u}reci},\ and\ \citenamefont {Koch}}]{houck2012chip}%
	\BibitemOpen
	\bibfield  {author} {\bibinfo {author} {\bibfnamefont {A.~A.}\ \bibnamefont
			{Houck}}, \bibinfo {author} {\bibfnamefont {H.~E.}\ \bibnamefont
			{T{\"u}reci}}, \ and\ \bibinfo {author} {\bibfnamefont {J.}~\bibnamefont
			{Koch}},\ }\href@noop {} {\bibfield  {journal} {\bibinfo  {journal} {Nat.
				Phys.}\ }\textbf {\bibinfo {volume} {8}},\ \bibinfo {pages} {292} (\bibinfo
		{year} {2012})}\BibitemShut {NoStop}%
	\bibitem [{\citenamefont {Le~Hur}\ \emph {et~al.}(2016)\citenamefont {Le~Hur},
		\citenamefont {Henriet}, \citenamefont {Petrescu}, \citenamefont {Plekhanov},
		\citenamefont {Roux},\ and\ \citenamefont {Schir{\'o}}}]{lehur2016many}%
	\BibitemOpen
	\bibfield  {author} {\bibinfo {author} {\bibfnamefont {K.}~\bibnamefont
			{Le~Hur}}, \bibinfo {author} {\bibfnamefont {L.}~\bibnamefont {Henriet}},
		\bibinfo {author} {\bibfnamefont {A.}~\bibnamefont {Petrescu}}, \bibinfo
		{author} {\bibfnamefont {K.}~\bibnamefont {Plekhanov}}, \bibinfo {author}
		{\bibfnamefont {G.}~\bibnamefont {Roux}}, \ and\ \bibinfo {author}
		{\bibfnamefont {M.}~\bibnamefont {Schir{\'o}}},\ }\href@noop {} {\bibfield
		{journal} {\bibinfo  {journal} {C. R. Phys.}\ }\textbf {\bibinfo {volume}
			{17}},\ \bibinfo {pages} {808} (\bibinfo {year} {2016})}\BibitemShut
	{NoStop}%
	\bibitem [{\citenamefont {Fitzpatrick}\ \emph {et~al.}(2017)\citenamefont
		{Fitzpatrick}, \citenamefont {Sundaresan}, \citenamefont {Li}, \citenamefont
		{Koch},\ and\ \citenamefont {Houck}}]{fitzpatrick2017observation}%
	\BibitemOpen
	\bibfield  {author} {\bibinfo {author} {\bibfnamefont {M.}~\bibnamefont
			{Fitzpatrick}}, \bibinfo {author} {\bibfnamefont {N.~M.}\ \bibnamefont
			{Sundaresan}}, \bibinfo {author} {\bibfnamefont {A.~C.~Y.}\ \bibnamefont {Li}},
		\bibinfo {author} {\bibfnamefont {J.}~\bibnamefont {Koch}}, \ and\ \bibinfo
		{author} {\bibfnamefont {A.~A.}\ \bibnamefont {Houck}},\ }\href@noop {}
	{\bibfield  {journal} {\bibinfo  {journal} {Phys. Rev. X}\ }\textbf {\bibinfo
			{volume} {7}},\ \bibinfo {pages} {011016} (\bibinfo {year}
		{2017})}\BibitemShut {NoStop}%
	\bibitem [{\citenamefont {Ma}\ \emph {et~al.}(2019)\citenamefont {Ma},
		\citenamefont {Saxberg}, \citenamefont {Owens}, \citenamefont {Leung},
		\citenamefont {Lu}, \citenamefont {Simon},\ and\ \citenamefont
		{Schuster}}]{ma2019dissipatively}%
	\BibitemOpen
	\bibfield  {author} {\bibinfo {author} {\bibfnamefont {R.}~\bibnamefont
			{Ma}}, \bibinfo {author} {\bibfnamefont {B.}~\bibnamefont {Saxberg}},
		\bibinfo {author} {\bibfnamefont {C.}~\bibnamefont {Owens}}, \bibinfo
		{author} {\bibfnamefont {N.}~\bibnamefont {Leung}}, \bibinfo {author}
		{\bibfnamefont {Y.}~\bibnamefont {Lu}}, \bibinfo {author} {\bibfnamefont
			{J.}~\bibnamefont {Simon}}, \ and\ \bibinfo {author} {\bibfnamefont {D.~I.}\
			\bibnamefont {Schuster}},\ }\href {\doibase 10.1038/s41586-019-0897-9}
	{\bibfield  {journal} {\bibinfo  {journal} {Nature}\ }\textbf {\bibinfo
			{volume} {566}},\ \bibinfo {pages} {51} (\bibinfo {year} {2019})}\BibitemShut
	{NoStop}%
	\bibitem [{\citenamefont {Klinder}\ \emph {et~al.}(2015)\citenamefont
		{Klinder}, \citenamefont {Ke{\ss}ler}, \citenamefont {Bakhtiari},
		\citenamefont {Thorwart},\ and\ \citenamefont
		{Hemmerich}}]{klinder2015observation}%
	\BibitemOpen
	\bibfield  {author} {\bibinfo {author} {\bibfnamefont {J.}~\bibnamefont
			{Klinder}}, \bibinfo {author} {\bibfnamefont {H.}~\bibnamefont {Ke{\ss}ler}},
		\bibinfo {author} {\bibfnamefont {M.~R.}\ \bibnamefont {Bakhtiari}}, \bibinfo
		{author} {\bibfnamefont {M.}~\bibnamefont {Thorwart}}, \ and\ \bibinfo
		{author} {\bibfnamefont {A.}~\bibnamefont {Hemmerich}},\ }\href@noop {}
	{\bibfield  {journal} {\bibinfo  {journal} {Phys. Rev. Lett.}\ }\textbf
		{\bibinfo {volume} {115}},\ \bibinfo {pages} {230403} (\bibinfo {year}
		{2015})}\BibitemShut {NoStop}%
	\bibitem [{\citenamefont {Landig}\ \emph {et~al.}(2016)\citenamefont {Landig},
		\citenamefont {Hruby}, \citenamefont {Dogra}, \citenamefont {Landini},
		\citenamefont {Mottl}, \citenamefont {Donner},\ and\ \citenamefont
		{Esslinger}}]{landig2016quantum}%
	\BibitemOpen
	\bibfield  {author} {\bibinfo {author} {\bibfnamefont {R.}~\bibnamefont
			{Landig}}, \bibinfo {author} {\bibfnamefont {L.}~\bibnamefont {Hruby}},
		\bibinfo {author} {\bibfnamefont {N.}~\bibnamefont {Dogra}}, \bibinfo
		{author} {\bibfnamefont {M.}~\bibnamefont {Landini}}, \bibinfo {author}
		{\bibfnamefont {R.}~\bibnamefont {Mottl}}, \bibinfo {author} {\bibfnamefont
			{T.}~\bibnamefont {Donner}}, \ and\ \bibinfo {author} {\bibfnamefont
			{T.}~\bibnamefont {Esslinger}},\ }\href@noop {} {\bibfield  {journal}
		{\bibinfo  {journal} {Nature}\ }\textbf {\bibinfo {volume} {532}},\ \bibinfo
		{pages} {476} (\bibinfo {year} {2016})}\BibitemShut {NoStop}%
	\bibitem [{\citenamefont {Roux}\ \emph {et~al.}(2020)\citenamefont {Roux},
		\citenamefont {Konishi}, \citenamefont {Helson},\ and\ \citenamefont
		{Brantut}}]{roux2020strongly}%
	\BibitemOpen
	\bibfield  {author} {\bibinfo {author} {\bibfnamefont {K.}~\bibnamefont
			{Roux}}, \bibinfo {author} {\bibfnamefont {H.}~\bibnamefont {Konishi}},
		\bibinfo {author} {\bibfnamefont {V.}~\bibnamefont {Helson}}, \ and\ \bibinfo
		{author} {\bibfnamefont {J.-P.}\ \bibnamefont {Brantut}},\ }\href {\doibase
		10.1038/s41467-020-16767-8} {\bibfield  {journal} {\bibinfo  {journal} {Nat.
				Commun.}\ }\textbf {\bibinfo {volume} {11}},\ \bibinfo {pages} {2974}
		(\bibinfo {year} {2020})}\BibitemShut {NoStop}%
	\bibitem [{\citenamefont {Todorov}\ \emph {et~al.}(2010)\citenamefont
		{Todorov}, \citenamefont {Andrews}, \citenamefont {Colombelli}, \citenamefont
		{De~Liberato}, \citenamefont {Ciuti}, \citenamefont {Klang}, \citenamefont
		{Strasser},\ and\ \citenamefont {Sirtori}}]{todorov2010ultrastrong}%
	\BibitemOpen
	\bibfield  {author} {\bibinfo {author} {\bibfnamefont {Y.}~\bibnamefont
			{Todorov}}, \bibinfo {author} {\bibfnamefont {A.~M.}\ \bibnamefont
			{Andrews}}, \bibinfo {author} {\bibfnamefont {R.}~\bibnamefont {Colombelli}},
		\bibinfo {author} {\bibfnamefont {S.}~\bibnamefont {De~Liberato}}, \bibinfo
		{author} {\bibfnamefont {C.}~\bibnamefont {Ciuti}}, \bibinfo {author}
		{\bibfnamefont {P.}~\bibnamefont {Klang}}, \bibinfo {author} {\bibfnamefont
			{G.}~\bibnamefont {Strasser}}, \ and\ \bibinfo {author} {\bibfnamefont
			{C.}~\bibnamefont {Sirtori}},\ }\href {\doibase
		10.1103/PhysRevLett.105.196402} {\bibfield  {journal} {\bibinfo  {journal}
			{Phys. Rev. Lett.}\ }\textbf {\bibinfo {volume} {105}},\ \bibinfo {pages}
		{196402} (\bibinfo {year} {2010})}\BibitemShut {NoStop}%
	\bibitem [{\citenamefont {Scalari}\ \emph {et~al.}(2012)\citenamefont
		{Scalari}, \citenamefont {Maissen}, \citenamefont {Tur{\v{c}}inkov{\'a}},
		\citenamefont {Hagenm{\"u}ller}, \citenamefont {De~Liberato}, \citenamefont
		{Ciuti}, \citenamefont {Reichl}, \citenamefont {Schuh}, \citenamefont
		{Wegscheider}, \citenamefont {Beck},\ and\ \citenamefont
		{Faist}}]{scalari2012ultrastrong}%
	\BibitemOpen
	\bibfield  {author} {\bibinfo {author} {\bibfnamefont {G.}~\bibnamefont
			{Scalari}}, \bibinfo {author} {\bibfnamefont {C.}~\bibnamefont {Maissen}},
		\bibinfo {author} {\bibfnamefont {D.}~\bibnamefont {Tur{\v{c}}inkov{\'a}}},
		\bibinfo {author} {\bibfnamefont {D.}~\bibnamefont {Hagenm{\"u}ller}},
		\bibinfo {author} {\bibfnamefont {S.}~\bibnamefont {De~Liberato}}, \bibinfo
		{author} {\bibfnamefont {C.}~\bibnamefont {Ciuti}}, \bibinfo {author}
		{\bibfnamefont {C.}~\bibnamefont {Reichl}}, \bibinfo {author} {\bibfnamefont
			{D.}~\bibnamefont {Schuh}}, \bibinfo {author} {\bibfnamefont
			{W.}~\bibnamefont {Wegscheider}}, \bibinfo {author} {\bibfnamefont
			{M.}~\bibnamefont {Beck}}, \ and\ \bibinfo {author} {\bibfnamefont
			{J.}~\bibnamefont {Faist}},\ }\href@noop {} {\bibfield  {journal} {\bibinfo
			{journal} {Science}\ }\textbf {\bibinfo {volume} {335}},\ \bibinfo {pages}
		{1323} (\bibinfo {year} {2012})}\BibitemShut {NoStop}%
	\bibitem [{\citenamefont {Zhang}\ \emph {et~al.}(2016)\citenamefont {Zhang},
		\citenamefont {Lou}, \citenamefont {Li}, \citenamefont {Reno}, \citenamefont
		{Pan}, \citenamefont {Watson}, \citenamefont {Manfra},\ and\ \citenamefont
		{Kono}}]{zhang2016collective}%
	\BibitemOpen
	\bibfield  {author} {\bibinfo {author} {\bibfnamefont {Q.}~\bibnamefont
			{Zhang}}, \bibinfo {author} {\bibfnamefont {M.}~\bibnamefont {Lou}}, \bibinfo
		{author} {\bibfnamefont {X.}~\bibnamefont {Li}}, \bibinfo {author}
		{\bibfnamefont {J.~L.}\ \bibnamefont {Reno}}, \bibinfo {author}
		{\bibfnamefont {W.}~\bibnamefont {Pan}}, \bibinfo {author} {\bibfnamefont
			{J.~D.}\ \bibnamefont {Watson}}, \bibinfo {author} {\bibfnamefont {M.~J.}\
			\bibnamefont {Manfra}}, \ and\ \bibinfo {author} {\bibfnamefont
			{J.}~\bibnamefont {Kono}},\ }\href {\doibase 10.1038/nphys3850} {\bibfield
		{journal} {\bibinfo  {journal} {Nat. Phys.}\ }\textbf {\bibinfo {volume}
			{12}},\ \bibinfo {pages} {1005} (\bibinfo {year} {2016})}\BibitemShut
	{NoStop}%
	\bibitem [{\citenamefont {Slootsky}\ \emph {et~al.}(2014)\citenamefont
		{Slootsky}, \citenamefont {Liu}, \citenamefont {Menon},\ and\ \citenamefont
		{Forrest}}]{slootsky2014room}%
	\BibitemOpen
	\bibfield  {author} {\bibinfo {author} {\bibfnamefont {M.}~\bibnamefont
			{Slootsky}}, \bibinfo {author} {\bibfnamefont {X.}~\bibnamefont {Liu}},
		\bibinfo {author} {\bibfnamefont {V.~M.}\ \bibnamefont {Menon}}, \ and\
		\bibinfo {author} {\bibfnamefont {S.~R.}\ \bibnamefont {Forrest}},\
	}\href@noop {} {\bibfield  {journal} {\bibinfo  {journal} {Phys. Rev. Lett.}\
		}\textbf {\bibinfo {volume} {112}},\ \bibinfo {pages} {076401} (\bibinfo
		{year} {2014})}\BibitemShut {NoStop}%
	\bibitem [{\citenamefont {Liu}\ \emph {et~al.}(2015)\citenamefont {Liu},
		\citenamefont {Galfsky}, \citenamefont {Sun}, \citenamefont {Xia},
		\citenamefont {Lin}, \citenamefont {Lee}, \citenamefont {K{\'e}na-Cohen},\
		and\ \citenamefont {Menon}}]{liu2015strong}%
	\BibitemOpen
	\bibfield  {author} {\bibinfo {author} {\bibfnamefont {X.}~\bibnamefont
			{Liu}}, \bibinfo {author} {\bibfnamefont {T.}~\bibnamefont {Galfsky}},
		\bibinfo {author} {\bibfnamefont {Z.}~\bibnamefont {Sun}}, \bibinfo {author}
		{\bibfnamefont {F.}~\bibnamefont {Xia}}, \bibinfo {author} {\bibfnamefont
			{E.-c.}\ \bibnamefont {Lin}}, \bibinfo {author} {\bibfnamefont {Y.-H.}\
			\bibnamefont {Lee}}, \bibinfo {author} {\bibfnamefont {S.}~\bibnamefont
			{K{\'e}na-Cohen}}, \ and\ \bibinfo {author} {\bibfnamefont {V.~M.}\
			\bibnamefont {Menon}},\ }\href@noop {} {\bibfield  {journal} {\bibinfo
			{journal} {Nat. Photonics}\ }\textbf {\bibinfo {volume} {9}},\ \bibinfo
		{pages} {30} (\bibinfo {year} {2015})}\BibitemShut {NoStop}%
	\bibitem [{\citenamefont {Basov}\ \emph {et~al.}(2016)\citenamefont {Basov},
		\citenamefont {Fogler},\ and\ \citenamefont
		{De~Abajo}}]{basov2016polaritons}%
	\BibitemOpen
	\bibfield  {author} {\bibinfo {author} {\bibfnamefont {D.}~\bibnamefont
			{Basov}}, \bibinfo {author} {\bibfnamefont {M.}~\bibnamefont {Fogler}}, \
		and\ \bibinfo {author} {\bibfnamefont {F.~G.}\ \bibnamefont {De~Abajo}},\
	}\href@noop {} {\bibfield  {journal} {\bibinfo  {journal} {Science}\ }\textbf
		{\bibinfo {volume} {354}} (\bibinfo {year} {2016})}\BibitemShut {NoStop}%
	\bibitem [{\citenamefont {Orgiu}\ \emph {et~al.}(2015)\citenamefont {Orgiu},
		\citenamefont {George}, \citenamefont {Hutchison}, \citenamefont {Devaux},
		\citenamefont {Dayen}, \citenamefont {Doudin}, \citenamefont {Stellacci},
		\citenamefont {Genet}, \citenamefont {Schachenmayer}, \citenamefont {Genes},
		\citenamefont {Pupillo}, \citenamefont {Samor{\`\i}},\ and\ \citenamefont
		{Ebbesen}}]{orgiu2015conductivity}%
	\BibitemOpen
	\bibfield  {author} {\bibinfo {author} {\bibfnamefont {E.}~\bibnamefont
			{Orgiu}}, \bibinfo {author} {\bibfnamefont {J.}~\bibnamefont {George}},
		\bibinfo {author} {\bibfnamefont {J.}~\bibnamefont {Hutchison}}, \bibinfo
		{author} {\bibfnamefont {E.}~\bibnamefont {Devaux}}, \bibinfo {author}
		{\bibfnamefont {J.}~\bibnamefont {Dayen}}, \bibinfo {author} {\bibfnamefont
			{B.}~\bibnamefont {Doudin}}, \bibinfo {author} {\bibfnamefont
			{F.}~\bibnamefont {Stellacci}}, \bibinfo {author} {\bibfnamefont
			{C.}~\bibnamefont {Genet}}, \bibinfo {author} {\bibfnamefont
			{J.}~\bibnamefont {Schachenmayer}}, \bibinfo {author} {\bibfnamefont
			{C.}~\bibnamefont {Genes}}, \bibinfo {author} {\bibfnamefont
			{G.}~\bibnamefont {Pupillo}}, \bibinfo {author} {\bibfnamefont
			{P.}~\bibnamefont {Samor{\`\i}}}, \ and\ \bibinfo {author} {\bibfnamefont
			{T.~W.}\ \bibnamefont {Ebbesen}},\ }\href@noop {} {\bibfield  {journal}
		{\bibinfo  {journal} {Nat. Mater.}\ }\textbf {\bibinfo {volume} {14}},\
		\bibinfo {pages} {1123} (\bibinfo {year} {2015})}\BibitemShut {NoStop}%
	\bibitem [{\citenamefont {Mergenthaler}\ \emph {et~al.}(2017)\citenamefont
		{Mergenthaler}, \citenamefont {Liu}, \citenamefont {Le~Roy}, \citenamefont
		{Ares}, \citenamefont {Thompson}, \citenamefont {Bogani}, \citenamefont
		{Luis}, \citenamefont {Blundell}, \citenamefont {Lancaster}, \citenamefont
		{Ardavan}, \citenamefont {Briggs}, \citenamefont {Leek},\ and\ \citenamefont
		{Laird}}]{mergenthaler2017strong}%
	\BibitemOpen
	\bibfield  {author} {\bibinfo {author} {\bibfnamefont {M.}~\bibnamefont
			{Mergenthaler}}, \bibinfo {author} {\bibfnamefont {J.}~\bibnamefont {Liu}},
		\bibinfo {author} {\bibfnamefont {J.~J.}\ \bibnamefont {Le~Roy}}, \bibinfo
		{author} {\bibfnamefont {N.}~\bibnamefont {Ares}}, \bibinfo {author}
		{\bibfnamefont {A.~L.}\ \bibnamefont {Thompson}}, \bibinfo {author}
		{\bibfnamefont {L.}~\bibnamefont {Bogani}}, \bibinfo {author} {\bibfnamefont
			{F.}~\bibnamefont {Luis}}, \bibinfo {author} {\bibfnamefont {S.~J.}\
			\bibnamefont {Blundell}}, \bibinfo {author} {\bibfnamefont {T.}~\bibnamefont
			{Lancaster}}, \bibinfo {author} {\bibfnamefont {A.}~\bibnamefont {Ardavan}},
		\bibinfo {author} {\bibfnamefont {G.~A.~D.}\ \bibnamefont {Briggs}}, \bibinfo
		{author} {\bibfnamefont {P.~J.}\ \bibnamefont {Leek}}, \ and\ \bibinfo
		{author} {\bibfnamefont {E.~A.}\ \bibnamefont {Laird}},\ }\href {\doibase
		10.1103/PhysRevLett.119.147701} {\bibfield  {journal} {\bibinfo  {journal}
			{Phys. Rev. Lett.}\ }\textbf {\bibinfo {volume} {119}},\ \bibinfo {pages}
		{147701} (\bibinfo {year} {2017})}\BibitemShut {NoStop}%
	\bibitem [{\citenamefont {Thomas}\ \emph {et~al.}(2019)\citenamefont {Thomas},
		\citenamefont {Devaux}, \citenamefont {Nagarajan}, \citenamefont {Chervy},
		\citenamefont {Seidel}, \citenamefont {Hagenm{\"u}ller}, \citenamefont
		{Sch{\"u}tz}, \citenamefont {Schachenmayer}, \citenamefont {Genet},
		\citenamefont {Pupillo},\ and\ \citenamefont
		{Ebbesen}}]{thomas2019exploring}%
	\BibitemOpen
	\bibfield  {author} {\bibinfo {author} {\bibfnamefont {A.}~\bibnamefont
			{Thomas}}, \bibinfo {author} {\bibfnamefont {E.}~\bibnamefont {Devaux}},
		\bibinfo {author} {\bibfnamefont {K.}~\bibnamefont {Nagarajan}}, \bibinfo
		{author} {\bibfnamefont {T.}~\bibnamefont {Chervy}}, \bibinfo {author}
		{\bibfnamefont {M.}~\bibnamefont {Seidel}}, \bibinfo {author} {\bibfnamefont
			{D.}~\bibnamefont {Hagenm{\"u}ller}}, \bibinfo {author} {\bibfnamefont
			{S.}~\bibnamefont {Sch{\"u}tz}}, \bibinfo {author} {\bibfnamefont
			{J.}~\bibnamefont {Schachenmayer}}, \bibinfo {author} {\bibfnamefont
			{C.}~\bibnamefont {Genet}}, \bibinfo {author} {\bibfnamefont
			{G.}~\bibnamefont {Pupillo}}, \ and\ \bibinfo {author} {\bibfnamefont
			{T.~W.}\ \bibnamefont {Ebbesen}},\ }\href@noop {} {\bibfield  {journal}
		{\bibinfo  {journal} {arXiv preprint arXiv:1911.01459}\ } (\bibinfo {year}
		{2019})}\BibitemShut {NoStop}%
	\bibitem [{\citenamefont {Laplace}\ \emph {et~al.}(2016)\citenamefont
		{Laplace}, \citenamefont {Fernandez-Pena}, \citenamefont {Gariglio},
		\citenamefont {Triscone},\ and\ \citenamefont
		{Cavalleri}}]{laplace2016proposed}%
	\BibitemOpen
	\bibfield  {author} {\bibinfo {author} {\bibfnamefont {Y.}~\bibnamefont
			{Laplace}}, \bibinfo {author} {\bibfnamefont {S.}~\bibnamefont
			{Fernandez-Pena}}, \bibinfo {author} {\bibfnamefont {S.}~\bibnamefont
			{Gariglio}}, \bibinfo {author} {\bibfnamefont {J.-M.}\ \bibnamefont
			{Triscone}}, \ and\ \bibinfo {author} {\bibfnamefont {A.}~\bibnamefont
			{Cavalleri}},\ }\href@noop {} {\bibfield  {journal} {\bibinfo  {journal}
			{Phys. Rev. B}\ }\textbf {\bibinfo {volume} {93}},\ \bibinfo {pages} {075152}
		(\bibinfo {year} {2016})}\BibitemShut {NoStop}%
	\bibitem [{\citenamefont {Kiffner}\ \emph {et~al.}(2019)\citenamefont
		{Kiffner}, \citenamefont {Coulthard}, \citenamefont {Schlawin}, \citenamefont
		{Ardavan},\ and\ \citenamefont {Jaksch}}]{kiffner2019mott}%
	\BibitemOpen
	\bibfield  {author} {\bibinfo {author} {\bibfnamefont {M.}~\bibnamefont
			{Kiffner}}, \bibinfo {author} {\bibfnamefont {J.}~\bibnamefont {Coulthard}},
		\bibinfo {author} {\bibfnamefont {F.}~\bibnamefont {Schlawin}}, \bibinfo
		{author} {\bibfnamefont {A.}~\bibnamefont {Ardavan}}, \ and\ \bibinfo
		{author} {\bibfnamefont {D.}~\bibnamefont {Jaksch}},\ }\href@noop {}
	{\bibfield  {journal} {\bibinfo  {journal} {New J. Phys}\ }\textbf {\bibinfo
			{volume} {21}},\ \bibinfo {pages} {073066} (\bibinfo {year}
		{2019})}\BibitemShut {NoStop}%
	\bibitem [{\citenamefont {Schachenmayer}\ \emph {et~al.}(2015)\citenamefont
		{Schachenmayer}, \citenamefont {Genes}, \citenamefont {Tignone},\ and\
		\citenamefont {Pupillo}}]{schachenmayer2015cavity}%
	\BibitemOpen
	\bibfield  {author} {\bibinfo {author} {\bibfnamefont {J.}~\bibnamefont
			{Schachenmayer}}, \bibinfo {author} {\bibfnamefont {C.}~\bibnamefont
			{Genes}}, \bibinfo {author} {\bibfnamefont {E.}~\bibnamefont {Tignone}}, \
		and\ \bibinfo {author} {\bibfnamefont {G.}~\bibnamefont {Pupillo}},\
	}\href@noop {} {\bibfield  {journal} {\bibinfo  {journal} {Phys. Rev. Lett.}\
		}\textbf {\bibinfo {volume} {114}},\ \bibinfo {pages} {196403} (\bibinfo
		{year} {2015})}\BibitemShut {NoStop}%
	\bibitem [{\citenamefont {Hagenm{\"u}ller}\ \emph {et~al.}(2018)\citenamefont
		{Hagenm{\"u}ller}, \citenamefont {Sch{\"u}tz}, \citenamefont {Schachenmayer},
		\citenamefont {Genes},\ and\ \citenamefont
		{Pupillo}}]{hagenmuller2018cavity}%
	\BibitemOpen
	\bibfield  {author} {\bibinfo {author} {\bibfnamefont {D.}~\bibnamefont
			{Hagenm{\"u}ller}}, \bibinfo {author} {\bibfnamefont {S.}~\bibnamefont
			{Sch{\"u}tz}}, \bibinfo {author} {\bibfnamefont {J.}~\bibnamefont
			{Schachenmayer}}, \bibinfo {author} {\bibfnamefont {C.}~\bibnamefont
			{Genes}}, \ and\ \bibinfo {author} {\bibfnamefont {G.}~\bibnamefont
			{Pupillo}},\ }\href@noop {} {\bibfield  {journal} {\bibinfo  {journal} {Phys.
				Rev. B}\ }\textbf {\bibinfo {volume} {97}},\ \bibinfo {pages} {205303}
		(\bibinfo {year} {2018})}\BibitemShut {NoStop}%
	\bibitem [{\citenamefont {Schlawin}\ \emph {et~al.}(2019)\citenamefont
		{Schlawin}, \citenamefont {Cavalleri},\ and\ \citenamefont
		{Jaksch}}]{schlawin2019cavity}%
	\BibitemOpen
	\bibfield  {author} {\bibinfo {author} {\bibfnamefont {F.}~\bibnamefont
			{Schlawin}}, \bibinfo {author} {\bibfnamefont {A.}~\bibnamefont {Cavalleri}},
		\ and\ \bibinfo {author} {\bibfnamefont {D.}~\bibnamefont {Jaksch}},\
	}\href@noop {} {\bibfield  {journal} {\bibinfo  {journal} {Phys. Rev. Lett.}\
		}\textbf {\bibinfo {volume} {122}},\ \bibinfo {pages} {133602} (\bibinfo
		{year} {2019})}\BibitemShut {NoStop}%
	\bibitem [{\citenamefont {Sentef}\ \emph {et~al.}(2018)\citenamefont {Sentef},
		\citenamefont {Ruggenthaler},\ and\ \citenamefont
		{Rubio}}]{sentef2018cavity}%
	\BibitemOpen
	\bibfield  {author} {\bibinfo {author} {\bibfnamefont {M.~A.}\ \bibnamefont
			{Sentef}}, \bibinfo {author} {\bibfnamefont {M.}~\bibnamefont
			{Ruggenthaler}}, \ and\ \bibinfo {author} {\bibfnamefont {A.}~\bibnamefont
			{Rubio}},\ }\href@noop {} {\bibfield  {journal} {\bibinfo  {journal} {Science
				advances}\ }\textbf {\bibinfo {volume} {4}},\ \bibinfo {pages} {eaau6969}
		(\bibinfo {year} {2018})}\BibitemShut {NoStop}%
	\bibitem [{\citenamefont {Sheikhan}\ and\ \citenamefont
		{Kollath}(2019)}]{sheikhan2019cavity}%
	\BibitemOpen
	\bibfield  {author} {\bibinfo {author} {\bibfnamefont {A.}~\bibnamefont
			{Sheikhan}}\ and\ \bibinfo {author} {\bibfnamefont {C.}~\bibnamefont
			{Kollath}},\ }\href@noop {} {\bibfield  {journal} {\bibinfo  {journal} {Phys.
				Rev. A}\ }\textbf {\bibinfo {volume} {99}},\ \bibinfo {pages} {053611}
		(\bibinfo {year} {2019})}\BibitemShut {NoStop}%
	\bibitem [{\citenamefont {Curtis}\ \emph {et~al.}(2019)\citenamefont {Curtis},
		\citenamefont {Raines}, \citenamefont {Allocca}, \citenamefont {Hafezi},\
		and\ \citenamefont {Galitski}}]{curtis2019cavity}%
	\BibitemOpen
	\bibfield  {author} {\bibinfo {author} {\bibfnamefont {J.~B.}\ \bibnamefont
			{Curtis}}, \bibinfo {author} {\bibfnamefont {Z.~M.}\ \bibnamefont {Raines}},
		\bibinfo {author} {\bibfnamefont {A.~A.}\ \bibnamefont {Allocca}}, \bibinfo
		{author} {\bibfnamefont {M.}~\bibnamefont {Hafezi}}, \ and\ \bibinfo {author}
		{\bibfnamefont {V.~M.}\ \bibnamefont {Galitski}},\ }\href@noop {} {\bibfield
		{journal} {\bibinfo  {journal} {Phys. Rev. Lett.}\ }\textbf {\bibinfo
			{volume} {122}},\ \bibinfo {pages} {167002} (\bibinfo {year}
		{2019})}\BibitemShut {NoStop}%
	\bibitem [{\citenamefont {Allocca}\ \emph {et~al.}(2019)\citenamefont
		{Allocca}, \citenamefont {Raines}, \citenamefont {Curtis},\ and\
		\citenamefont {Galitski}}]{allocca2019cavity}%
	\BibitemOpen
	\bibfield  {author} {\bibinfo {author} {\bibfnamefont {A.~A.}\ \bibnamefont
			{Allocca}}, \bibinfo {author} {\bibfnamefont {Z.~M.}\ \bibnamefont {Raines}},
		\bibinfo {author} {\bibfnamefont {J.~B.}\ \bibnamefont {Curtis}}, \ and\
		\bibinfo {author} {\bibfnamefont {V.~M.}\ \bibnamefont {Galitski}},\
	}\href@noop {} {\bibfield  {journal} {\bibinfo  {journal} {Phys. Rev. B}\
		}\textbf {\bibinfo {volume} {99}},\ \bibinfo {pages} {020504 (R)} (\bibinfo
		{year} {2019})}\BibitemShut {NoStop}%
	\bibitem [{\citenamefont {Li}\ and\ \citenamefont
		{Eckstein}(2020)}]{li2020manipulating}%
	\BibitemOpen
	\bibfield  {author} {\bibinfo {author} {\bibfnamefont {J.}~\bibnamefont
			{Li}}\ and\ \bibinfo {author} {\bibfnamefont {M.}~\bibnamefont {Eckstein}},\
	}\href@noop {} {\bibfield  {journal} {\bibinfo  {journal} {arXiv preprint
				arXiv:2005.07643}\ } (\bibinfo {year} {2020})}\BibitemShut {NoStop}%
	\bibitem [{\citenamefont {Ashida}\ \emph
		{et~al.}(2020{\natexlab{a}})\citenamefont {Ashida}, \citenamefont {Imamoglu},
		\citenamefont {Faist}, \citenamefont {Jaksch}, \citenamefont {Cavalleri},\
		and\ \citenamefont {Demler}}]{ashida2020quantum}%
	\BibitemOpen
	\bibfield  {author} {\bibinfo {author} {\bibfnamefont {Y.}~\bibnamefont
			{Ashida}}, \bibinfo {author} {\bibfnamefont {A.}~\bibnamefont {Imamoglu}},
		\bibinfo {author} {\bibfnamefont {J.}~\bibnamefont {Faist}}, \bibinfo
		{author} {\bibfnamefont {D.}~\bibnamefont {Jaksch}}, \bibinfo {author}
		{\bibfnamefont {A.}~\bibnamefont {Cavalleri}}, \ and\ \bibinfo {author}
		{\bibfnamefont {E.}~\bibnamefont {Demler}},\ }\href@noop {} {\bibfield
		{journal} {\bibinfo  {journal} {arXiv preprint arXiv:2003.13695}\ } (\bibinfo
		{year} {2020}{\natexlab{a}})}\BibitemShut {NoStop}%
	\bibitem [{\citenamefont {Gul{\'a}csi}\ and\ \citenamefont
		{D{\'o}ra}(2015)}]{gulacsi2015floquet}%
	\BibitemOpen
	\bibfield  {author} {\bibinfo {author} {\bibfnamefont {B.}~\bibnamefont
			{Gul{\'a}csi}}\ and\ \bibinfo {author} {\bibfnamefont {B.}~\bibnamefont
			{D{\'o}ra}},\ }\href@noop {} {\bibfield  {journal} {\bibinfo  {journal}
			{Phys. Rev. Lett.}\ }\textbf {\bibinfo {volume} {115}},\ \bibinfo {pages}
		{160402} (\bibinfo {year} {2015})}\BibitemShut {NoStop}%
	\bibitem [{\citenamefont {Mazza}\ and\ \citenamefont
		{Georges}(2019)}]{mazza2019superradiant}%
	\BibitemOpen
	\bibfield  {author} {\bibinfo {author} {\bibfnamefont {G.}~\bibnamefont
			{Mazza}}\ and\ \bibinfo {author} {\bibfnamefont {A.}~\bibnamefont
			{Georges}},\ }\href@noop {} {\bibfield  {journal} {\bibinfo  {journal} {Phys.
				Rev. Lett.}\ }\textbf {\bibinfo {volume} {122}},\ \bibinfo {pages} {017401}
		(\bibinfo {year} {2019})}\BibitemShut {NoStop}%
	\bibitem [{\citenamefont {Nataf}\ and\ \citenamefont
		{Ciuti}(2010)}]{nataf2010no}%
	\BibitemOpen
	\bibfield  {author} {\bibinfo {author} {\bibfnamefont {P.}~\bibnamefont
			{Nataf}}\ and\ \bibinfo {author} {\bibfnamefont {C.}~\bibnamefont {Ciuti}},\
	}\href@noop {} {\bibfield  {journal} {\bibinfo  {journal} {Nat. Commun.}\
		}\textbf {\bibinfo {volume} {1}},\ \bibinfo {pages} {72} (\bibinfo {year}
		{2010})}\BibitemShut {NoStop}%
	\bibitem [{\citenamefont {Viehmann}\ \emph {et~al.}(2011)\citenamefont
		{Viehmann}, \citenamefont {von Delft},\ and\ \citenamefont
		{Marquardt}}]{viehmann2011superradiant}%
	\BibitemOpen
	\bibfield  {author} {\bibinfo {author} {\bibfnamefont {O.}~\bibnamefont
			{Viehmann}}, \bibinfo {author} {\bibfnamefont {J.}~\bibnamefont {von Delft}},
		\ and\ \bibinfo {author} {\bibfnamefont {F.}~\bibnamefont {Marquardt}},\
	}\href@noop {} {\bibfield  {journal} {\bibinfo  {journal} {Phys. Rev. Lett.}\
		}\textbf {\bibinfo {volume} {107}},\ \bibinfo {pages} {113602} (\bibinfo
		{year} {2011})}\BibitemShut {NoStop}%
	\bibitem [{\citenamefont {
		}\ \emph {et~al.}(2019)\citenamefont
		{Andolina}, \citenamefont {Pellegrino}, \citenamefont {Giovannetti},
		\citenamefont {MacDonald},\ and\ \citenamefont
		{Polini}}]{andolina2019cavity}%
	\BibitemOpen
	\bibfield  {author} {\bibinfo {author} {\bibfnamefont {G.~M.}~\bibnamefont
			{Andolina}}, \bibinfo {author} {\bibfnamefont {F.~M.~D.}~\bibnamefont
			{Pellegrino}}, \bibinfo {author} {\bibfnamefont {V.}~\bibnamefont
			{Giovannetti}}, \bibinfo {author} {\bibfnamefont {A.~H.}~\bibnamefont
			{MacDonald}}, \ and\ \bibinfo {author} {\bibfnamefont {M.}~\bibnamefont
			{Polini}},\ }\href@noop {} {\bibfield  {journal} {\bibinfo  {journal} {Phys.
				Rev. B}\ }\textbf {\bibinfo {volume} {100}},\ \bibinfo {pages} {121109 (R)}
		(\bibinfo {year} {2019})}\BibitemShut {NoStop}%
	\bibitem [{\citenamefont {Nataf}\ \emph {et~al.}(2019)\citenamefont {Nataf},
		\citenamefont {Champel}, \citenamefont {Blatter},\ and\ \citenamefont
		{Basko}}]{nataf2019rashba}%
	\BibitemOpen
	\bibfield  {author} {\bibinfo {author} {\bibfnamefont {P.}~\bibnamefont
			{Nataf}}, \bibinfo {author} {\bibfnamefont {T.}~\bibnamefont {Champel}},
		\bibinfo {author} {\bibfnamefont {G.}~\bibnamefont {Blatter}}, \ and\
		\bibinfo {author} {\bibfnamefont {D.~M.}\ \bibnamefont {Basko}},\ }\href
	{\doibase 10.1103/PhysRevLett.123.207402} {\bibfield  {journal} {\bibinfo
			{journal} {Phys. Rev. Lett.}\ }\textbf {\bibinfo {volume} {123}},\ \bibinfo
		{pages} {207402} (\bibinfo {year} {2019})}\BibitemShut {NoStop}%
	\bibitem [{\citenamefont {Guerci}\ \emph {et~al.}(2020)\citenamefont {Guerci},
		\citenamefont {Simon},\ and\ \citenamefont {Mora}}]{guerci2020superradiant}%
	\BibitemOpen
	\bibfield  {author} {\bibinfo {author} {\bibfnamefont {D.}~\bibnamefont
			{Guerci}}, \bibinfo {author} {\bibfnamefont {P.}~\bibnamefont {Simon}}, \
		and\ \bibinfo {author} {\bibfnamefont {C.}~\bibnamefont {Mora}},\ }\href@noop
	{} {\bibfield  {journal} {\bibinfo  {journal} {Phys. Rev. Lett.}\ }\textbf
		{\bibinfo {volume} {125}},\ \bibinfo {pages} {257604} (\bibinfo {year}
		{2020})}\BibitemShut {NoStop}%
	\bibitem [{\citenamefont {Andolina}\ \emph {et~al.}(2020)\citenamefont
		{Andolina}, \citenamefont {Pellegrino}, \citenamefont {Giovannetti},
		\citenamefont {MacDonald},\ and\ \citenamefont
		{Polini}}]{andolina2020theory}%
	\BibitemOpen
	\bibfield  {author} {\bibinfo {author} {\bibfnamefont {G.~M.}~\bibnamefont
			{Andolina}}, \bibinfo {author} {\bibfnamefont {F.~M.~D.}~\bibnamefont
			{Pellegrino}}, \bibinfo {author} {\bibfnamefont {V.}~\bibnamefont
			{Giovannetti}}, \bibinfo {author} {\bibfnamefont {A.~H.}~\bibnamefont
			{MacDonald}}, \ and\ \bibinfo {author} {\bibfnamefont {M.}~\bibnamefont
			{Polini}},\ }\href@noop {} {\bibfield  {journal} {\bibinfo  {journal} {Phys.
				Rev. B}\ }\textbf {\bibinfo {volume} {102}},\ \bibinfo {pages} {125137}
		(\bibinfo {year} {2020})}\BibitemShut {NoStop}%
	\bibitem [{\citenamefont {Li}\ \emph {et~al.}(2020)\citenamefont {Li},
		\citenamefont {Golez}, \citenamefont {Mazza}, \citenamefont {Millis},
		\citenamefont {Georges},\ and\ \citenamefont
		{Eckstein}}]{li2020electromagnetic}%
	\BibitemOpen
	\bibfield  {author} {\bibinfo {author} {\bibfnamefont {J.}~\bibnamefont
			{Li}}, \bibinfo {author} {\bibfnamefont {D.}~\bibnamefont {Golez}}, \bibinfo
		{author} {\bibfnamefont {G.}~\bibnamefont {Mazza}}, \bibinfo {author}
		{\bibfnamefont {A.~J.}\ \bibnamefont {Millis}}, \bibinfo {author}
		{\bibfnamefont {A.}~\bibnamefont {Georges}}, \ and\ \bibinfo {author}
		{\bibfnamefont {M.}~\bibnamefont {Eckstein}},\ }\href@noop {} {\bibfield
		{journal} {\bibinfo  {journal} {Phys. Rev. B}\ }\textbf {\bibinfo {volume}
			{101}},\ \bibinfo {pages} {205140} (\bibinfo {year} {2020})}\BibitemShut
	{NoStop}%
	\bibitem [{\citenamefont {De~Bernardis}\ \emph
		{et~al.}(2018{\natexlab{a}})\citenamefont {De~Bernardis}, \citenamefont
		{Jaako},\ and\ \citenamefont {Rabl}}]{de2018cavity}%
	\BibitemOpen
	\bibfield  {author} {\bibinfo {author} {\bibfnamefont {D.}~\bibnamefont
			{De~Bernardis}}, \bibinfo {author} {\bibfnamefont {T.}~\bibnamefont {Jaako}},
		\ and\ \bibinfo {author} {\bibfnamefont {P.}~\bibnamefont {Rabl}},\ }\href
	{\doibase 10.1103/PhysRevA.97.043820} {\bibfield  {journal} {\bibinfo
			{journal} {Phys. Rev. A}\ }\textbf {\bibinfo {volume} {97}},\ \bibinfo
		{pages} {043820} (\bibinfo {year} {2018}{\natexlab{a}})}\BibitemShut
	{NoStop}%
	\bibitem [{\citenamefont {De~Bernardis}\ \emph
		{et~al.}(2018{\natexlab{b}})\citenamefont {De~Bernardis}, \citenamefont
		{Pilar}, \citenamefont {Jaako}, \citenamefont {De~Liberato},\ and\
		\citenamefont {Rabl}}]{de2018breakdown}%
	\BibitemOpen
	\bibfield  {author} {\bibinfo {author} {\bibfnamefont {D.}~\bibnamefont
			{De~Bernardis}}, \bibinfo {author} {\bibfnamefont {P.}~\bibnamefont {Pilar}},
		\bibinfo {author} {\bibfnamefont {T.}~\bibnamefont {Jaako}}, \bibinfo
		{author} {\bibfnamefont {S.}~\bibnamefont {De~Liberato}}, \ and\ \bibinfo
		{author} {\bibfnamefont {P.}~\bibnamefont {Rabl}},\ }\href@noop {} {\bibfield
		{journal} {\bibinfo  {journal} {Phys. Rev. A}\ }\textbf {\bibinfo {volume}
			{98}},\ \bibinfo {pages} {053819} (\bibinfo {year}
		{2018}{\natexlab{b}})}\BibitemShut {NoStop}%
	\bibitem [{\citenamefont {Stokes}\ and\ \citenamefont
		{Nazir}(2019)}]{stokes2019gauge}%
	\BibitemOpen
	\bibfield  {author} {\bibinfo {author} {\bibfnamefont {A.}~\bibnamefont
			{Stokes}}\ and\ \bibinfo {author} {\bibfnamefont {A.}~\bibnamefont {Nazir}},\
	}\href {\doibase 10.1038/s41467-018-08101-0} {\bibfield  {journal} {\bibinfo
			{journal} {Nat. Commun.}\ }\textbf {\bibinfo {volume} {10}},\ \bibinfo
		{pages} {499} (\bibinfo {year} {2019})}\BibitemShut {NoStop}%
	\bibitem [{\citenamefont {Ciuti}\ \emph {et~al.}(2005)\citenamefont {Ciuti},
		\citenamefont {Bastard},\ and\ \citenamefont {Carusotto}}]{ciuti2005quantum}%
	\BibitemOpen
	\bibfield  {author} {\bibinfo {author} {\bibfnamefont {C.}~\bibnamefont
			{Ciuti}}, \bibinfo {author} {\bibfnamefont {G.}~\bibnamefont {Bastard}}, \
		and\ \bibinfo {author} {\bibfnamefont {I.}~\bibnamefont {Carusotto}},\ }\href
	{\doibase 10.1103/PhysRevB.72.115303} {\bibfield  {journal} {\bibinfo
			{journal} {Phys. Rev. B}\ }\textbf {\bibinfo {volume} {72}},\ \bibinfo
		{pages} {115303} (\bibinfo {year} {2005})}\BibitemShut {NoStop}%
	\bibitem [{\citenamefont {Frisk~Kockum}\ \emph {et~al.}(2019)\citenamefont
		{Frisk~Kockum}, \citenamefont {Miranowicz}, \citenamefont {De~Liberato},
		\citenamefont {Savasta},\ and\ \citenamefont {Nori}}]{frisk2019ultrastrong}%
	\BibitemOpen
	\bibfield  {author} {\bibinfo {author} {\bibfnamefont {A.}~\bibnamefont
			{Frisk~Kockum}}, \bibinfo {author} {\bibfnamefont {A.}~\bibnamefont
			{Miranowicz}}, \bibinfo {author} {\bibfnamefont {S.}~\bibnamefont
			{De~Liberato}}, \bibinfo {author} {\bibfnamefont {S.}~\bibnamefont
			{Savasta}}, \ and\ \bibinfo {author} {\bibfnamefont {F.}~\bibnamefont
			{Nori}},\ }\href {\doibase 10.1038/s42254-018-0006-2} {\bibfield  {journal}
		{\bibinfo  {journal} {Nat. Rev. Phys.}\ }\textbf {\bibinfo {volume} {1}},\
		\bibinfo {pages} {19} (\bibinfo {year} {2019})}\BibitemShut {NoStop}%
	\bibitem [{\citenamefont {Forn-D\'{\i}az}\ \emph {et~al.}(2019)\citenamefont
		{Forn-D\'{\i}az}, \citenamefont {Lamata}, \citenamefont {Rico}, \citenamefont
		{Kono},\ and\ \citenamefont {Solano}}]{forndiaz2019ultrastrong}%
	\BibitemOpen
	\bibfield  {author} {\bibinfo {author} {\bibfnamefont {P.}~\bibnamefont
			{Forn-D\'{\i}az}}, \bibinfo {author} {\bibfnamefont {L.}~\bibnamefont
			{Lamata}}, \bibinfo {author} {\bibfnamefont {E.}~\bibnamefont {Rico}},
		\bibinfo {author} {\bibfnamefont {J.}~\bibnamefont {Kono}}, \ and\ \bibinfo
		{author} {\bibfnamefont {E.}~\bibnamefont {Solano}},\ }\href {\doibase
		10.1103/RevModPhys.91.025005} {\bibfield  {journal} {\bibinfo  {journal}
			{Rev. Mod. Phys.}\ }\textbf {\bibinfo {volume} {91}},\ \bibinfo {pages}
		{025005} (\bibinfo {year} {2019})}\BibitemShut {NoStop}%
	\bibitem [{\citenamefont {Settineri}\ \emph {et~al.}(2019)\citenamefont
		{Settineri}, \citenamefont {Di~Stefano}, \citenamefont {Zueco}, \citenamefont
		{Hughes}, \citenamefont {Savasta},\ and\ \citenamefont
		{Nori}}]{settineri2020gauge}%
	\BibitemOpen
	\bibfield  {author} {\bibinfo {author} {\bibfnamefont {A.}~\bibnamefont
			{Settineri}}, \bibinfo {author} {\bibfnamefont {O.}~\bibnamefont
			{Di~Stefano}}, \bibinfo {author} {\bibfnamefont {D.}~\bibnamefont {Zueco}},
		\bibinfo {author} {\bibfnamefont {S.}~\bibnamefont {Hughes}}, \bibinfo
		{author} {\bibfnamefont {S.}~\bibnamefont {Savasta}}, \ and\ \bibinfo
		{author} {\bibfnamefont {F.}~\bibnamefont {Nori}},\ }\href@noop {} {\bibfield
		{journal} {\bibinfo  {journal} {arXiv preprint arXiv:1912.08548}\ }
		(\bibinfo {year} {2019})}\BibitemShut {NoStop}%
	\bibitem [{\citenamefont {Ashida}\ \emph
		{et~al.}(2020{\natexlab{b}})\citenamefont {Ashida}, \citenamefont
		{Imamoglu},\ and\ \citenamefont {Demler}}]{ashida2020cavity}%
	\BibitemOpen
	\bibfield  {author} {\bibinfo {author} {\bibfnamefont {Y.}~\bibnamefont
			{Ashida}}, \bibinfo {author} {\bibfnamefont {A.}~\bibnamefont {Imamoglu}}, \
		and\ \bibinfo {author} {\bibfnamefont {E.}~\bibnamefont {Demler}},\
	}\href@noop {} {\bibfield  {journal} {\bibinfo  {journal} {arXiv preprint
				arXiv:2010.03583}\ } (\bibinfo {year} {2020}{\natexlab{b}})}\BibitemShut
	{NoStop}%
	\bibitem [{\citenamefont {Sch{\"a}fer}\ \emph {et~al.}(2020)\citenamefont
		{Sch{\"a}fer}, \citenamefont {Ruggenthaler}, \citenamefont {Rokaj},\ and\
		\citenamefont {Rubio}}]{schafer2020relevance}%
	\BibitemOpen
	\bibfield  {author} {\bibinfo {author} {\bibfnamefont {C.}~\bibnamefont
			{Sch{\"a}fer}}, \bibinfo {author} {\bibfnamefont {M.}~\bibnamefont
			{Ruggenthaler}}, \bibinfo {author} {\bibfnamefont {V.}~\bibnamefont {Rokaj}},
		\ and\ \bibinfo {author} {\bibfnamefont {A.}~\bibnamefont {Rubio}},\
	}\href@noop {} {\bibfield  {journal} {\bibinfo  {journal} {ACS photonics}\
		}\textbf {\bibinfo {volume} {7}},\ \bibinfo {pages} {975} (\bibinfo {year}
		{2020})}\BibitemShut {NoStop}%
	\bibitem [{\citenamefont {Di~Stefano}\ \emph {et~al.}(2019)\citenamefont
		{Di~Stefano}, \citenamefont {Settineri}, \citenamefont {Macr{\`\i}},
		\citenamefont {Garziano}, \citenamefont {Stassi}, \citenamefont {Savasta},\
		and\ \citenamefont {Nori}}]{di2019resolution}%
	\BibitemOpen
	\bibfield  {author} {\bibinfo {author} {\bibfnamefont {O.}~\bibnamefont
			{Di~Stefano}}, \bibinfo {author} {\bibfnamefont {A.}~\bibnamefont
			{Settineri}}, \bibinfo {author} {\bibfnamefont {V.}~\bibnamefont
			{Macr{\`\i}}}, \bibinfo {author} {\bibfnamefont {L.}~\bibnamefont
			{Garziano}}, \bibinfo {author} {\bibfnamefont {R.}~\bibnamefont {Stassi}},
		\bibinfo {author} {\bibfnamefont {S.}~\bibnamefont {Savasta}}, \ and\
		\bibinfo {author} {\bibfnamefont {F.}~\bibnamefont {Nori}},\ }\href@noop {}
	{\bibfield  {journal} {\bibinfo  {journal} {Nat. Phys.}\ }\textbf {\bibinfo
			{volume} {15}},\ \bibinfo {pages} {803} (\bibinfo {year} {2019})}\BibitemShut
	{NoStop}%
	\bibitem [{\citenamefont {Garziano}\ \emph {et~al.}(2020)\citenamefont
		{Garziano}, \citenamefont {Settineri}, \citenamefont {Di~Stefano},
		\citenamefont {Savasta},\ and\ \citenamefont {Nori}}]{garziano2020gauge}%
	\BibitemOpen
	\bibfield  {author} {\bibinfo {author} {\bibfnamefont {L.}~\bibnamefont
			{Garziano}}, \bibinfo {author} {\bibfnamefont {A.}~\bibnamefont {Settineri}},
		\bibinfo {author} {\bibfnamefont {O.}~\bibnamefont {Di~Stefano}}, \bibinfo
		{author} {\bibfnamefont {S.}~\bibnamefont {Savasta}}, \ and\ \bibinfo
		{author} {\bibfnamefont {F.}~\bibnamefont {Nori}},\ }\href {\doibase
		10.1103/PhysRevA.102.023718} {\bibfield  {journal} {\bibinfo  {journal}
			{Phys. Rev. A}\ }\textbf {\bibinfo {volume} {102}},\ \bibinfo {pages}
		{023718} (\bibinfo {year} {2020})}\BibitemShut {NoStop}%
	\bibitem [{\citenamefont {Reiche}\ and\ \citenamefont
		{Thomas}(1925)}]{reiche1925uber}%
	\BibitemOpen
	\bibfield  {author} {\bibinfo {author} {\bibfnamefont {F.}~\bibnamefont
			{Reiche}}\ and\ \bibinfo {author} {\bibfnamefont {W.}~\bibnamefont
			{Thomas}},\ }\href {\doibase 10.1007/BF01328494} {\bibfield  {journal}
		{\bibinfo  {journal} {Zeitschrift f{\"u}r Physik}\ }\textbf {\bibinfo
			{volume} {34}},\ \bibinfo {pages} {510} (\bibinfo {year} {1925})}\BibitemShut
	{NoStop}%
	\bibitem [{\citenamefont {Kuhn}(1925)}]{kuhn1925unter}%
	\BibitemOpen
	\bibfield  {author} {\bibinfo {author} {\bibfnamefont {W.}~\bibnamefont
			{Kuhn}},\ }\href {\doibase 10.1007/BF01328322} {\bibfield  {journal}
		{\bibinfo  {journal} {Zeitschrift f{\"u}r Physik}\ }\textbf {\bibinfo
			{volume} {33}},\ \bibinfo {pages} {408} (\bibinfo {year} {1925})}\BibitemShut
	{NoStop}%
	\bibitem [{\citenamefont {Savasta}\ \emph
		{et~al.}(2020{\natexlab{a}})\citenamefont {Savasta}, \citenamefont
		{Di~Stefano},\ and\ \citenamefont {Nori}}]{savasta2020trk}%
	\BibitemOpen
	\bibfield  {author} {\bibinfo {author} {\bibfnamefont {S.}~\bibnamefont
			{Savasta}}, \bibinfo {author} {\bibfnamefont {O.}~\bibnamefont {Di~Stefano}},
		\ and\ \bibinfo {author} {\bibfnamefont {F.}~\bibnamefont {Nori}},\
	}\href@noop {} {\bibfield  {journal} {\bibinfo  {journal} {arXiv preprint
				arXiv:2002.02139}\ } (\bibinfo {year} {2020}{\natexlab{a}})}\BibitemShut
	{NoStop}%
	\bibitem [{\citenamefont {Cottet}\ \emph {et~al.}(2015)\citenamefont {Cottet},
		\citenamefont {Kontos},\ and\ \citenamefont
		{Dou{\c{c}}ot}}]{cottet2015electron}%
	\BibitemOpen
	\bibfield  {author} {\bibinfo {author} {\bibfnamefont {A.}~\bibnamefont
			{Cottet}}, \bibinfo {author} {\bibfnamefont {T.}~\bibnamefont {Kontos}}, \
		and\ \bibinfo {author} {\bibfnamefont {B.}~\bibnamefont {Dou{\c{c}}ot}},\
	}\href@noop {} {\bibfield  {journal} {\bibinfo  {journal} {Phys. Rev. B}\
		}\textbf {\bibinfo {volume} {91}},\ \bibinfo {pages} {205417} (\bibinfo
		{year} {2015})}\BibitemShut {NoStop}%
	\bibitem [{\citenamefont {Babiker}\ and\ \citenamefont
		{Loudon}(1983)}]{babiker1983derivation}%
	\BibitemOpen
	\bibfield  {author} {\bibinfo {author} {\bibfnamefont {M.}~\bibnamefont
			{Babiker}}\ and\ \bibinfo {author} {\bibfnamefont {R.}~\bibnamefont
			{Loudon}},\ }\href@noop {} {\bibfield  {journal} {\bibinfo  {journal}
			{Proceedings of the Royal Society of London. A. Mathematical and Physical
				Sciences}\ }\textbf {\bibinfo {volume} {385}},\ \bibinfo {pages} {439}
		(\bibinfo {year} {1983})}\BibitemShut {NoStop}%
	\bibitem [{\citenamefont {Starace}(1971)}]{starace1971length}%
	\BibitemOpen
	\bibfield  {author} {\bibinfo {author} {\bibfnamefont {A.~F.}\ \bibnamefont
			{Starace}},\ }\href {\doibase 10.1103/PhysRevA.3.1242} {\bibfield  {journal}
		{\bibinfo  {journal} {Phys. Rev. A}\ }\textbf {\bibinfo {volume} {3}},\
		\bibinfo {pages} {1242} (\bibinfo {year} {1971})}\BibitemShut {NoStop}%
	\bibitem [{\citenamefont {Bassani}\ \emph {et~al.}(1977)\citenamefont
		{Bassani}, \citenamefont {Forney},\ and\ \citenamefont
		{Quattropani}}]{bassani1977choice}%
	\BibitemOpen
	\bibfield  {author} {\bibinfo {author} {\bibfnamefont {F.}~\bibnamefont
			{Bassani}}, \bibinfo {author} {\bibfnamefont {J.~J.}\ \bibnamefont {Forney}},
		\ and\ \bibinfo {author} {\bibfnamefont {A.}~\bibnamefont {Quattropani}},\
	}\href {\doibase 10.1103/PhysRevLett.39.1070} {\bibfield  {journal} {\bibinfo
			{journal} {Phys. Rev. Lett.}\ }\textbf {\bibinfo {volume} {39}},\ \bibinfo
		{pages} {1070} (\bibinfo {year} {1977})}\BibitemShut {NoStop}%
	\bibitem [{\citenamefont {Girlanda}\ \emph {et~al.}(1981)\citenamefont
		{Girlanda}, \citenamefont {Quattropani},\ and\ \citenamefont
		{Schwendimann}}]{girlanda1981twophoton}%
	\BibitemOpen
	\bibfield  {author} {\bibinfo {author} {\bibfnamefont {R.}~\bibnamefont
			{Girlanda}}, \bibinfo {author} {\bibfnamefont {A.}~\bibnamefont
			{Quattropani}}, \ and\ \bibinfo {author} {\bibfnamefont {P.}~\bibnamefont
			{Schwendimann}},\ }\href {\doibase 10.1103/PhysRevB.24.2009} {\bibfield
		{journal} {\bibinfo  {journal} {Phys. Rev. B}\ }\textbf {\bibinfo {volume}
			{24}},\ \bibinfo {pages} {2009} (\bibinfo {year} {1981})}\BibitemShut
	{NoStop}%
	\bibitem [{\citenamefont {Ismail-Beigi}\ \emph {et~al.}(2001)\citenamefont
		{Ismail-Beigi}, \citenamefont {Chang},\ and\ \citenamefont
		{Louie}}]{ismail2001coupling}%
	\BibitemOpen
	\bibfield  {author} {\bibinfo {author} {\bibfnamefont {S.}~\bibnamefont
			{Ismail-Beigi}}, \bibinfo {author} {\bibfnamefont {E.~K.}\ \bibnamefont
			{Chang}}, \ and\ \bibinfo {author} {\bibfnamefont {S.~G.}\ \bibnamefont
			{Louie}},\ }\href {\doibase 10.1103/PhysRevLett.87.087402} {\bibfield
		{journal} {\bibinfo  {journal} {Phys. Rev. Lett.}\ }\textbf {\bibinfo
			{volume} {87}},\ \bibinfo {pages} {087402} (\bibinfo {year}
		{2001})}\BibitemShut {NoStop}%
	\bibitem [{\citenamefont {Boykin}\ \emph {et~al.}(2001)\citenamefont {Boykin},
		\citenamefont {Bowen},\ and\ \citenamefont
		{Klimeck}}]{boykin2001electromagnetic}%
	\BibitemOpen
	\bibfield  {author} {\bibinfo {author} {\bibfnamefont {T.~B.}\ \bibnamefont
			{Boykin}}, \bibinfo {author} {\bibfnamefont {R.~C.}\ \bibnamefont {Bowen}}, \
		and\ \bibinfo {author} {\bibfnamefont {G.}~\bibnamefont {Klimeck}},\
	}\href@noop {} {\bibfield  {journal} {\bibinfo  {journal} {Phys. Rev. B}\
		}\textbf {\bibinfo {volume} {63}},\ \bibinfo {pages} {245314} (\bibinfo
		{year} {2001})}\BibitemShut {NoStop}%
	\bibitem [{\citenamefont {Marzari}\ \emph {et~al.}(2012)\citenamefont
		{Marzari}, \citenamefont {Mostofi}, \citenamefont {Yates}, \citenamefont
		{Souza},\ and\ \citenamefont {Vanderbilt}}]{marzari2012maximally}%
	\BibitemOpen
	\bibfield  {author} {\bibinfo {author} {\bibfnamefont {N.}~\bibnamefont
			{Marzari}}, \bibinfo {author} {\bibfnamefont {A.~A.}\ \bibnamefont
			{Mostofi}}, \bibinfo {author} {\bibfnamefont {J.~R.}\ \bibnamefont {Yates}},
		\bibinfo {author} {\bibfnamefont {I.}~\bibnamefont {Souza}}, \ and\ \bibinfo
		{author} {\bibfnamefont {D.}~\bibnamefont {Vanderbilt}},\ }\href {\doibase
		10.1103/RevModPhys.84.1419} {\bibfield  {journal} {\bibinfo  {journal} {Rev.
				Mod. Phys.}\ }\textbf {\bibinfo {volume} {84}},\ \bibinfo {pages} {1419}
		(\bibinfo {year} {2012})}\BibitemShut {NoStop}%
	\bibitem [{\citenamefont {Lechermann}\ \emph {et~al.}(2006)\citenamefont
		{Lechermann}, \citenamefont {Georges}, \citenamefont {Poteryaev},
		\citenamefont {Biermann}, \citenamefont {Posternak}, \citenamefont
		{Yamasaki},\ and\ \citenamefont {Andersen}}]{lecherman2006dynamical}%
	\BibitemOpen
	\bibfield  {author} {\bibinfo {author} {\bibfnamefont {F.}~\bibnamefont
			{Lechermann}}, \bibinfo {author} {\bibfnamefont {A.}~\bibnamefont {Georges}},
		\bibinfo {author} {\bibfnamefont {A.}~\bibnamefont {Poteryaev}}, \bibinfo
		{author} {\bibfnamefont {S.}~\bibnamefont {Biermann}}, \bibinfo {author}
		{\bibfnamefont {M.}~\bibnamefont {Posternak}}, \bibinfo {author}
		{\bibfnamefont {A.}~\bibnamefont {Yamasaki}}, \ and\ \bibinfo {author}
		{\bibfnamefont {O.~K.}\ \bibnamefont {Andersen}},\ }\href {\doibase
		10.1103/PhysRevB.74.125120} {\bibfield  {journal} {\bibinfo  {journal} {Phys.
				Rev. B}\ }\textbf {\bibinfo {volume} {74}},\ \bibinfo {pages} {125120}
		(\bibinfo {year} {2006})}\BibitemShut {NoStop}%
	\bibitem [{\citenamefont {Amadon}\ \emph {et~al.}(2008)\citenamefont {Amadon},
		\citenamefont {Lechermann}, \citenamefont {Georges}, \citenamefont {Jollet},
		\citenamefont {Wehling},\ and\ \citenamefont
		{Lichtenstein}}]{amadon2008plane}%
	\BibitemOpen
	\bibfield  {author} {\bibinfo {author} {\bibfnamefont {B.}~\bibnamefont
			{Amadon}}, \bibinfo {author} {\bibfnamefont {F.}~\bibnamefont {Lechermann}},
		\bibinfo {author} {\bibfnamefont {A.}~\bibnamefont {Georges}}, \bibinfo
		{author} {\bibfnamefont {F.}~\bibnamefont {Jollet}}, \bibinfo {author}
		{\bibfnamefont {T.~O.}\ \bibnamefont {Wehling}}, \ and\ \bibinfo {author}
		{\bibfnamefont {A.~I.}\ \bibnamefont {Lichtenstein}},\ }\href {\doibase
		10.1103/PhysRevB.77.205112} {\bibfield  {journal} {\bibinfo  {journal} {Phys.
				Rev. B}\ }\textbf {\bibinfo {volume} {77}},\ \bibinfo {pages} {205112}
		(\bibinfo {year} {2008})}\BibitemShut {NoStop}%
	\bibitem [{\citenamefont {Savasta}\ \emph
		{et~al.}(2020{\natexlab{b}})\citenamefont {Savasta}, \citenamefont
		{Di~Stefano}, \citenamefont {Settineri}, \citenamefont {Zueco}, \citenamefont
		{Hughes},\ and\ \citenamefont {Nori}}]{savasta2020gauge}%
	\BibitemOpen
	\bibfield  {author} {\bibinfo {author} {\bibfnamefont {S.}~\bibnamefont
			{Savasta}}, \bibinfo {author} {\bibfnamefont {O.}~\bibnamefont {Di~Stefano}},
		\bibinfo {author} {\bibfnamefont {A.}~\bibnamefont {Settineri}}, \bibinfo
		{author} {\bibfnamefont {D.}~\bibnamefont {Zueco}}, \bibinfo {author}
		{\bibfnamefont {S.}~\bibnamefont {Hughes}}, \ and\ \bibinfo {author}
		{\bibfnamefont {F.}~\bibnamefont {Nori}},\ }\href@noop {} {\bibfield
		{journal} {\bibinfo  {journal} {arXiv preprint arXiv:2006.06583}\ } (\bibinfo
		{year} {2020}{\natexlab{b}})}\BibitemShut {NoStop}%
	\bibitem [{\citenamefont {Paul}\ and\ \citenamefont
		{Kotliar}(2003)}]{paul2003thermal}%
	\BibitemOpen
	\bibfield  {author} {\bibinfo {author} {\bibfnamefont {I.}~\bibnamefont
			{Paul}}\ and\ \bibinfo {author} {\bibfnamefont {G.}~\bibnamefont {Kotliar}},\
	}\href@noop {} {\bibfield  {journal} {\bibinfo  {journal} {Phys. Rev. B}\
		}\textbf {\bibinfo {volume} {67}},\ \bibinfo {pages} {115131} (\bibinfo
		{year} {2003})}\BibitemShut {NoStop}%
	\bibitem [{\citenamefont {Millis}(2001)}]{millis2001optical}%
	\BibitemOpen
	\bibfield  {author} {\bibinfo {author} {\bibfnamefont {A.~J.}\ \bibnamefont
			{Millis}},\ }\href@noop {} {\bibfield  {journal} {\bibinfo  {journal} {J.
				Electron Spectrosc. Relat. Phenomen.}\ }\textbf {\bibinfo {volume}
			{114-116}},\ \bibinfo {pages} {669} (\bibinfo {year} {2001})}\BibitemShut
	{NoStop}%
	\bibitem [{\citenamefont {Tomczak}\ and\ \citenamefont
		{Biermann}(2009)}]{tomczak2009optical}%
	\BibitemOpen
	\bibfield  {author} {\bibinfo {author} {\bibfnamefont {J.~M.}\ \bibnamefont
			{Tomczak}}\ and\ \bibinfo {author} {\bibfnamefont {S.}~\bibnamefont
			{Biermann}},\ }\href {\doibase 10.1103/PhysRevB.80.085117} {\bibfield
		{journal} {\bibinfo  {journal} {Phys. Rev. B}\ }\textbf {\bibinfo {volume}
			{80}},\ \bibinfo {pages} {085117} (\bibinfo {year} {2009})}\BibitemShut
	{NoStop}%
	\bibitem [{\citenamefont {Wissgott}\ \emph {et~al.}(2012)\citenamefont
		{Wissgott}, \citenamefont {Kune\ifmmode~\check{s}\else \v{s}\fi{}},
		\citenamefont {Toschi},\ and\ \citenamefont {Held}}]{wissgott2012dipole}%
	\BibitemOpen
	\bibfield  {author} {\bibinfo {author} {\bibfnamefont {P.}~\bibnamefont
			{Wissgott}}, \bibinfo {author} {\bibfnamefont {J.}~\bibnamefont
			{Kune\ifmmode~\check{s}\else \v{s}\fi{}}}, \bibinfo {author} {\bibfnamefont
			{A.}~\bibnamefont {Toschi}}, \ and\ \bibinfo {author} {\bibfnamefont
			{K.}~\bibnamefont {Held}},\ }\href {\doibase 10.1103/PhysRevB.85.205133}
	{\bibfield  {journal} {\bibinfo  {journal} {Phys. Rev. B}\ }\textbf {\bibinfo
			{volume} {85}},\ \bibinfo {pages} {205133} (\bibinfo {year}
		{2012})}\BibitemShut {NoStop}%
	\bibitem [{\citenamefont {Sch{\"u}ler}\ \emph {et~al.}(2021)\citenamefont
		{Sch{\"u}ler}, \citenamefont {Marks}, \citenamefont {Murakami}, \citenamefont
		{Jia},\ and\ \citenamefont {Devereaux}}]{schuler2021gauge}%
	\BibitemOpen
	\bibfield  {author} {\bibinfo {author} {\bibfnamefont {M.}~\bibnamefont
			{Sch{\"u}ler}}, \bibinfo {author} {\bibfnamefont {J.~A.}\ \bibnamefont
			{Marks}}, \bibinfo {author} {\bibfnamefont {Y.}~\bibnamefont {Murakami}},
		\bibinfo {author} {\bibfnamefont {C.}~\bibnamefont {Jia}}, \ and\ \bibinfo
		{author} {\bibfnamefont {T.~P.}\ \bibnamefont {Devereaux}},\ }\href@noop {}
	{\bibfield  {journal} {\bibinfo  {journal} {arXiv preprint arXiv:2101.01143}\
		} (\bibinfo {year} {2021})}\BibitemShut {NoStop}%
	\bibitem [{\citenamefont {Skolimowski}\ \emph {et~al.}(2020)\citenamefont
		{Skolimowski}, \citenamefont {Amaricci},\ and\ \citenamefont
		{Fabrizio}}]{misuse2020skolimowski}%
	\BibitemOpen
	\bibfield  {author} {\bibinfo {author} {\bibfnamefont {J.}~\bibnamefont
			{Skolimowski}}, \bibinfo {author} {\bibfnamefont {A.}~\bibnamefont
			{Amaricci}}, \ and\ \bibinfo {author} {\bibfnamefont {M.}~\bibnamefont
			{Fabrizio}},\ }\href {\doibase 10.1103/PhysRevB.101.121104} {\bibfield
		{journal} {\bibinfo  {journal} {Phys. Rev. B}\ }\textbf {\bibinfo {volume}
			{101}},\ \bibinfo {pages} {121104} (\bibinfo {year} {2020})}\BibitemShut
	{NoStop}%
	\bibitem [{\citenamefont {Todorov}\ and\ \citenamefont
		{Sirtori}(2012)}]{todorov2012intersubband}%
	\BibitemOpen
	\bibfield  {author} {\bibinfo {author} {\bibfnamefont {Y.}~\bibnamefont
			{Todorov}}\ and\ \bibinfo {author} {\bibfnamefont {C.}~\bibnamefont
			{Sirtori}},\ }\href {\doibase 10.1103/PhysRevB.85.045304} {\bibfield
		{journal} {\bibinfo  {journal} {Phys. Rev. B}\ }\textbf {\bibinfo {volume}
			{85}},\ \bibinfo {pages} {045304} (\bibinfo {year} {2012})}\BibitemShut
	{NoStop}%
	\bibitem [{\citenamefont {Lenk}\ and\ \citenamefont
		{Eckstein}(2020)}]{lenk2020collective}%
	\BibitemOpen
	\bibfield  {author} {\bibinfo {author} {\bibfnamefont {K.}~\bibnamefont
			{Lenk}}\ and\ \bibinfo {author} {\bibfnamefont {M.}~\bibnamefont
			{Eckstein}},\ }\href@noop {} {\bibfield  {journal} {\bibinfo  {journal}
			{Phys. Rev. B}\ }\textbf {\bibinfo {volume} {102}},\ \bibinfo {pages}
		{205129} (\bibinfo {year} {2020})}\BibitemShut {NoStop}%
	\bibitem [{\citenamefont {Sentef}\ \emph {et~al.}(2020)\citenamefont {Sentef},
		\citenamefont {Li}, \citenamefont {K\"unzel},\ and\ \citenamefont
		{Eckstein}}]{sentef2020quantum}%
	\BibitemOpen
	\bibfield  {author} {\bibinfo {author} {\bibfnamefont {M.~A.}\ \bibnamefont
			{Sentef}}, \bibinfo {author} {\bibfnamefont {J.}~\bibnamefont {Li}}, \bibinfo
		{author} {\bibfnamefont {F.}~\bibnamefont {K\"unzel}}, \ and\ \bibinfo
		{author} {\bibfnamefont {M.}~\bibnamefont {Eckstein}},\ }\href {\doibase
		10.1103/PhysRevResearch.2.033033} {\bibfield  {journal} {\bibinfo  {journal}
			{Phys. Rev. Research}\ }\textbf {\bibinfo {volume} {2}},\ \bibinfo {pages}
		{033033} (\bibinfo {year} {2020})}\BibitemShut {NoStop}%
	\bibitem [{\citenamefont {Lerose}\ \emph {et~al.}(2019)\citenamefont {Lerose},
		\citenamefont {{\v{Z}}unkovi{\v{c}}}, \citenamefont {Marino}, \citenamefont
		{Gambassi},\ and\ \citenamefont {Silva}}]{lerose2019impact}%
	\BibitemOpen
	\bibfield  {author} {\bibinfo {author} {\bibfnamefont {A.}~\bibnamefont
			{Lerose}}, \bibinfo {author} {\bibfnamefont {B.}~\bibnamefont
			{{\v{Z}}unkovi{\v{c}}}}, \bibinfo {author} {\bibfnamefont {J.}~\bibnamefont
			{Marino}}, \bibinfo {author} {\bibfnamefont {A.}~\bibnamefont {Gambassi}}, \
		and\ \bibinfo {author} {\bibfnamefont {A.}~\bibnamefont {Silva}},\
	}\href@noop {} {\bibfield  {journal} {\bibinfo  {journal} {Phys. Rev. B}\
		}\textbf {\bibinfo {volume} {99}},\ \bibinfo {pages} {045128} (\bibinfo
		{year} {2019})}\BibitemShut {NoStop}%
	\bibitem [{\citenamefont {De~Liberato}(2014)}]{deliberato2014light}%
	\BibitemOpen
	\bibfield  {author} {\bibinfo {author} {\bibfnamefont {S.}~\bibnamefont
			{De~Liberato}},\ }\href@noop {} {\bibfield  {journal} {\bibinfo  {journal}
			{Phys. Rev. Lett.}\ }\textbf {\bibinfo {volume} {112}},\ \bibinfo {pages}
		{016401} (\bibinfo {year} {2014})}\BibitemShut {NoStop}%
\end{thebibliography}
\end{document}